\documentclass[prb,showpacs,preprintnumbers,amsfonts,amssymb,amsmath,floats,twocolumn,aps]{revtex4}

\usepackage{graphicx}
\usepackage{dcolumn}
\usepackage{bm}
\usepackage[final,dvips]{epsfig}
\usepackage{bm}
\usepackage{color}

\newcommand{\be}{\begin{equation}}
\newcommand{\ee}{\end{equation}}
\newcommand{\ba}{\begin{eqnarray}}
\newcommand{\ea}{\end{eqnarray}}
\newcommand{\ff}[1]{{\bm #1}}
\newcommand{\tr}{\mbox{tr}}
\newcommand{\Tr}{\mbox{Tr}}
\newcommand{\refeq}[1]{Eq.\ (\ref{eq:#1})}
\newcommand{\labeq}[1]{\label{eq:#1}}

\begin{document} 
  
\title[SFT for systems of interacting electrons with disorder]
      {Self-energy-functional theory for systems of interacting electrons with disorder} 

\author{Michael Potthoff and Matthias Balzer}

\affiliation{
Institut f\"ur Theoretische Physik und Astrophysik, 
Universit\"at W\"urzburg, Am Hubland, D-97074 W\"urzburg, Germany
}
 
\begin{abstract}
Based on a functional-integral formalism, a generalization of the self-energy-functional
theory (SFT) is proposed which is applicable to systems of interacting electrons with 
disorder.
Similar to the pure case without disorder, a variational principle is set up which gives 
the physical (disorder) self-energy as a stationary point of the (averaged) grand potential. 
Although the resulting self-energy functional turns out to be more complicated, the 
formal structure of the theory can be retained since the unknown part of the functional
is universal.
This allows to construct non-perturbative and thermodynamically consistent approximations
via searching for a stationary point on a restricted domain of the functional.
The theory and the possible approximations are worked out for models with local 
interactions and local disorder. 
This results in a derivation of different mean-field approaches and various cluster 
extensions, including well-known concepts as the statistical dynamical mean-field 
theory, the molecular coherent-potential approximation and the dynamical cluster 
approximation. 
Due to the common formal framework provided by the SFT, one achieves a general 
systematization of dynamical approaches, i.e.\ approaches based on the spectrum of 
one-particle excitations.
New mean-field and new cluster schemes naturally appear in this framework and complement
the existing ones.
Their prospects for future applications are discussed.
\end{abstract} 
 
%71.10.-w 	Theories and models of many-electron systems
%71.10.Fd 	Lattice fermion models (Hubbard model, etc.)
%71.23.-k 	Electronic structure of disordered solids

\maketitle 

\section{Introduction}
\label{sec:intro}
 
The combined influence of electron-electron interaction and of disorder on 
material properties represents a central question of solid-state theory.
In diluted magnetic semiconductors \cite{Die02} like Ga$_{1-x}$Mn$_x$As or
Zn$_{1-x}$Mn$_x$Se, the magnetic properties, such as the Curie temperature, 
sensitively depend on the (random) distribution of the Mn ions as well as
on the type and the strength of their effective magnetic interaction which
results from strong Coulomb interaction among the Mn 3d valence electrons.
Several transition-metal oxides \cite{IFT98} with partially filled metal 
3d shells (e.g.\ manganites or cuprates) are antiferromagnetic Mott or 
charge-transfer insulators and exhibit a rich phase diagram upon doping with 
charge carriers. 
The disorder potential introduced due to the substitution process considerably 
affects their magnetic, charge and orbital ordering.

While these examples show the need for a comprehensive theory of interacting
and disordered electron systems, they also demonstrate the immense complexity
one faces in any theoretical approach. 
Even strongly simplified (Anderson-Hubbard-type) models with local interactions 
and local disorder only, are highly non-trivial if studied in a regime which 
excludes a simple perturbative treatment.
\cite{LR85,VW92,BK94,GO03,MD05}

For three (and higher) dimensions, one may in first place focus on the {\em local} 
charge and spin dynamics of the electrons and, complementary to scaling theories,
\cite{LR85} disregard the long-wavelength modes which govern the immediate vicinity 
of a phase transition.
In this context, mean-field approaches and cluster extensions are well justified. 
The mean-field concept is formally valid in the limit of high dimensions.
Subsequent cluster extensions are suited to reincorporate short-range correlations
which are neglected in the purely local mean-field approach.
Clearly, a mean-field treatment excludes important effects such as the destruction 
of long-range order due to thermal order-parameter fluctuations \cite{MW66} or Anderson 
localization, \cite{And58} for example. 
Nevertheless, tractable mean-field theories can be valuable tools for an understanding 
of interacting and disordered systems with different competing orders
and complex phase diagrams.

A mean-field theory can be formulated on the level of Hamiltonians and electronic 
states. 
This yields simple approaches such as the Hartree-Fock appoximation to treat the 
interaction part and the virtual-crystal approximation to treat the disorder part 
of the problem. \cite{FW71,Gon92}
These are completely static theories which in addition to spatial also neglect 
temporal fluctuations.

Temporal degrees of freedom can be taken into account in a mean-field theory when
this is based on the spectrum of excitations. 
Placing the one-particle Green's function in the center of interest, results in
a mean-field theory which is distinguished by the fact that it yields the {\em exact} 
result in the limit of infinite spatial dimensions.
With a proper scaling of the model parameters this limit preserves a highly non-trivial
dynamics. \cite{MV89,VV92}
This distinguished mean-field theory, for the interaction part of the problem, is 
the dynamical mean-field theory (DMFT). \cite{GK92a,Jar92,GKKR96,KV04}
It gives the exact (local) interaction self-energy of the prototypical Hubbard model 
\cite{Hub63,Gut63,Kan63} in the $D=\infty$ limit. 
For the disorder part, the coherent-potential approximation (CPA)
\cite{Sov67,Tay67,VKE68,EKL74} gives the exact (local) disorder self-energy of the
disorder Anderson model \cite{And58} in the $D=\infty$ limit. 

Phenomena depending on dimension are missed in a local mean-field approach but can
be restored step by step using cluster expansions. \cite{MJPH05,Gon92}
Using a single-site mean-field theory as a starting point for a systematic expansion
is surely an inadequate approach to include long-wavelength modes and their effects.
The main motivation for the subsequent inclusion of spatial correlations in cluster
theories is rather the expected rapid convergence of {\em local} observables.

The purpose of the present paper is to contribute to a systematization of mean-field
approaches and their cluster extensions in the combined case of interactions and 
disorder and to explore new approximation schemes.
The strategy is to seek for a proper generalization of the self-energy-functional
theory (SFT) developed recently. \cite{Pot03a,Pot03b}
For the pure (disorder-free) case, it has been shown that different mean-field and
cluster approaches are recovered, and new approximations can be constructed in a 
systematic way which guarantees thermodynamical consistency. \cite{PAD03,Pot05}
Here, we describe a novel derivation of the SFT which is non-perturbative, i.e.\ a 
formulation which does not refer to formal sums of skeleton diagrams.
This formulation is well suited for an extension of the theory to disordered (and 
interacting) systems.
The generalized SFT is worked out in detail. 
It is shown that it makes contact with (i.e.\ rederives) a number of previous approaches: 

(i) the DMFT+CPA put forward by Jani\u{s} and Vollhardt \cite{JV92b} and by 
Dobrosavljevi\'c{} and Kotliar, \cite{DK93b,DK94} which has recently been used to
study metal-insulator transitions \cite{BHV04} at non-integer filling and the effects
of disorder on magnetism, \cite{UJV95,BUV03,BU05}

(ii) the local distribution approach of Abou-Chacra, Anderson and Thouless \cite{AAT73}
which has recently been evaluated numerically by Alvermann and Fehske \cite{AF04}
and which is the conceptual basis for 

(iii) the statistical DMFT proposed by Dobrosavljevi\'c{} and Kotliar \cite{DK97,DK98}
with several recent applications, \cite{BAF04} e.g.\ to strongly coupled disordered 
electron-phonon systems, 
 
(iv) the molecular CPA \cite{Gon92} and its combination with the cellular DMFT (C-DMFT)
\cite{KSPB01,LK00} of Kotliar et al.\ and Lichtenstein and Katsnelson,

(v) the dynamical cluster approximation (DCA) for disordered systems as introduced
by Jarrell, Krishnamurthy and Maier, \cite{JK01,MJ02b}

(vi) the disorder analog of a simplified DCA recently introduced 
by Minh-Tien \cite{MT06}

as well as with several variants of these approaches -- such as the typical medium theory (TMT)
\cite{DPN03} involving the geometrical averaging of the local density of states which has been 
suggested by Dobrosavljevi\'c{}, Pastor and Nikoli\'c and applied in combination with DMFT by 
Byczuk, Hofstetter and Vollhardt. \cite{BHV05,Byc05}

The construction of a generalized SFT provides {\em a unified theoretical framework} 
which is able to rederive and thereby to classify the above-mentioned approximations
(i)--(iv).
This procedure automatically discloses the view on new approximations:
Generalizations of the periodized cellular DMFT (PC-DMFT) \cite{BPK04} and of the 
cluster-perturbation theory (CPT) \cite{GV93,SPPL00} as well as the variational cluster
approach (VCA) \cite{PAD03,DAH+04,AEvdLP+04} are suggested for disordered (and interacting) 
systems.

The main intention of the paper is to work out the formal concepts.
The benchmarking and application of the different approaches requires a numerical
implementation which is beyond the present scope but intended for the future.

The paper is organized as follows:
The next Sec.\ \ref{sec:ham} introduces a number of basic quantities needed
for the subsequent construction of the self-energy functionals. 
In Sec.\ \ref{sec:funcint} the self-energy functional for pure systems is derived
non-perturbatively within the functional-integral formalism.
A brief {\em general} discussion of approximations follows in Sec.\ \ref{sec:appfix} for
the case of a fixed disorder configuration.
This provides the basis for the statistical SFT (statSFT) in Sec.\ \ref{sec:statsft} and
for the statistical DMFT in particular. 
The main ideas for the construction of the generalized self-energy functional of configuration
independent self-energies are provided in Sec.\ \ref{sec:funcdis} while
Sec.\ \ref{sec:app} shows how to generate consistent approximations.
The specialization to limiting cases, in particular to the disordered but non-interacting 
electron system is given in Sec.\ \ref{sec:lim}. This conclude the general build-up of the
theory. The case of disorder in the interaction part is briefly sketched in the 
Appendix \ref{sec:udis}.
In the rest of the paper several concrete approximations are derived and classified. 
This includes well-known but also new approximation schemes.
Mean-field approximations are disussed in Sec.\ \ref{sec:sys},
cluster approximations in Sec.\ \ref{sec:cluster}.
A summary and a discussion of general topics in Sec.\ \ref{sec:dis} concludes the paper.

\section{Hamiltonian and dynamic quantities}
\label{sec:ham}

We consider a system of fermions in equilibrium at temperature $T$ and
chemical potential $\mu$. 
In the grand-canonical ensemble the macrostate of the system is given 
by the density operator 
\be
  \rho = \frac
  {\exp(-(H - \mu N)/T) }
  {  \tr \exp(-(H - \mu N)/T) } \: ,
\ee
where $N$ is the total particle-number operator, and $H$ is the Hamiltonian. 
$H$ is assumed to consist of a free (bilinear) part $H_0$ which exhibits the
(discrete) translational symmetries of an underlying $D$-dimensional lattice, 
a disorder potential $H_{\rm dis}$, and an interaction part $H_{\rm int}$:
\be 
  H = H(\ff t, \ff \eta, \ff U) = H_0(\ff t) + H_{\rm dis}(\ff \eta) 
    + H_{\rm int}(\ff U) \: .
\ee

The free part
\be
  H_0(\ff t) = \sum_{\alpha\beta} t_{\alpha\beta} \:
  c^\dagger_\alpha c_\beta 
\ee
is characterized by a set of hopping parameters $t_{\alpha\beta}$
where an index $\alpha$ labels the states of an orthonormal one-particle basis
$\{ | \alpha \rangle \}$. 
Typically, $\alpha$ refers to the sites $\ff x$ of the lattice as well as to some 
local degrees of freedom (e.g.\ spin projection $\sigma=\uparrow,\downarrow$), 
i.e.\ $\alpha=(\ff x,\sigma)$.
The full hopping matrix with elements $t_{\alpha\beta}$ is denoted by $\ff t$.

The interaction part
\be
  H_{\rm int}(\ff U) = \frac{1}{2} 
  \sum_{\alpha\beta\gamma\delta} U_{\alpha\beta\delta\gamma} \:
  c^\dagger_\alpha c^\dagger_\beta c_\gamma c_\delta 
\ee
is a four-fermion point interaction and is specified by the (Coulomb) 
interaction parameters $U_{\alpha\beta\delta\gamma}$. 
The full set of interaction parameters is written as $\ff U$ for short.

The disorder potential
\be
  H_{\rm dis}(\ff \eta) = \sum_{\alpha\beta} \eta_{\alpha\beta} \:
  c^\dagger_\alpha c_\beta 
\ee
is bilinear and given in terms of parameters $\eta_{\alpha\beta}$ which are random
numbers with a joint probability distribution $P(\ff \eta)$ with $P(\ff \eta) \ge 0$
and $\int d \ff \eta \, P(\ff \eta) = 1$.
The configurational average for any quantity $A_{\ff \eta}$ depending on $\ff \eta$ is:
\be
  \langle A \rangle_{P} = \int d \ff \eta \: P(\ff \eta) A_{\ff \eta} \: .
\ee
For the theoretical setup, $H_{\rm dis}(\ff \eta)$ and $H_{\rm int}(\ff U)$ are taken
to be completely general. 
The construction of mean-field approximations will be most convenient for a 
local (Hubbard-type) interaction $\ff U$ and a diagonal (local) disorder potential,
$\eta_{\alpha\beta} = \delta_{\alpha\beta} \eta_\alpha$, with independent energies:
$P(\ff \eta) = \prod_\alpha p(\eta_\alpha)$.

Using the functional-integral formalism, \cite{NO88} the grand potential,
\be
  \Omega_{\ff t, \ff \eta, \ff U} = - T \ln  Z_{\ff t,\ff \eta,\ff U} \: ,
\labeq{ooo}  
\ee
and the partition function for a given configuration $\ff \eta$,
\ba
  Z_{\ff t,\ff \eta,\ff U} &=& \tr \exp(-(H(\ff t, \ff \eta, \ff U) - \mu N)/T)
\nonumber \\  
  & = &
  \int D \xi D \xi^\ast 
  \exp\left( A_{\ff t, \ff \eta, \ff U,\xi\xi^\ast} \right) \: ,
\labeq{zzz}  
\ea
depend on the model parameters via the action
\ba
   &&
   A_{\ff t, \ff \eta, \ff U,\xi\xi^\ast} 
\nonumber \\   
   &&=\sum_{n,\alpha\beta} \xi_\alpha^\ast(i\omega_n) 
   ((i\omega_n + \mu)\delta_{\alpha\beta} - t_{\alpha\beta} - \eta_{\alpha\beta})
   \xi_\beta(i\omega_n) 
\nonumber \\   
   &&- 
   \frac{1}{2} \sum_{\alpha\beta\gamma\delta} U_{\alpha\beta\delta\gamma} \int_0^{1/T} 
   \!\! d\tau \:
   \xi_\alpha^\ast(\tau) 
   \xi_\beta^\ast(\tau)
   \xi_\gamma(\tau) 
   \xi_\delta(\tau) \: .
\labeq{act}
\ea
Here $\xi_\alpha(i\omega_n)= T^{1/2} \int_0^{1/T} d\tau \: e^{i\omega_n\tau} \xi_\alpha(\tau)$ 
($\xi^\ast_\alpha(i\omega_n)= T^{1/2} \int_0^{1/T} d\tau \: e^{-i\omega_n\tau} \xi^\ast_\alpha(\tau)$) 
are Grassmann fields at the fermionic Matsubara frequencies $i\omega_n= i (2n+1) \pi T$ (with
$n = 0, \pm 1, ...$).
The configurational average of the grand potential is given by:
\be
  \Omega_{\ff t, P, \ff U}
  =
  \int d \ff \eta \: P(\ff \eta) \Omega_{\ff t, \ff \eta, \ff U} 
  =
  \left\langle 
  \Omega_{\ff t, \ff \eta, \ff U}
  \right\rangle_P \: .
\ee
The subscript $P$ indicates the dependence on the probability distribution. 

For later purposes we need the free one-particle Green's function,
\be
  \ff G_{\ff t,0,0} = \frac{1}{i\omega_n + \mu - \ff t} \: ,
\labeq{gt00}  
\ee
which is a matrix with the elements $G_{\ff t, 0,0, \alpha\beta}(i\omega_n)$.
The dependence of $\ff G_{\ff t,0,0}$ on $\ff t$ is indicated by the subscript.
Dependencies on the chemical potential $\mu$ and the temperature $T$ will not be
indicated,
$\mu$ and $T$ are assumed to be fixed.
Similarly,
\be
  \ff G_{\ff t,\ff \eta,0} = \frac{1}{i\omega_n + \mu - \ff t - \ff \eta}
\labeq{gfreeeta}
\ee
denotes the free Green's function in the presence of the disorder potential. 
The action determines the full Green's function $\ff G_{\ff t, \ff \eta, \ff U}$ 
the elements of which read:
\ba
&&
  G_{\ff t, \ff \eta, \ff U, \alpha\beta}(i\omega_n) 
\nonumber \\ && = \frac{-1}{Z_{\ff t,\ff \eta,\ff U}}
  \int D \xi D \xi^\ast 
  \xi_\alpha(i\omega_n) \xi^\ast_\beta(i\omega_n)
  \exp\left( A_{\ff t, \ff \eta, \ff U,\xi\xi^\ast} \right) \: .
\nonumber \\  
\labeq{intg}
\ea
Finally, we introduce the (interaction) self-energy 
\be
  \ff \Sigma_{\ff t, \ff \eta, \ff U} = 
  \ff G_{\ff t,0,0}^{-1} - \ff \eta -
  \ff G_{\ff t, \ff \eta, \ff U}^{-1}
  = 
  \ff G_{\ff t,\ff \eta,0}^{-1} -
  \ff G_{\ff t, \ff \eta, \ff U}^{-1} \: .
\labeq{defsig}
\ee
$\ff \Sigma_{\ff t, \ff \eta, \ff U}$ depends on the configuration $\ff \eta$. 
The configuration independent (full) self-energy 
\be
  \ff S_{\ff t, P, \ff U} = 
  \ff G_{\ff t,0,0}^{-1} - 
  \ff \Gamma_{\ff t, P, \ff U}^{-1}
\labeq{defs}
\ee
is defined with the help of the averaged Green's function
\be
  {\ff \Gamma}_{\ff t, P, \ff U}  = 
  \langle \ff G_{\ff t, \ff \eta, \ff U} \rangle_{P} \: .
\labeq{defg}
\ee

\section{Configuration-dependent self-energy functional}
\label{sec:funcint}

The main idea of the self-energy-functional theory (SFT) is to express a thermodynamical 
potential as a functional of the (interaction) self-energy which is stationary 
at the physical self-energy of the system. 
Variation of the self-energy is achieved by taking trial self-energies from
an (exactly solvable) reference system and varying its parameters.
To be able to evaluate the self-energy functional (which in most cases is defined
only formally), it is of crucial importance that the reference system shares with
the original system the non-trivial part of the functional so that this can be
eliminated. 
Details of the SFT are described in Refs.\ \onlinecite{Pot03a,Pot03b,PAD03,Pot05}. 
In the following, we present a construction of the self-energy functional for a 
fixed configuration $\ff \eta$.
The construction is non-perturbative (i.e.\ does not refer to formal sums of skeleton 
diagrams) and allows for a generalization in the case of disorder (see Sec.\ 
\ref{sec:funcdis}).

To start with, we note that the action can be considered as a functional of the 
(inverse) free Green's function (\refeq{gfreeeta}):
\ba
   &&
   \widehat{A}_{\ff U,\xi\xi^\ast} [\ff G_0^{-1}]
   =\sum_{n,\alpha\beta} \xi_\alpha^\ast(i\omega_n) 
   G_{0,\alpha\beta}^{-1}(i\omega_n)
   \xi_\beta(i\omega_n) 
\nonumber \\   
   &&- 
   \frac{1}{2} \sum_{\alpha\beta\gamma\delta} U_{\alpha\beta\delta\gamma} \int_0^{1/T} 
   \!\! d\tau \:
   \xi_\alpha^\ast(\tau) 
   \xi_\beta^\ast(\tau)
   \xi_\gamma(\tau) 
   \xi_\delta(\tau) \: .
\ea
Here $\ff G_0^{-1}$ is considered to be a free ``variable''. 
The physical action $A_{\ff t, \ff \eta, \ff U,\xi\xi^\ast}$ (\refeq{act}) is 
obtained by evaluating the functional $\widehat{A}_{\ff U,\xi\xi^\ast} [...]$ at the 
physical inverse free Green's function $\ff G_0^{-1}=\ff G_{\ff t, \ff \eta, 0}^{-1}$
(\refeq{gfreeeta}), i.e.:
\be
  A_{\ff t, \ff \eta, \ff U,\xi\xi^\ast} = 
  \widehat{A}_{\ff U,\xi\xi^\ast} [\ff G_{\ff t, \ff \eta, 0}^{-1}] \: .
\ee
Note that a hat is used to distinguish functionals from physical quantities.
Additional dependencies of a functional (parameters) are indicated by subscripts.

In the same way, via \refeq{ooo} and \refeq{zzz}, the grand potential can be 
considered as a functional of $\ff G_0^{-1}$, and one has:
\be
\Omega_{\ff t, \ff \eta, \ff U} = 
\widehat{\Omega}_{\ff U} [\ff G_{\ff t, \ff \eta, 0}^{-1}]
\; ,
\qquad
\Omega_{\ff t, 0, \ff U} = 
\widehat{\Omega}_{\ff U} [\ff G_{\ff t, 0, 0}^{-1}] \: .
\labeq{omofg0}
\ee
Again, one has to distinguish clearly e.g.\ between $\Omega_{\ff t, \ff \eta, \ff U}$, 
the exact grand potential of the model $H(\ff t, \ff \eta, \ff U)$ on the one hand, and
$\widehat{\Omega}_{\ff U}[\ff G_0^{-1}]$, a functional of the variable $\ff G_0^{-1}$
on the other. 
The latter only acquires the value $\Omega_{\ff t, \ff \eta, \ff U}$ if evaluated at 
$\ff G_0^{-1} = \ff G_{\ff t,  \ff \eta, 0}^{-1}$.

The functional derivative 
\be
\frac{1}{T} \frac{\delta \widehat{\Omega}_{\ff U}\left[\ff G_0^{-1}\right]}
{\delta \ff G_0^{-1}} 
= - \: \frac{1}{\widehat{Z}_{\ff U}\left[\ff G_0^{-1}\right]} 
\frac{\delta \widehat{Z}_{\ff U}\left[\ff G_0^{-1}\right]}
{\delta \ff G_0^{-1}} 
= - \widehat{\cal \ff G}_{\ff U}\left[\ff G_0^{-1}\right]
\labeq{omder}
\ee
defines a functional $\widehat{\cal \ff G}_{\ff U}$ as
\ba
&& \widehat{\cal \ff G}_{\ff U,\alpha\beta}\left[\ff G_0^{-1}\right]
\nonumber \\
&& = - \frac{1}{\widehat{Z}_{\ff U}\left[\ff G_0^{-1}\right]} 
\int D \xi D \xi^\ast 
   \xi_\alpha(i\omega_n)\xi_\beta^\ast(i\omega_n) 
   e^{\widehat{A}_{\ff U,\xi\xi^\ast}\left[\ff G_0^{-1}\right] }
\nonumber \\
\ea
which has the property
\be
  \widehat{\cal \ff G}_{\ff U}\left[\ff G_{\ff t, \ff \eta, 0}^{-1}\right] 
  = \ff G_{\ff t, \ff \eta, \ff U}
\; ,
\qquad
  \widehat{\cal \ff G}_{\ff U}\left[\ff G_{\ff t, 0, 0}^{-1}\right] 
  = \ff G_{\ff t, 0, \ff U} \: .
\ee
Namely, at the physical inverse free Green's function the functional integral in
\refeq{omder} defines the physical interacting Green's function, see \refeq{intg}.

Up to this point the derivations are standard. 
The decisive point in the construction of the self-energy functional is the following
equation:
\be
  \widehat{\cal \ff G}_{\ff U}\left[\ff G^{-1} + \ff \Sigma \right] = \ff G \: .
\labeq{rel}  
\ee
The only purpose of this is to constitute a relation between the variables $\ff G$ 
and $\ff \Sigma$ which may formally be solved for $\ff G$. 
This formal solution ${\ff G}=\widehat{\ff G}_{\ff U}[\ff \Sigma]$ then defines a 
functional $\widehat{\ff G}_{\ff U}[\ff \Sigma]$ (which parametrically depends on 
$\ff U$), i.e.\ we have
\be
  \widehat{\cal \ff G}_{\ff U}\left[ \widehat{\ff G}_{\ff U}[\ff \Sigma]^{-1} + 
  \ff \Sigma \right] = 
  \widehat{\ff G}_{\ff U}[\ff \Sigma]
\labeq{gfunc}
\ee
for any $\ff \Sigma$ by construction.
It is important to note that the functional $\widehat{\ff G}_{\ff U}[\ff \Sigma]$ 
is universal, i.e.\ it does neither depend on $\ff t$ nor $\ff \eta$.
If evaluated at the physical self-energy, the functional yields the exact Green's function
\be
  \widehat{\ff G}_{\ff U}\left[ \ff \Sigma_{\ff t, \ff \eta, \ff U}\right] = 
  \ff G_{\ff t, \ff \eta, \ff U}
\labeq{gex}
\ee
since, by definition, $\widehat{\ff G}_{\ff U}\left[ \ff \Sigma_{\ff t, \ff \eta, \ff U}\right]$
solves \refeq{gfunc} if 
$\widehat{\ff G}_{\ff U}\left[ \ff \Sigma_{\ff t, \ff \eta, \ff U}\right]^{-1}
= \ff G_{\ff t, \ff \eta, 0}^{-1} - \ff \Sigma_{\ff t, \ff \eta, \ff U}$.

The final step is to use the (universal, i.e.\ $\ff t$ and $\ff \eta$ independent) 
functionals 
$\widehat{\Omega}_{\ff U}\left[\ff G_0^{-1}\right]$
and
$\widehat{\ff G}_{\ff U}[\ff \Sigma]$ 
to express the grand potential as a functional of the self-energy. 
We define the functional $\widehat{F}_{\ff U}[\ff \Sigma]$ as
\be
  \widehat{F}_{\ff U}[\ff \Sigma] = 
  \widehat{\Omega}_{\ff U}\left[ 
  \widehat{\ff G}_{\ff U}[\ff \Sigma]^{-1} + \ff \Sigma
  \right]
  - 
  \Tr \ln \widehat{\ff G}_{\ff U}[\ff \Sigma]
\labeq{fdef}
\ee
where $\Tr \ff A \equiv T \sum_n \sum_\alpha e^{i\omega_n0^+} A_{\alpha\alpha}(i\omega_n)$.
\refeq{omder} and \refeq{gfunc} imply
\be
\frac{1}{T} \frac{\delta \widehat{F}_{\ff U}[\ff \Sigma]}{\delta \ff \Sigma} 
=
- \widehat{\ff G}_{\ff U}[\ff \Sigma] \: .
\ee
Hence, $\widehat{\ff G}_{\ff U}[\ff \Sigma]$ is a ``gradient'' of the ``scalar'' 
self-energy functional $\widehat{F}_{\ff U}[\ff \Sigma]$.
The physical meaning of $\widehat{F}_{\ff U}[\ff \Sigma]$ is obvious when comparing 
with the original derivation of the SFT (cf.\ Ref.\ \onlinecite{Pot03a}): 
$\widehat{F}_{\ff U}[\ff \Sigma]$ is the Legendre transform of the Luttinger-Ward
functional. \cite{LW60}

Now, the grand potential can be considered as a functional of the self-energy:
\be
  \widehat{\Omega}_{\ff t, \ff \eta, \ff U}[\ff \Sigma] = 
  \Tr \ln \frac{1}{\ff G_{\ff t, \ff \eta, 0}^{-1} - \ff \Sigma}
  + \widehat{F}_{\ff U}[\ff \Sigma] \: .
\labeq{sef}
\ee
Two properties of this functional are very useful:

First, at the exact self-energy $\ff \Sigma=\ff \Sigma_{\ff t, \ff \eta, \ff U}$
the self-energy functional yields the exact grand potential:
\be
  \widehat{\Omega}_{\ff t, \ff \eta, \ff U}[\ff \Sigma_{\ff t,\ff \eta, \ff U}]
  = {\Omega}_{\ff t, \ff \eta, \ff U} 
\labeq{omex}  
\ee
since
\be
  \widehat{\Omega}_{\ff t, \ff \eta, \ff U}[\ff \Sigma_{\ff t, \ff \eta,\ff U}]
  =
  \Tr \ln \ff G_{\ff t, \ff \eta, \ff U} + \widehat{F}_{\ff U}[\ff \Sigma_{\ff t, \ff \eta, \ff U}]
\ee
from \refeq{sef} and \refeq{defsig} and 
\be
  \widehat{F}_{\ff U}[\ff \Sigma_{\ff t, \ff \eta, \ff U}]
  =
  \widehat{\Omega}_{\ff U}\left[ 
  \ff G^{-1}_{\ff t, \ff \eta, 0}
  \right]
  - 
  \Tr \ln \ff G_{\ff t, \ff \eta, \ff U} 
\ee
from \refeq{fdef} and \refeq{gex}. 
Hence, \refeq{omex} follows from \refeq{omofg0}.

Second, consider the derivative:
\be
  \frac{1}{T} \frac{\delta \widehat{\Omega}_{\ff t, \ff \eta, \ff U}[\ff \Sigma]}
  {\delta \ff \Sigma} = 
  \frac{1}{\ff G_{\ff t, \ff \eta, 0}^{-1} - \ff \Sigma} - 
  \widehat{G}_{\ff U}[\ff \Sigma] \: .
\ee
The equation 
\be
  \widehat{\ff G}_{\ff U}[\ff \Sigma] = \frac{1}{\ff G_{\ff t, \ff \eta, 0}^{-1} - \ff \Sigma}
\label{eq:sig}
\ee
is a (highly non-linear) conditional equation in the variable $\ff \Sigma$ 
with parameters $\ff t$, $\ff \eta$, $\ff U$ which is solved by the physical
self-energy $\ff \Sigma = \ff \Sigma_{\ff t, \ff \eta, \ff U}$.
It is by no means straightforward to find a solution, however, since the functional 
$\widehat{\ff G}_{\ff U}[\ff \Sigma]$ is not known explicitly but was constructed 
in a formal way only. 
Obviously, this is equivalent to a search for the stationary point of the grand potential
as a functional of the self-energy:
\be
  \frac{\delta \widehat{\Omega}_{\ff t, \ff \eta, \ff U}[\ff \Sigma]}
  {\delta \ff \Sigma} = 0 \: .
\ee
This establishes a very general variational principle without the need for an
expansion in powers of the interaction strength, i.e.\ the construction is
non-perturbative.

\section{Approximations for a fixed configuration}
\label{sec:appfix}

For the discussion of possible approximations, we first consider a fixed configuration 
$\ff \eta$. 
Then $\ff t + \ff \eta$ is a fixed matrix of hopping parameters but without translational 
symmetries.
The idea of the SFT is to construct approximations by searching for the stationary
point of the functional \refeq{sef} on a {\em restricted} domain of trial self-energies.
Trial self-energies are chosen from a reference system which shares with the original 
system the same interaction part $H_{\rm int}(\ff U)$.
In the Hamiltonian of the reference system,
\be 
  H' = H_0(\ff t') + H_{\rm int}(\ff U) \: ,
\ee
the bilinear part $H_0(\ff t')$ is varied arbitrarily. 
We have set $\ff \eta'=0$ for the reference system. 
However, no translational symmetry is assumed for the hopping $\ff t'$. 

Since the interaction part is the same and due to the universality of the functional  
$\widehat{F}_{\ff U}[\ff \Sigma]$, only the first term of the self-energy functional
of the reference system,
\be
  \widehat{\Omega}_{\ff t', 0, \ff U}[\ff \Sigma] = 
  \Tr \ln \frac{1}{\ff G_{\ff t', 0, 0}^{-1} - \ff \Sigma}
  + \widehat{F}_{\ff U}[\ff \Sigma] \: ,
\labeq{sefr}
\ee
differs from the functional of the original system \refeq{sef}.
Combination of the functionals \refeq{sef} and \refeq{sefr} therefore gives
\ba
  \widehat{\Omega}_{\ff t, \ff \eta, \ff U}[\ff \Sigma]
  &=&
  \widehat{\Omega}_{\ff t', 0, \ff U}[\ff \Sigma]
  +
  \mbox{Tr} \ln \frac{1}{\ff G_{\ff t,\ff \eta,0}^{-1} - \ff \Sigma}
\nonumber \\ 
  &-&
  \mbox{Tr} \ln \frac{1}{\ff G_{\ff t',0,0}^{-1} - \ff \Sigma} \: .
\ea
The not explicitly known functional $\widehat{F}_{\ff U}[\ff \Sigma]$ cancels out.

To search for the stationary point of the self-energy functional of the original system, 
we insert as trial self-energies the exact self-energies of the reference system:
$\ff \Sigma = \ff \Sigma_{\ff t', 0, \ff U}$.
This yields a function of $\ff t'$,
\be
  \Omega_{\ff t, \ff \eta, \ff U}(\ff t') \equiv
  \widehat{\Omega}_{\ff t, \ff \eta, \ff U}[\ff \Sigma_{\ff t', 0, \ff U}]
\labeq{fct}  
\ee
Searching for the stationary point of $\Omega_{\ff t, \ff \eta, \ff U}(\ff t')$ as a
function of $\ff t'$ means to search for the stationary point of the exact 
self-energy functional \refeq{sef} on the restricted set of trial self-energies 
generated by the reference system with parameters $\ff t'$.
From \refeq{omex} and \refeq{defsig} for the reference system we have: 
\ba
  \Omega_{\ff t, \ff \eta, \ff U}(\ff t') &=&
  \Omega_{\ff t', 0, \ff U}
  +
  \mbox{Tr} \ln \frac{1}{\ff G_{\ff t,\ff \eta,0}^{-1} - \ff \Sigma_{\ff t', 0, \ff U}}
\nonumber \\
  &-&
  \mbox{Tr} \ln \ff G_{\ff t', 0, \ff U} \: .
\labeq{rest}
\ea
The important point is that the r.h.s.\ can be computed exactly if the reference system 
is an exactly solvable model.
Specifying a certain reference system means to generate a particular approximation. 
Typically, a suitable reference system can be found for lattice models by tiling the 
original lattice into clusters of finite size and by neglecting the inter-cluster
hopping. 

\section{Statistical SFT}
\label{sec:statsft}

To make contact with the statistical DMFT, \cite{DK97,DK98} we consider a system with 
local interaction and local (and uncorrelated) disorder. 
The statistical DMFT treats the disorder part of the problem exactly while the 
(dynamical) mean-field approximation is used for the interaction part.
Within the framework of the SFT, a mean-field approximation is generated by a 
reference system in which all sites are decoupled. 
This implies that spatial correlations are neglected altogether in the computation
of the self-energy. 
The local (temporal) dynamics, however, can be optimized by introducing additional 
local degrees of freedom in the reference system. 
For the Hamiltonian of the reference system this means to introduce additional 
uncorrelated sites (``bath sites''), the on-site energies of which as well as their 
hybridizations with the original correlated sites are treated as variational 
parameters.

\begin{figure}[t]
  \includegraphics[width=0.85\columnwidth]{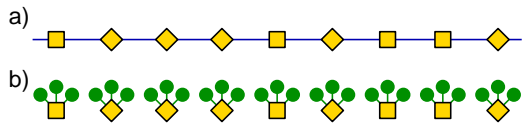}
\caption{(Color online)
a) Representation of the Anderson-Hubbard model in one dimension. 
Squares/diamonds: fixed configuration of sites with local (Hubbard) interaction 
and local binary-alloy disorder.
Lines: nearest-neighbor hopping. 
b) Reference system generating a mean-field approximation ($n_s=4$).
Circles: bath sites (no interaction) with on-site energies and hybridizations to
original sites (square, diamonds) to be treated as independent variational parameters.
}
\label{fig:mf}
\end{figure}

To be explicit, the discussion is restricted to the Anderson-Hubbard model
\be
  H = \sum_{\ff x\ff x'\sigma} t_{\ff x\ff x'} c_{\ff x\sigma}^\dagger c_{\ff x'\sigma} 
    + \sum_{\ff x\sigma} \eta_{\ff x} n_{\ff x\sigma} 
    + U \sum_{\ff x} n_{\ff x\uparrow} n_{\ff x\downarrow} \: .
\labeq{ahm}
\ee
Here, $\ff x$ refers to the sites of a lattice, $n_{\ff x\sigma} = c_{\ff x\sigma}^\dagger 
c_{\ff x\sigma}$, and $\eta_{\ff x}$ are independent random numbers distributed according to 
\be
  P(\ff \eta) = \prod_{\ff x} p(\eta_\ff x)
\labeq{locdis}
\ee
with some density $p(\eta_{\ff x})$.
A reference system generating a mean-field approximation is
$H'=\sum_{\ff x} H'_{\ff x}$ with (see Fig.\ \ref{fig:mf})
\ba
  H'_{\ff x} &=& \sum_{\sigma} t'_{\ff x\ff x} c_{\ff x\sigma}^\dagger c_{\ff x\sigma} 
       + U \sum_{\ff x} n_{\ff x\uparrow} n_{\ff x\downarrow} 
\nonumber \\       
       &+& \sum_{\sigma} \sum_{i=2}^{n_{s}}
	   \varepsilon^{(\ff x)}_i a_{\ff xi\sigma}^\dagger a_{\ff xi\sigma}
\nonumber \\       
       &+& \sum_{\sigma} \sum_{i=2}^{n_{s}}	   
           V^{(\ff x)}_i (a_{\ff xi\sigma}^\dagger c_{\ff x\sigma} + c^\dagger_{\ff x\sigma} 
	   a_{\ff xi\sigma}) 
	   \: .
\nonumber \\ 
\labeq{iahm}      
\ea
It consists of effective impurity models with $n_s$ sites each: The correlated
site $\ff x$ (with $U\ne 0$) and $n_s-1$ bath sites (with $U=0$) labeled by $i$.
The effective impurity models can be solved independently to get a trial self-energy.

%*********************************************************************************
%*********************************************************************************

There is an indirect coupling, however, via the optimization of the variational 
parameters $\ff t' = (t'_{\ff x\ff x}, \varepsilon^{(\ff x)}_i, V^{(\ff x)}_i)$:
The Euler equation,
\be
\frac{\partial}{\partial \ff t'} {\Omega}_{\ff t, \ff \eta, \ff U}(\ff t') = 0 \: ,
\labeq{eu}
\ee
simplifies due to the fact that the trial self-energy (and its derivative w.r.t $\ff t'$, 
see Ref.\ \onlinecite{Pot06}) is necessarily local.
Using \refeq{rest} and carrying out the derivative, \refeq{eu} can be written as:
\ba
  \sum_{n,\ff x}
  \left(
  \frac{1}{\ff G^{-1}_{\ff t, \ff \eta, 0} - \ff \Sigma_{\ff t', 0 ,\ff U}} - \ff G_{\ff t',0,\ff U}
  \right)_{n,\ff x\ff x}
  \frac{\partial \Sigma_{\ff x\ff x}(i\omega_n)}{\partial \ff t'} = 0 \: .
  \nonumber \\
\ea
Now, the variation of the one-particle parameters of the impurity model at site $\ff x$
does not affect the self-energy of the impurity model at site $\ff x'$ ($\ff x'\ne \ff x$). 
Therefore, the Euler equation simplifies to:
\ba
  \sum_{n}
  \left(
  \frac{1}{\ff G^{-1}_{\ff t, \ff \eta, 0} - \ff \Sigma_{\ff t', 0 ,\ff U}} - \ff G_{\ff t',0,\ff U}
  \right)_{n,\ff x\ff x}
  \frac{\partial \Sigma_{\ff x\ff x}(i\omega_n)}{\partial t_{\ff x\ff x}'} = 0 \: ,
  \nonumber \\
\labeq{eu1}
\ea
where $t_{\ff x\ff x}'$ denotes the variational parameters at the correlated site $\ff x$.
This is a set of equations labeled by the site index $\ff x$.
Due to the matrix inversion in \refeq{eu1}, however, the equations are coupled. 
This implies that for a generic configuration, the individual self-energies of the 
effective impurity models are different at the stationary point as the sites of the 
original lattice model are inequivalent.

%*********************************************************************************
\begin{figure}[t]
  \includegraphics[width=0.85\columnwidth]{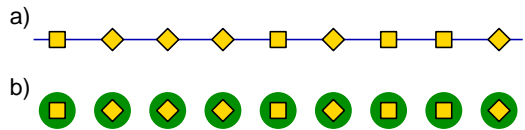}
\caption{(Color online)
The same as Fig.\ \ref{fig:mf} but for a continuum of bath sites ($n_s \to \infty$),
represented by big circles. The reference system generates the generalized or 
statistical DMFT (see text).
}
\label{fig:gendmft}
\end{figure}
%*********************************************************************************

For $n_s \to \infty$, i.e.\ for a continuous bath (see Fig.\ \ref{fig:gendmft}), 
an additional formal simplification is possible: 
As it is obvious from \refeq{eu1}, a solution $\ff t'$ of the coupled equations
\be
  \left(
  \frac{1}{\ff G^{-1}_{\ff t, \ff \eta, 0}(i\omega_n) - \ff \Sigma_{\ff t', 0 ,\ff U}(i\omega_n)}
  \right)_{\ff x\ff x}
  = 
  (\ff G_{\ff t',0,\ff U}(i\omega_n) )_{\ff x\ff x} \: ,
\labeq{dmftreduced}  
\ee
yields a stationary point of the self-energy functional.
Note that for any {\em finite} $n_s < \infty$ there is no solution:
The r.h.s.\ of \refeq{dmftreduced} is the Green's function of a finite system 
which exhibits a finite set of poles on the real frequency axis (after analytical 
continuation). 
Contrary, the l.h.s.\ represents an approximate lattice Green's function
which in the thermodynamical limit $L\to \infty$ has branch cuts on the real axis 
induced by the branch cut of the free Green's function.

In the case of systems with a few inequivalent sites, i.e.\ for inhomogeneous systems 
with a somewhat reduced translational symmetry, the equations (\ref{eq:dmftreduced}) 
exactly recover a generalization of the DMFT which has been put forward to describe 
correlation effects at surfaces and in thin films. \cite{PN99c}
They are just the self-consistency equations of this generalized DMFT.
Typically, only a few effective impurity models have to be considered in this approach. 
\cite{PN99a,PN99d}

For disordered systems without any translational symmetry, the self-consistency equations 
(\ref{eq:dmftreduced}) constitute the statistical DMFT as introduced by Dobrosavljevi\'c{} 
and Kotliar. \cite{DK97,DK98}
In principle, the Eqs.\ (\ref{eq:dmftreduced}) can be solved iteratively.
For any iteration in the self-consistency cycle, one then needs the local Green's function
at each site of the system which in each case requires the solution of an interacting 
impurity problem.
Eqs.\ (\ref{eq:dmftreduced}) and the absence of translational symmetry also imply 
the need for an inversion of matrices with dimension given by the system size. 
For these reasons, the statistical DMFT is a numerically extremely expensive method.

Choosing a reference system consisting of decoupled effective impurity models with
finite (and actually small) $n_s$ could thus be an interesting alternative. 
Calculations based on such a statistical dynamical impurity approximation (stat-DIA), 
however, have not yet been performed.
Since all physical quantities derive from an explicit though approximate expression
for a thermodynamical potential, the stat-DIA is a thermodynamically consistent 
approximation.
This can be seen as an advantage compared to stat-DMFT approaches which
employ additional approximations to render practical calculations possible. 

The self-consistency equations (\ref{eq:dmftreduced}) allow for a stochastic reinterpretation:
For a given configuration of on-site energies $\{ \eta_{\ff x} \}$, the local Green's function
$G_{\ff x\ff x}$ will be site-dependent.
The {\em distribution} of the local Green's function at a site $\ff x$ (generated by all 
configurations), however, will be the same as the distribution at a site $\ff x' \ne \ff x$
since the on-site energies $\eta_{\ff x}$ have been assumed to be independent random 
numbers distributed according to the {\em same} density $p(\eta_{\ff x})$ for each $\ff x$.
Moreover, the distribution of the local Green's function at a site $\ff x$, generated by 
all configurations $\{ \eta_{\ff x} \}$, is identical to the distribution of the local Green's 
function on all lattice sites for one fixed typical configuration of on-site energies.
Hence, the index $\ff x$ in the self-consistency equations (\ref{eq:dmftreduced}) can be 
viewed as a label for a particular realization of the random variable $G_{\ff x\ff x}$.

For a Bethe lattice, the equations (\ref{eq:dmftreduced}) can be reinterpreted
as stochastic recursion equations. 
Starting from an arbitrary initial sample for the local Green's function, $\{ G_{\ff x\ff x} \}$, 
the equations recursively generate a sequence of samples which converges to a sample which
is representative for the distribution of the local Green's function.
The practical advantage of this reinterpretation consists in the fact that a Gibbs-sampling 
Monte-Carlo algorithm for the calculation of marginal distributions can be applied 
(see Ref.\ \onlinecite{BAF04}, for an example). 
The iterative solution of the self-consistency equations (\ref{eq:dmftreduced}) for 
a given typical configuration of on-site energies is then equivalent with the recursive 
update of a sample of local Green's functions.
Furthermore, the matrix inversion required in Eq.\ (\ref{eq:dmftreduced}) can be avoided
in the case of a Bethe lattice.
For a general (e.g.\ cubic) lattice, however, the stochastic scheme breaks down, 
and one has to return to the site interpretation again.

\section{Configuation-independent self-energy functional}
\label{sec:funcdis}

The above discussion has shown the practical needs to construct more simple 
approximations.
An intuitive strategy is to consider quantities involving configurational 
averages and to search for sensible approximations of the averaged quantities
instead of considering full distributions. 
The simplest and most natural average is the arithmetical average 
$\langle \cdots \rangle_P$ which has been introduced in Sec.\ \ref{sec:ham}.
It shall be understood that one has to be extremely careful when discussing
{\em transport properties} in terms of the averaged one-particle Green's function
$\ff \Gamma_{\ff t,P,\ff U}$:
Close to Anderson localization the distribution of the local Green's function
(at $\omega=0$) can exhibit an extreme asymmetry and a long tail such that
the average is of no physical meaning and can by no means serve as an order
parameter for a metal-insulator transition. \cite{AF04}

Here our goal is to construct non-perturbative and thermodynamically consistent
approximations for averaged quantities which give information on thermodynamic 
properties and one-particle excitations.
Within the framework of the self-energy-functional approach, this can be achieved
by introducing functionals that involve quantities averaged according to the given 
probability distribution $P$.
In particular, we consider functionals of the configuration-independent (full) 
self-energy $\ff S$ and the configuration-dependent (interaction) self-energy
$\ff \Sigma_{\ff \eta}$ as defined at the end of Sec.\ \ref{sec:ham}.

Analogous to \refeq{rel}, the equation
\be
  \left\langle
  \frac{1}
  {\ff \Gamma^{-1} + \ff S - \ff \eta - \ff \Sigma_{\ff \eta}}
  \right\rangle_P = \ff \Gamma
\label{eq:disfunc}  
\ee
constitutes a relation between the averaged Green's function $\ff \Gamma$ on the 
one hand and $\ff S$ and $\ff \Sigma_{\ff \eta}$ on the other.
The functional $\widehat{\cal \ff G}_{\ff U}\left[ \cdots \right]$ in \refeq{rel} 
is replaced by the functional $\left\langle 1 / (\cdots - \ff \eta) \right\rangle_P$, 
and the probability distribution $P$, instead of the interaction parameters $\ff U$,
plays the role of the external parameters.
Contrary to \refeq{rel}, the above relation is diagonal in the frequency $i\omega_n$
(which is suppressed in notations).

Assume that $\ff S$ and (for any configuration $\ff \eta$) $\ff \Sigma_{\ff \eta}$ 
are given. 
Then, the equation can formally be solved for $\ff \Gamma$. 
This defines a functional $\widehat{\ff \Gamma}_{P}[\cdots,\cdots]$ which assigns 
the averaged Green's function
$\ff \Gamma = \widehat{\ff \Gamma}_{P}[\ff S , \{ \ff \Sigma_{\ff \eta} \}]$
to any $\ff S$ and $\ff \Sigma_{\ff \eta}$. 
This functional plays a role analogous to the functional 
$\widehat{\ff G}_{\ff U}\left[ \ff \Sigma \right]$ in Sec.\ \ref{sec:funcint}.

Analogous to \refeq{gex}, we have:
\be
  \widehat{\ff \Gamma}_{P}[
  \ff S_{\ff t, P, \ff U},
  \{ \ff \Sigma_{\ff t, \ff \eta, \ff U} \} 
  ]
  =
  {\ff \Gamma}_{\ff t, P, \ff U}
\labeq{disfunceval}
\ee
since \refeq{disfunc} holds when evaluated for 
$\ff S=\ff S_{\ff t, P, \ff U}$,
$\ff \Sigma_{\ff \eta} = \ff \Sigma_{\ff t,\ff \eta, \ff U}$,
and
$\ff \Gamma={\ff \Gamma}_{\ff t, P, \ff U}$
as it is obvious from Eqs.\ (\ref{eq:defsig}), (\ref{eq:defs}) and (\ref{eq:defg}). 
From \refeq{sig} we have
\be
  \widehat{\ff G}_{\ff U}[\ff \Sigma_{\ff \eta}] 
  = 
  \frac{1}
  {\ff G_{\ff t,0,0}^{-1} - \ff \eta - \ff \Sigma_{\ff \eta}} 
\labeq{tosolve1}  
\ee
for any $\ff \eta$.
This equation and 
\be
  \widehat{\ff \Gamma}_{P}[
  \ff S,
  \{ \ff \Sigma_{\ff \eta} \}]
  =
  \frac{1}{\ff G_{\ff t,0,0}^{-1} - \ff S} 
\labeq{tosolve2}  
\ee
form a (highly non-linear) system of conditional equations for the variables 
$\ff S(i\omega_n)$ and $\ff \Sigma_{\ff \eta}(i\omega_n)$.
The external parameters $\ff t$, $P$, $\ff U$ specify the model under consideration.
Eqs.\ (\ref{eq:tosolve1}) and (\ref{eq:tosolve2}) are satisfied for the exact self-energies
$\ff S(i\omega_n) = \ff S_{\ff t, P, \ff U}(i\omega_n)$ and
$\ff \Sigma_{\ff \eta}(i\omega_n) = \ff \Sigma_{\ff t, \ff \eta, \ff U}(i\omega_n)$.

In the following we show that the conditional equations \refeq{tosolve1} and 
\refeq{tosolve2} can be considered as stationarity conditions of the averaged 
grand potential as a functional of the self-energies.
We define the self-energy functional
\ba
&&
\widehat{\Omega}_{\ff t, P, \ff U}[\ff S , \{ \ff \Sigma_{\ff \eta} \}]
  = \mbox{Tr} \ln \frac{1}{\ff G_{\ff t,0,0}^{-1} - \ff S}
\nonumber \\  
  && +
 \left\langle \mbox{Tr} \ln 
  \frac{1}{ \widehat{\ff \Gamma}_{P}[\ff S , \{ \ff \Sigma_{\ff \eta} \}]^{-1}
  + \ff S - \ff \eta - \ff \Sigma_{\ff \eta}
  }
  \right\rangle_P
\nonumber \\  
&& - \mbox{Tr} \ln \widehat{\ff \Gamma}_{P}[
  \ff S,
  \{ \ff \Sigma_{\ff \eta} \}]
  + \left\langle \widehat{F}_{\ff U}[\ff \Sigma_{\ff \eta}] \right\rangle_P
\labeq{func}
\ea
The sum of the second and third term on the r.h.s.\ is a functional which is universal, 
i.e.\ it is independent of $\ff t$
(note that the terms do not cancel each other as the operations $\ln(\cdots)$ and 
$\langle \cdots \rangle_P$ do not commute).
With Eqs.\ (\ref{eq:defsig}), (\ref{eq:defs}), (\ref{eq:defg}) and Eqs.\
(\ref{eq:disfunc}), (\ref{eq:disfunceval}), the evaluation of the functional 
at the exact self-energies yields:
\ba 
  \widehat{\Omega}_{\ff t, P, \ff U}
  [\ff S_{\ff t, P, \ff U}, \{ \ff \Sigma_{\ff t, \ff \eta, \ff U} \}]
  &=&
  \left\langle 
  \mbox{Tr} \ln \ff G_{\ff t, \ff \eta, \ff U}
  \right\rangle_P
\nonumber \\
  &+ &
  \left\langle \widehat{F}_{\ff U}[\ff \Sigma_{\ff t, \ff \eta, \ff U}] \right\rangle_P
\nonumber \\
  & =& 
  \left\langle 
  \Omega_{\ff t, \ff \eta, \ff U}
  \right\rangle_P
  = 
  \Omega_{\ff t, P, \ff U} \: ,
\nonumber \\
\labeq{omexdis}
\ea
i.e.\ the exact averaged grand potential.
The functional derivatives are readily calculated:
\be
\frac{1}{T}
\frac{\delta \widehat{\Omega}_{\ff t, P, \ff U}[\ff S , \{ \ff \Sigma_{\ff \eta} \}]}
{\delta \ff S} 
=
\frac{1}{\ff G_{\ff t,0,0}^{-1} - \ff S}
-
\widehat{\ff \Gamma}_{P}[\ff S , \{ \ff \Sigma_{\ff \eta} \}]
\ee
and
\ba
\frac{1}{T}
\frac{\delta \widehat{\Omega}_{\ff t, P, \ff U}[\ff S , \{ \ff \Sigma_{\ff \eta} \}]}
{\delta \ff \Sigma_{\ff \eta}} 
&=&
\Big(
\frac{1}{
\widehat{\ff \Gamma}_{P}[\ff S , \{ \ff \Sigma_{\ff \eta} \}]^{-1}
+ \ff S - \ff \eta - \ff \Sigma_{\ff \eta}
}
\nonumber \\
&-&
\widehat{\ff G}_{\ff U}[\ff \Sigma_{\ff \eta}]
\Big) P(\ff \eta) \: .
\ea
Hence, setting the functional derivatives to zero, yields two equations
equivalent with \refeq{tosolve1} and \refeq{tosolve2}.
Therefore, the functional is stationary at the exact self-energies:
\ba
\frac{\delta \widehat{\Omega}_{\ff t, P, \ff U}[\ff S_{\ff t, P, \ff U},
  \{ \ff \Sigma_{\ff t, \ff \eta, \ff U} \}]}
{\delta \ff S} 
&=& 0 \; ,
\nonumber \\
\frac{\delta \widehat{\Omega}_{\ff t, P, \ff U}[\ff S_{\ff t, P, \ff U},
  \{ \ff \Sigma_{\ff t, \ff \eta, \ff U} \} ]}
{\delta \ff \Sigma_{\ff \eta}} 
&=& 0 \; .
\ea
The self-energy functional \refeq{func} represents a generalization of 
the self-energy functional \refeq{sef} for interacting systems with disorder.
It is completely general and provides an exact variational principle.

\section{Consistent approximations}
\label{sec:app}

In the spirit of the SFT for pure systems, approximations shall be constructed by restricting
the domain of self-energies in the functional \refeq{func} while retaining the exact 
functional dependence.
We consider both, the full as well as the interaction self-energy. 
Trial self-energies are taken from a reference system which is a system in the
same macroscopic state, i.e.\ with the same temperature $T$ and the same chemical 
potential $\mu$ as the original system, but has different one-particle parameters $\ff t'$. 
The Hamiltonian of the reference system reads:
\be 
  H' = H(\ff t', \ff \eta, \ff U) = H_0(\ff t') + H_{\rm dis}(\ff \eta) + H_{\rm int}(\ff U) \: .
\ee
$H'$ has the same interaction part as compared to the original system. 
Likewise the disorder potential, i.e.\ the distribution $P(\ff \eta)$, is assumed to 
be unchanged.
Hence, the self-energy functional of the reference system is given by
\ba
  && \widehat{\Omega}_{\ff t', P, \ff U}[\ff S , \{ \ff \Sigma_{\ff \eta} \}]
  = \mbox{Tr} \ln \frac{1}{\ff G_{\ff t',0,0}^{-1} - \ff S}
\nonumber \\ 
  && +
  \left\langle \mbox{Tr} \ln 
  \frac{1}{ \widehat{\ff \Gamma}_{P}[\ff S , \{ \ff \Sigma_{\ff \eta} \}]^{-1}
  + \ff S - \ff \eta - \ff \Sigma_{\ff \eta}
  }
  \right\rangle_P
  \nonumber \\
  && - \mbox{Tr} \ln \widehat{\ff \Gamma}_{P}[\ff S, \{ \ff \Sigma_{\ff \eta} \}]
  + \left\langle \widehat{F}_{\ff U}[\ff \Sigma_{\ff \eta}] \right\rangle_P \: .
\labeq{funcp}
\ea
Only the first term on the r.h.s.\ is different as compared to the functional
for the original system \refeq{func}.
Combining \refeq{func} and \refeq{funcp}, the last three terms on the respective
r.h.s.\ cancel out, and one is left with
\ba
  \widehat{\Omega}_{\ff t, P, \ff U}[\ff S , \{ \ff \Sigma_{\ff \eta} \}]
  &=&
  \widehat{\Omega}_{\ff t', P, \ff U}[\ff S , \{ \ff \Sigma_{\ff \eta} \}]
  +
  \mbox{Tr} \ln \frac{1}{\ff G_{\ff t,0,0}^{-1} - \ff S}
\nonumber \\ 
  &-&
  \mbox{Tr} \ln \frac{1}{\ff G_{\ff t',0,0}^{-1} - \ff S} \: .
\labeq{func1}  
\ea
Note that the full and the interaction self-energies are considered as variables at 
this point, and that for the cancellation of the functionals it is of crucial 
importance to choose the reference system to have the same interaction and disorder.
The self-energy functional \refeq{func1} is still exact.

As trial self-energies we insert the exact self-energies of the reference system:
$\ff S = \ff S_{\ff t', P, \ff U}$ and 
$\ff \Sigma_{\ff \eta} = \ff \Sigma_{\ff t', \ff \eta, \ff U}$.
Searching for the stationary point of the exact self-energy functional \refeq{func} 
on the subspace of trial self-energies taken from $H'$ and parameterized by $\ff t'$,
means to search for the stationary point of a function of $\ff t'$:
\be
  \Omega_{\ff t, P, \ff U}(\ff t') \equiv
  \widehat{\Omega}_{\ff t, P, \ff U}[\ff S_{\ff t', P, \ff U},\{ \ff \Sigma_{\ff t', \ff \eta, \ff U} \}]
  \: ,
\ee
where $\ff t$, $P$ and $\ff U$ are fixed by the original system.
From \refeq{omexdis} and \refeq{defs} for the reference system we get the comparatively 
simple result:
\ba
  \Omega_{\ff t, P, \ff U}(\ff t') &=&
  \Omega_{\ff t', P, \ff U}
  +
  \mbox{Tr} \ln \frac{1}{\ff G_{\ff t,0,0}^{-1} - \ff S_{\ff t', P, \ff U}}
\nonumber \\
  &-&
  \mbox{Tr} \ln \ff \Gamma_{\ff t',P,\ff U} \: .
\labeq{res}
\ea
This result is formally very similar to \refeq{rest} for pure systems.
Again, the important point is that the r.h.s.\ can be computed exactly if the reference 
system is an exactly solvable model.
The only difference consists in the fact that the grand potential, the Green's function
and the self-energy of the reference system on the r.h.s.\ are replaced by the corresponding
averaged quantities and the configuration independent (full) self-energy.

A certain approximation may be constructed along the following steps:
(i) 
A reference system is specified with $P$ and $\ff U$ fixed as in the original system. 
The hopping part, i.e.\ $\ff t'$, however, is fully at one's disposal and should be
used to simplify the problem posed by the reference system.
(ii) 
For a given set of variational parameters $\ff t'$, the reference system's Hamiltonian 
$H' = H(\ff t', \ff \eta, \ff U)$ is diagonalized for any configuration $\ff \eta$ to 
get the (many-body) eigenenergies and eigenstates.
(iii) 
The grand potential $\Omega_{\ff t', \ff \eta, \ff U}$ and, from the Lehmann representation, 
the Green's function $\ff G_{\ff t', \ff \eta, \ff U}$ are obtained for any $\ff \eta$.
(iv) 
Averaging yields $\Omega_{\ff t', P, \ff U} = \langle \Omega_{\ff t', \ff \eta, \ff U} \rangle_P$
and ${\ff \Gamma}_{\ff t', P, \ff U}  = \langle \ff G_{\ff t', \ff \eta, \ff U} \rangle_{P}$. 
The self-energy is computed via
$\ff S_{\ff t', P, \ff U} = \ff G_{\ff t',0,0}^{-1} - \ff \Gamma_{\ff t', P, \ff U}^{-1}$.
(v) 
Inserting these results as well as the free Green's function of the original model 
into \refeq{res} yields $\Omega_{\ff t,P,\ff U}(\ff t')$.
(vi) 
Steps (ii) - (v) are repeated for different $\ff t'$ to find the stationary point 
$\ff t'_{\rm s}$ given by 
\be
  \frac{\partial \Omega_{\ff t, P, \ff U}(\ff t'_{\rm s})}{\partial \ff t'} = 0
  \: .
\labeq{parder}
\ee

This approximation strategy shares a number of advantageous features with the 
corresponding strategy for pure systems:
Any approximation constructed in this way is a thermodynamically consistent one
since the theromdynamics as well as the (averaged) one-particle excitation properties
both derive from an explicit expression for the approximate averaged grand potential 
$\Omega_{\ff t, P, \ff U}(\ff t'_{\rm s})$ at the stationary point (see the
discussion in Ref.\ \onlinecite{AAPH06a}).
The only approximation consists in the restriction of the domain of the self-energy
functional. 
The approach is systematic as an enlarged domain leads to an improved approximation
(see the discussion in Ref.\ \onlinecite{Pot06}).
As the exact functional form is retained, approximations are non-perturbative by 
construction. 

\section{Limiting cases}
\label{sec:lim}

Eq.\ (\ref{eq:func}) gives the self-energy functional 
$\widehat{\Omega}_{\ff t, P, \ff U}[\ff S ,\{\ff \Sigma_{\ff \eta}\}]$
for a disordered and interacting system.
To discuss the limiting cases of the pure, of the non-interacting and of the 
pure and non-interacting system, we first have to specify the domain of this
functional.
This also applies to the functionals $\widehat{\ff \Gamma}_{P}[\ff S,\{\ff \Sigma_{\ff \eta}\}]$,
$\widehat{\ff G}_{\ff U}[\ff \Sigma]$, etc. 

For a given set of interaction parameters $\ff U$ and for a given probability
distribution $P$, the domain ${\cal D} = {\cal D}_S \times {\cal D}_{\Sigma}$
of the self-energy functional 
$\widehat{\Omega}_{\ff t, P, \ff U}[\ff S ,\{\ff \Sigma_{\ff \eta}\}]$
shall consist of (full) self-energies $\ff S \in {\cal D}_S$ and (for any $\ff \eta$) 
interaction self-energies $\ff \Sigma_{\ff \eta} \in {\cal D}_{\Sigma}$ taken from
the reference system.
Namely, a (full) self-energy $\ff S$ belongs to ${\cal D}_S$, if there is some $\ff t'$
such that $\ff S = \ff S_{\ff t',P,\ff U}$, i.e.\ such that $\ff S$ is the exact self-energy
of the problem given by $H' = H(\ff t', \ff \eta, \ff U)$ and $P$ for some $\ff t'$.
Likewise, a set of interaction self-energies $\{\ff \Sigma_{\ff \eta}\}$ (for all possible
$\ff \eta$) belongs to ${\cal D}_{\Sigma}$, if there is some $\ff t'$, such that
$\ff \Sigma_{\eta} = \ff \Sigma_{\ff t',\ff \eta, \ff U}$, i.e.\ such that 
$\ff \Sigma_{\ff \eta}$ is the exact (interaction) self-energy of the problem given by 
$H' = H(\ff t', \ff \eta, \ff U)$ for some $\ff t'$.
Hence, the hopping parameters $\ff t'$ {\em span} the domain of the self-energy 
functional. 

This definition is very convenient as it automatically ensures the correct 
analytical and causal properties for any self-energy in the domain.
It also avoids formal difficulties for pure or non-interacting 
systems which arise from the fact that conditional equations such as Eq.\ (\ref{eq:disfunc})
become tautological in these limits and cannot serve to define a self-energy functional. 
With the above definition of the domain, however, this becomes irrelevant as for the 
cases of pure or non-interacting systems the domain consists of a single element 
or a null set only.

With these preparations, let us discuss the limits in detail:

(i) The pure and non-interacting case is given by $P(\ff \eta) =
\delta(\ff \eta - \ff \eta_0)$ and $\ff U=0$.
Note that we can set $\ff \eta_0=0$ for simplicity (in the absence of disorder, a non-zero
$\ff \eta_0$ merely implies a redefinition of the hopping: $\ff t + \ff \eta_0$).
The domain of the self-energy functional \refeq{func} shrinks to the point
$\ff S = \ff \Sigma_{\ff \eta} = 0$.
According to \refeq{fdef} and \refeq{gex}, this implies
\be
  \widehat{F}_{\ff U=0}[0] = \widehat{\Omega}_{\ff U=0}\left[ 
  \ff G_{\ff t, 0, 0}^{-1} \right]
  - 
  \Tr \ln \ff G_{\ff t, 0, 0} = 0 
\ee
as $\Tr \ln \ff G_{\ff t, 0, 0}$ is the grand potential of a system of non-interacting 
electrons with hopping $\ff t$.
For $\ff \eta = \ff \eta_0 = 0$ and $\ff S = \ff \Sigma_{\ff \eta} = 0$ the second and
the third term on the r.h.s.\ of \refeq{func} cancel, and thus one is
left with 
\be
  \widehat{\Omega}_{\ff t, P_0, 0}[0,0]
  = \mbox{Tr} \ln \ff G_{\ff t,0,0}
\ee
which is the correct result.

(ii) In the case of $P(\ff \eta) = \delta(\ff \eta)$ 
but finite interaction $\ff U \ne 0$, one still has 
$\ff \Sigma_{\ff \eta} = \ff \Sigma = \ff S$ on the domain, and
due to the cancellation of the second and the third term on the r.h.s.\ of 
\refeq{func} the self-energy functional reduces to \refeq{sef}, as expected.

(iii) For a system of non-interacting electrons ($\ff U = 0$) moving in a 
disorder potential with $P(\ff \eta) \ne \delta(\ff \eta)$, one has $\ff \Sigma_{\ff \eta} = 0$ 
on the domain of the self-energy functional and thus $\widehat{F}_{\ff U=0}[0] = 0$. 
With $\widehat{\ff \Gamma}_{P}[\ff S] \equiv
\widehat{\ff \Gamma}_{P}[\ff S , \ff \Sigma_{\ff \eta} = 0]$,
this yields the self-energy functional
$\widehat{\Omega}_{\ff t, P}[\ff S] 
\equiv
\widehat{\Omega}_{\ff t, P, \ff U=0}[\ff S , \ff \Sigma_{\ff \eta} = 0]$
with 
\ba
  \widehat{\Omega}_{\ff t, P}[\ff S] 
&=&
  \mbox{Tr} \ln \frac{1}{\ff G_{\ff t,0,0}^{-1} - \ff S}
- 
  \mbox{Tr} \ln \widehat{\ff \Gamma}_{P}[\ff S]
\nonumber \\
&+&
 \left\langle \mbox{Tr} \ln 
  \frac{1}{ \widehat{\ff \Gamma}_{P}[\ff S]^{-1}
  + \ff S - \ff \eta 
  }
  \right\rangle_P
\labeq{disonly}
\ea
for the problem with disorder only.
The last two terms play the same role as the functional 
$\widehat{F}_{\ff U,\ff \eta}[\ff \Sigma]$ for the problem with interaction only.
This functional is discussed in the next section.

\section{Mean-field approximations}
\label{sec:sys}

Mean-field approximations for systems with local disorder and local interactions 
represent a simple but instructive class of approximations within the framework of 
the self-energy-functional approach.
It is well known that any mean-field theory of disorder will be deficient in various 
ways.
Issues such as localization cannot be addressed, for example, by means of the famous
coherent-potential approximation (CPA). \cite{HM74}
Nevertheless, the mean-field concept represents an important benchmark and starting 
point for improvements and is in many cases the best we have at hand for practical
calculations.

We start by considering the functional (\ref{eq:disonly}) for the non-interacting,
disorder-only limit of the model (\ref{eq:ahm}).
This is the Anderson model
\be
   H = \sum_{\ff x\ff x'} t_{\ff x\ff x'} c^\dagger_{\ff x} c_{\ff x'} 
   + \sum_{\ff x} \eta_{\ff x} c^\dagger_{\ff x} c_{\ff x}
\labeq{and}
\ee
with local disorder given by \refeq{locdis} and some density $p(\eta_{\ff x})$
characterizing, for example, an alloy with $R$ components ($\sum_r p_r =1$):
\be
  p(\eta) = \sum_{r=1}^R p_r \delta(\eta - \eta_r) \: .
\ee
For simplicity, the spin index is suppressed.

\begin{figure}[t]
  \includegraphics[width=0.85\columnwidth]{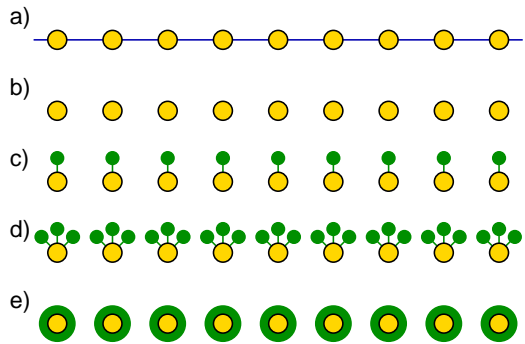}
\caption{(Color online)
a) Representation of the (disorder) Anderson model \refeq{and}. 
b) Reference system generating the atomic approximation (see text).
c) Two-site approximation. 
d) Improved mean-field approximation due to atomic reference system with more
bath sites ($n_s=4$).
e) The optimum mean-field approximation generated by a continuum of bath sites
is the coherent-potential approximation (CPA).
If a) represents the (pure) Hubbard model, b) yields a Hubbard-I-type atomic
approximation, d) is a typical mean-field approximation, and the optimum 
mean-field approximation e) is given by the DMFT. 
In case that a) represents the Anderson-Hubbard model \refeq{ahm}, the
reference system d) leads to the DMFT+CPA approach.
}
\label{fig:dmft}
\end{figure}

A mean-field or single-site approximation is generated by a reference system
consisting of decoupled sites (see Fig.\ \ref{fig:dmft}b), i.e.\ by switching
off the hopping term. 
For an alloy-type disorder, this reference system is exactly solvable, as one 
has to compute the Green's function for a finite number of $R$ configurations.
Models with a continuous distribution $p(\eta_{\ff x})$ have to be simulated by a 
finite but large $R$.
Adding ``bath'' sites to the reference system, i.e.\ sites with fixed,
configuration-independent on-site energies, enlarges the space of variational 
parameters $\ff t'$ and trial self-energies $\ff S= \ff S_{\ff t',P}$ and implies
an improved mean-field approximation (see Fig.\ \ref{fig:dmft}c,d).
Note that the disorder part is still the same as in the original model (\ref{eq:and}),
as it is required to justify \refeq{res}.

The Hamiltonian of the reference system reads:
\ba
   H' 
   &=&
   \sum_{\ff x} t'_{\ff x\ff x} c_{\ff x}^\dagger c_{\ff x} + \sum_{\ff x} \eta_{\ff x} 
   c^\dagger_{\ff x} c_{\ff x}
   \nonumber \\
   &+&
   \sum_{\ff x} \sum_i V_{i}^{(\ff x)} ( c^\dagger_{\ff x} a_{{\ff x}i} + \mbox{h.c.} ) 
   + \sum_{\ff x} \sum_i \varepsilon_i^{({\ff x})} a_{{\ff x}i}^\dagger a_{{\ff x}i} \: .
   \nonumber \\
\labeq{refsysdis}   
\ea
It consists of the local part of \refeq{and} and, for each site $\ff x$, includes bath
sites with (configuration-independent) energies $\varepsilon_i^{(\ff x)}$ hybridizing with 
the original sites via $V_{i}^{(\ff x)}$ where $i=2,...,n_s$.
As the Hamiltonian describes an impurity model with identical and decoupled replicas 
at any site $\ff x$, it is in fact sufficient to focus on one impurity model only.
The site index $\ff x$ can be suppressed in this case.
This reflects the translational symmetry of averaged quantities in the original model
(\ref{eq:and}).

Due to the decoupling of the original sites, the reference
system yields a trial self-energy which is local. 
Spatial correlations due to non-local contributions of the self-energy are
neglected.
Differences between different mean-field approximations are due to the temporal
correlations, i.e.\ due to additional bath sites.
Obviously, the optimum single-site approximation is obtained for a continuum 
of bath sites $n_s \to \infty$ (Fig.\ \ref{fig:dmft}e).

%*********************************************************************************
%*********************************************************************************

\subsection{Atomic approximation}
\label{sec:atomic}

The simplest approximation is obtained for $n_s=1$, i.e.\ no additional bath sites
(Fig.\ \ref{fig:dmft}b).
This case is instructive as it allows for a largly analytical treatment which elucidates
some general features of the disorder SFT.
The reference-system Hamiltonian consists of the first two terms on the r.h.s.\ of 
Eq.\ (\ref{eq:refsysdis}).
We consider the case of a binary alloy ($R=2$) with on-site energies $\eta_1$ and $\eta_2$
and corresponding probabilities $p_1$ and $p_2$.
The only variational parameter left is the (configuration-independent) on-site energy
$t'_0 \equiv t'_{\ff x\ff x}$.

For $T=0$ and $\mu = 0$ the averaged grand potential of the reference system 
$\Omega' \equiv \Omega_{\ff t',P,0}$ is easily calculated:
\begin{equation}
  \Omega' / L
  = 
  p_1 (t'_0 + \eta_1) \Theta(- t'_0 - \eta_1)
  + 
  p_2 (t'_0 + \eta_2) \Theta(- t'_0 - \eta_2)
\: .
\end{equation}
This is the first term on the r.h.s.\ of \refeq{res}.

The averaged Green's function of the reference system (Eq.\ (\ref{eq:defg})) is obtained 
immediately as $\Gamma'_{\ff x\ff x'}(\omega) = \delta_{\ff x \ff x'} \Gamma'(\omega)$ 
with
\begin{equation}
  \Gamma'(\omega) = 
  \frac{p_1}{\omega - t'_0 - \eta_1} 
  +
  \frac{p_2}{\omega - t'_0 - \eta_2} \: .
\end{equation}
Using the considerations of Ref.\ \onlinecite{Pot03b} (see Eq.\ (20) therein), only the poles 
$\omega'_{1,2}=t'_0+\eta_{1,2}$ enter the result for the third term on the r.h.s.\ of 
\refeq{res}:
\begin{equation}
  \mbox{Tr} \ln \ff \Gamma_{\ff t',P,0} / L
  =
  \sum_{r=1,2} \omega'_r \Theta(- \omega'_r)
  -
  R_S / L \: .
\end{equation}
The term $R_S$ (see Ref.\ \onlinecite{Pot03b}) cancels out later.

With the free Green's function of the reference system, $G'_0(\omega) = 1/(\omega - t'_0)$
we get from Eq.\ (\ref{eq:defs}) the self-energy as
$S_{\ff x \ff x'}(\omega) = \delta_{\ff x \ff x'} S(\omega)$ where
\begin{equation}
  S(\omega) = \langle \eta \rangle + \frac{\langle \eta^2 \rangle - \langle \eta \rangle^2}
  {\omega - t'_0 - \eta_1 - \eta_2 + \langle \eta \rangle} \: .
\label{eq:atomicsigma}
\end{equation}
Let $\varepsilon(\ff k)$ be the eigenvalues of the hopping matrix $\ff t$ and
let $\Gamma_{\ff x\ff x'}(\omega) = L^{-1} \sum_{\ff k} e^{i \ff k (\ff x -\ff x')} 
\Gamma_{\ff k}(\omega)$ with $L$ the number of lattice sites and
$\Gamma_{\ff k}(\omega) = 1/(\omega - \varepsilon(\ff k) - S(\omega))$ be the averaged 
Green's function of the original system as it appears in Eq.\ (\ref{eq:res}).
From \refeq{atomicsigma} we find
\begin{equation}
  \Gamma(\omega) = 
  \frac{\alpha_1}{\omega - \omega_1} 
  +
  \frac{\alpha_2}{\omega - \omega_2} \: 
\end{equation}
with poles $\omega_{1,2}=(\varepsilon(\ff k) + t'_0 + \eta_1 + \eta_2)/2 \pm
[(\varepsilon(\ff k) + 2 \langle \eta \rangle - \eta_1 - \eta_2 - t'_0)^2/4 
+ \langle \eta^2 \rangle - \langle \eta \rangle^2]^{1/2}$ and weights
$\alpha_1 = (\omega_1 - \eta_1 - \eta_2 + \langle \eta \rangle - t'_0) / (\omega_1 - \omega_2)$
and 
$\alpha_2 = (\omega_2 - \eta_1 - \eta_2 + \langle \eta \rangle - t'_0) / (\omega_2 - \omega_1)$.
The $\ff k$ dependence of the poles is only due to $\varepsilon(\ff k)$, i.e.\
$\omega_r = \omega_r(\ff k) = \omega_r(\varepsilon(\ff k))$.
Hence, using Eq.\ (21) of Ref.\ \onlinecite{Pot03b} and the definition of the free density
of states $\rho_0(z) = L^{-1} \sum_{\ff k} \delta(z -\varepsilon(\ff k))$, we get
\begin{eqnarray}
  \mbox{Tr} \ln \ff \Gamma_{\ff t,P,0} / L
  &=&
  \sum_{r=1,2} \int_{-\infty}^\infty dz \: \rho_0(z) \omega_r(z) \Theta(- \omega_r(z))
  \nonumber \\
  &-&
  R_S / L \: .
\end{eqnarray}
for the second term on the r.h.s.\ of \refeq{res}.

Adding the three contributions, one can search for a stationary point numerically.
Here, we restrict ourselves to the particle-hole symmetric case, i.e.\ we assume
$\rho_0(z)=\rho_0(-z)$, $\eta_1=\eta_2$ and $p_1=p_2=1/2$.
While the different contributions to the SFT grand potential \refeq{res} are asymmetric,
it is straightforward to see that their sum is symmetric with respect to a sign change 
of the variational parameter: $\Omega_{\ff t,P,0}(t'_0)=\Omega_{\ff t,P,0}(-t'_0)$. 
Furthermore, the dependence on $t'_0$ is smooth for $|t'_0| < |\eta_1|$.
Hence, the SFT grand potential is stationary at $t'_0=0$ which had to be expected by
virtue of particle-hole symmetry. 

The optimum self-energy is given by \refeq{atomicsigma} with $t'_0=0$.
It consists of the ``virtual-crystal'' potential $\langle \eta \rangle$ (note that 
$S(\omega) \equiv \langle \eta \rangle$ is the so-called virtual-crystal approximation 
\cite{Gon92}) and a frequency-dependent part with one simple pole at $\omega=0$.
Its weight is given by the disorder strength, namely by the variance
$\langle \eta^2 \rangle - \langle \eta \rangle^2=\eta_1^2$.
For any finite $\eta_1$ this leads to a splitting of the averaged local density of states 
into a lower and an upper alloy band. 
In the case of strong disorder, this result is qualitatively correct as could be expected
since the high-frequency behavior of $S(\omega)$ is correct up to the order $1/\omega^2$.
However, the widths of the alloy bands turns out to be too small. 

This atomic-like approximation is very much reminiscent of the Hubbard-I approximation
\cite{Hub63} for the pure but interacting system -- although the Hubbard-I self-consistency 
condition is somewhat different from the SFT Euler equation and leads to different results
away from the particle-hole symmetric point. 
The analogy between approximations for disordered but non-interacting and pure but interacting
systems relies on the same structure of the reference system.
In this analogy, the static part of the disorder self-energy, the virtual-crystal potential 
$\langle \eta \rangle$, corresponds to the static part of the Hubbard-I self-energy
which, for the Hubbard model, is given by $U \langle c^\dagger_{\ff x\sigma} 
c_{\ff x\sigma} \rangle$, i.e.\ to the Hartree(-Fock) approximation. 
A combination of both Hubbard-I-type approximations for the interacting and disordered system 
is straightforwardly set up with an atomic-like reference system including (local) interaction 
and disorder.

In view of the simplicity of the Hubbard-I-type approach, it is remarkable that the
variational optimization of the on-site hopping ensures thermodynamical consistency
with respect to the particle number, i.e.\ the averaged particle number as obtained 
from the (approximate) disorder-averaged grand potential as a $\mu$ derivative is
always the same as the averaged particle number calculated by integration of the
(approximate) disorder-averaged single-particle Green's function. 
The proof for this consistency is analogous to that given in Ref.\ \onlinecite{AAPH06a},
see Appendix \ref{sec:tdconsis}.
 
It is also interesting that the stationary point $t'=0$ actually (locally) {\em maximizes}
the SFT grand potential as has been verified by a simple numerical evaluation of \refeq{res}
for the particle-hole symmetric case.
In general, and for high-dimensional parameter spaces in particular, we expect that stationary 
points are saddles.

\subsection{Two-site approximation}

The two-site approximation (see Fig.\ \ref{fig:dmft}c) is the simplest mean-field
approach beyond the atomic approximation.
For the case of the pure Hubbard model without disorder is has proven to be very
instructive and successful. \cite{Pot03b,BP00,Pot01,IKSK05a,IKSK05b}
It provides a handy mean-field approach which is able to reproduce qualitatively 
the DMFT phase diagram for the Mott transition in the single-band model and which
has been employed to study more complex two- and multi-orbital systems.
A recent application \cite{BP05} to the Anderson-Hubbard model has shown that the 
approach can straightforwardly be extended to disordered (and interacting) systems
along the lines described here. The two-site approximation qualitatively reproduces
the results of the DMFT+CPA method but with a minimum computational effort.

\subsection{Coherent-potential approximation}
\label{sec:cpa}

The two-site approach can of course be improved by adding more bath sites
(Fig.\ \ref{fig:dmft}d).
The best mean-field approach is then obtained for $n_s \to \infty$, i.e.\
for a continuum of bath sites (Fig.\ \ref{fig:dmft}e).
Varying all parameters, i.e.\ $t'_0 \equiv t'_{\ff x\ff x}$, $\varepsilon_i=\varepsilon_i^{(\ff x)}$ 
and $V_i \equiv V_i^{(\ff x)}$, yields the optimum local self-energy as the
stationary point
\ba
  0 
  &=& 
  \frac{\partial}{\partial \ff t'} \widehat{\Omega}_{\ff t, P}[\ff S_{\ff t',P}] 
  =
  \sum_n \sum_{\ff x}
  \frac{\delta \widehat{\Omega}_{\ff t, P}[\ff S_{\ff t',P}]}{\delta S_{\ff x\ff x}(i\omega_n)}
  \frac{\partial S_{\ff x\ff x}(i\omega_n)}{\partial \ff t'}
  \nonumber \\
\label{eq:localeulerdisonly}
\ea
of the function $\Omega_{\ff t,P}(\ff t')\equiv\widehat{\Omega}_{\ff t, P}[\ff S_{\ff t',P}]$.
Here it has been used that the self-energy (and also its $\ff t'$-derivative) is local and
non-zero at the impurity site only.
From \refeq{disonly} we have:
\ba
  \Omega_{\ff t,P}(\ff t')
&=&
  \mbox{Tr} \ln \frac{1}{\ff G_{\ff t,0,0}^{-1} - \ff S_{\ff t',P}}
- 
  \mbox{Tr} \ln \widehat{\ff \Gamma}_{P}[\ff S_{\ff t',P}]
\nonumber \\
&+&
 \left\langle \mbox{Tr} \ln 
  \frac{1}{ \widehat{\ff \Gamma}_{P}[\ff S_{\ff t',P}]^{-1}
  + \ff S_{\ff t',P} - \ff \eta 
  }
  \right\rangle_P \: .
\labeq{disonlyin}
\ea
Note that the last term on the r.h.s.\ is just the averaged grand potential
of the reference system $\Omega_{\ff t',P}$.

The Euler equation (\ref{eq:localeulerdisonly}) is satisfied, if 
\be
  0 = \frac{\delta \widehat{\Omega}_{\ff t, P}[\ff S_{\ff t',P}]}{\delta S_{\ff x\ff x}(i\omega_n)}
\ee
for each site $\ff x$ and for each Matsubara frequency $\omega_n$.
Calculating the derivative of the functional \refeq{disonly}, we obtain:
\ba
&&
  \frac{1}{T}
  \frac{\delta \widehat{\Omega}_{\ff t, P,0}[\ff S_{\ff t',P}]}{\delta S_{\ff x\ff x}(i\omega_n)} 
  =
  \left( \frac{1}{i\omega_n + \mu - \ff t - \ff S(i\omega_n)} \right)_{\ff x\ff x}
\nonumber \\
&&
  - \left(\left\langle 
  \frac{1}{ \widehat{\ff \Gamma}_{P}[\ff S_{\ff t',P}]^{-1}
  + \ff S(i\omega_n) - \ff \eta 
  }
  \right\rangle_P \right)_{\ff x\ff x} \: .
\nonumber \\  
\labeq{eeu}
\ea
The first term on the r.h.s.\ is the local element of the averaged Green's 
function of the lattice model, $\Gamma_{\rm loc}(i\omega_n)$ which is 
calculated with the approximate local self-energy $S(i\omega_n)$
(exploiting translational symmetry, the site index can be supressed). 
The second term on the r.h.s.\ is the averaged Green's function of the 
reference system at the impurity site $\Gamma'_{\rm loc}(i\omega_n)$.
The optimum local disorder self-energy is thus determined by the
condition that the local averaged Green's function equals the 
averaged impurity Green's function of the reference system:
\be
  \Gamma_{\rm loc}(i\omega_n) = \Gamma'_{\rm loc}(i\omega_n) \: .
\labeq{cpaeq}
\ee
This is exactly the self-consistency condition of the coherent-potential
approximation.

To elucidate this point, we note that, for the reference system \refeq{refsysdis}, 
$\Gamma'_{\rm loc}(i\omega_n)$ is the average of the impurity Green's function for the 
different on-site energies, i.e.:
\be
  \Gamma'_{\rm loc}(i\omega_n) = \sum_r p_r 
  \frac{1}{i\omega_n + \mu - \eta_r - \Delta(i\omega_n)} \: ,
\labeq{gamhyb}
\ee
where $\Delta(i\omega_n) = \sum_i V_i^2 / (i\omega_n + \mu - \varepsilon_i)$ is 
the hybridization function.
On the other hand, from the definition of the disorder self-energy for 
the reference system, we have
\be
  \Gamma'_{\rm loc}(i\omega_n) = \frac{1}{i\omega_n + \mu - \Delta(i\omega_n) - S(i\omega_n)} \: .
\labeq{gamsig}
\ee
Eliminating the hybridization function, we get:
\be
  \Gamma'_{\rm loc}(i\omega_n) = \sum_r p_r 
  \frac{1}{ \frac{1}{\Gamma'_{\rm loc}(i\omega_n)} + S(i\omega_n) - \eta_r} \: .
\labeq{cpaeqs}
\ee
After a few manipulations, this equation can be cast into the form
\be
  \sum_r p_r \frac{\eta_r - S(i\omega_n)}
  {1 - \Gamma'_{\rm loc}(i\omega_n) ( \eta_r - S(i\omega_n) )} = 0
\ee
which makes contact with the original derivation of the CPA
where the averaged atomic scattering matrix is set to zero.
\cite{EKL74}
Introducing the free Bloch-band dispersion $\varepsilon(\ff k)$ as the 
Fourier transform of the hopping $\ff t$ and the free Bloch-density
of states $\rho_0(z) = (1/L) \sum_{\ff k} \delta(z - \varepsilon(\ff k))$,
we have
\be
  \Gamma_{\rm loc}(i\omega_n) = 
  \int \frac{\rho_0(z)dz}{i\omega_n + \mu - z - S(i\omega_n)} \: .
\labeq{gamdos}
\ee
In combination with \refeq{cpaeq}, this equation can be used to eliminate 
$\Gamma'_{\rm loc}(i\omega_n)$ from \refeq{cpaeqs} to obtain a single
conditional equation for $S(i\omega_n)$.
 
For a practical calculation, one may set up the following self-consistency
cycle:
Starting from a guess for the hybridization function $\Delta(i\omega_n)$, 
the averaged impurity Green's function $\Gamma'_{\rm loc}(i\omega_n)$ can 
be computed from \refeq{gamhyb}.
With the help of \refeq{gamsig} this determines $S(i\omega_n)$ which is then 
used in \refeq{gamdos} to get the CPA Green's function 
$\Gamma_{\rm loc}(i\omega_n)=\Gamma'_{\rm loc}(i\omega_n)$.
Using \refeq{gamsig} again, a new hybridization function can be found.

Obviously, the bath sites of the reference system play the role of an ``effective medium''.
The bath parameters or, equivalently, the hybridization function $\Delta(i\omega_n)$ parameterize 
the local disorder self-energy in the most general way consistent with causality requirements.
The present rederivation of the CPA therefore very clearly shows the CPA to be the best local 
approximation.

\subsection{Dynamical mean-field theory}

The reference systems shown in Fig.\ \ref{fig:dmft} can also be used in the context
of a {\em pure} system without disorder but with local interaction such as the Hubbard
model. In this case the theory reduces to the conventional SFT.
The reference system b) with the hopping between the correlated sites switched off
generates an atomic approximation very much the same as the Hubbard-I approximation
\cite{Hub63} but with the Hubbard-I self-consistency replaced by the SFT Euler equation
for that reference system which is different. 
The rather crude atomic-like approximation can be improved by adding uncorrelated bath sites.
This yields the reference system d) which is a set of disconnected single-impurity Anderson 
models with $n_s < \infty$ sites each.
Qualitatively, the results of the two-site dynamical-impurity approximation 
\cite{Pot03b,BP00,Pot01,IKSK05a,IKSK05b} (Fig.\ \ref{fig:dmft}c) are already close 
to the $n_s = \infty$ limit. 
The convergence with increasing $n_s$ is rapid (see Ref.\ \onlinecite{Poz04}, for example).

The most general local trial self-energy compatible with the requirements of causality is
generated in the limit $n_s = \infty$. 
Here, the calculation proceeds in a way analogous to the previous section and yields 
the Euler (or self-consistency) equation (see Ref.\ \onlinecite{Pot03a}):
\be
  G_{\rm loc}(i\omega_n) = G'_{\rm loc}(i\omega_n) \: .
\labeq{dmfteq}
\ee
Here $G_{\rm loc}(i\omega_n)$ is the on-site element of the lattice Green's function 
(Fig.\ \ref{fig:dmft}a) calculated approximately from the approximate local self-energy 
via the Dyson equation for the lattice model, and $G'_{\rm loc}(i\omega_n)$ is the exact
Green's function of the reference system (Fig.\ \ref{fig:dmft}e) at the impurity site.
Eq. \ (\ref{eq:dmfteq}) is just the self-consistency equation of the dynamical 
mean-field theory. \cite{MV89,GK92a,Jar92,GKKR96}

The CPA for a non-interacting system with local disorder has a formal structure which is 
very similar to the DMFT for a pure system with local interaction.
This is apparent when comparing the respective self-consistency conditions (\ref{eq:cpaeq}) 
and (\ref{eq:dmfteq}) and also the respective self-consistency cycles which serve to iteratively 
solve the mean-field equations.
As the DMFT, the CPA becomes exact in the limit of high spatial dimensions $D\to \infty$ as 
has been shown by Vlaming and Vollhardt. \cite{VV92,SS72}
Both approaches are characterized as approximations that yield the optimum local self-energy. 
All this is made very obvious within the framework of the SFT which discloses the formal
analogies between Green's-function-based approaches to disordered or interacting systems.

A self-evident idea that has been persued in the past is to derive the DMFT (or a different
many-body approach to pure interacting systems) by using the formal structure of the CPA. 
This requires, however, a transformation of a pure system with local interaction to a 
non-interacting system with local disorder to which the CPA can be applied.
Hubbard's alloy analogy \cite{Hub64b} and also a refined version \cite{HN96} represent such 
transformations.
The subsequent application of the CPA to the Hubbard's ficticious alloy yields the so-called
Hubbard-III approximation. 
The alloy analogy (the transformation) itself, however, must be seen as a rough 
approximation which fails to recover Fermi-liquid properties even for weak interactions.
The ``many-body CPA'' by Hirooka and Shimizu, \cite{HS77} the ``generalized CPA'' by
Jani\u{s} \cite{Jan89} as well as the ``dynamical CPA'' of Kakehashi \cite{Kak92} go beyond
a simple analogy.
The idea of the dynamical CPA is to perform the transformation to an effective one-particle 
Hamiltonian within the functional-integral formalism by means of a Hubbard-Stratonovich 
transformation and to recover the DMFT by the subsequent use of the CPA (see Ref.\ 
\onlinecite{Kak02} for a discussion).

In this context it is also worth mentioning the Falicov-Kimball model which can be considered as 
a variant of the Hubbard model with the hopping of one of the two spin species switched off.
The exact solution of the Falicov-Kimball model in the limit $D\to \infty$ has been worked
out by Brandt and Mielsch. \cite{BM89,BM90,BM91}
They could show that the local (interaction) self-energy of the mobile carriers is given 
by the CPA (disorder) self-energy -- no alloy analogy is necessary for this simplified 
model. It is tempting to understand the dynamics of the mobile carriers as the scattering 
of non-interacting particles from the (local and uncorrelated binary) disorder potential 
generated by the immobile ones.

\subsection{DMFT+CPA}
\label{sec:dmftcpa}

For a system with Hubbard-type interactions and local disorder, e.g.\ the prototypical
Anderson-Hubbard model \refeq{ahm}, the optimum mean-field theory is generated by the 
reference system shown in Fig.\ \ref{fig:dmft}e. 
Note that the reference system shares with the original system the same interaction part 
and the same disorder potential. 
The continuum of bath sites is uncorrelated and configuration independent.

We start from \refeq{res}.
For the present case the Euler equation
$\partial \Omega_{\ff t, P, \ff U}(\ff t') / \partial \ff t' = 0$
is satisfied if
\ba
  \frac{1}{T}
  \frac{\delta \widehat{\Omega}_{\ff t, P,\ff U}
  [\ff S_{\ff t',P,\ff U},\{ \ff \Sigma_{\ff t',\ff \eta,\ff U }\}]}
  {\delta S_{\ff x\ff x}(i\omega_n)} 
  = 
  0
\ea
for the local elements of the self-energy.
This is the Euler equation which fixes the variational paramters $\ff t'$.
Analogous to \refeq{eeu} we then obtain 
\be
  \left( \frac{1}{i\omega_n + \mu - \ff t - \ff S_{\ff t',P,\ff U} } \right)_{\ff x\ff x}
=
  \widehat{\ff \Gamma}_{P}[\ff S_{\ff t',P,\ff U},\{ \ff \Sigma_{\ff t',\ff \eta,\ff U}\}]
_{\ff x\ff x} \: ,
\labeq{eeeu}
\ee
i.e.\ the local averaged Green's function equals the averaged impurity Green's function of the 
reference system, $\Gamma_{\rm loc}(i\omega_n) = \Gamma'_{\rm loc}(i\omega_n)$, as in the 
non-interacting case with disorder.

Analogous to the procedure described in Sec.\ \ref{sec:cpa}, this self-consistency equation
can be solved in an iterative manner: 
Starting with a guess for the variational parameters $\ff t'$ or, equivalently, for the 
hybridization function $\Delta(i\omega_n) = \sum_i V_i^2 / (i\omega_n + \mu - \varepsilon_i)$,
the interacting impurity Green's function is calculated for any (local) configuration
of the reference single-impurity Anderson model and averaged over the configurations 
to get $\Gamma'_{\rm loc}(i\omega_n)$.
The self-energy is obtained from \refeq{gamsig} which is the defining equation 
\refeq{defs} for the self-energy also in the case of an interacting impurity model.
$S(i\omega_n)$ is then used in \refeq{gamdos} to get the averaged lattice Green's function 
$\Gamma_{\rm loc}(i\omega_n)$.
Via the self-consistency equation, this gives us $\Gamma'_{\rm loc}(i\omega_n)$ and,
using \refeq{gamsig} again, a new hybridization function.
Note that for a continuum of bath sites the optimum on-site element of the 
hopping $t'_0$ is always given by $t'_0=t_{\ff x\ff x}$. 
This can be shown in essentially the same way as has been done in Ref.\ \onlinecite{AAPH06a} 
for the pure case.

The self-consistency condition \refeq{eeeu} and the cycle desribed above constitutes
what is known as the DMFT+CPA approach which has been put forward by Jani\u{s} and 
Vollhardt \cite{JV92b} and by Dobrosavljevi\'c{} and Kotliar \cite{DK93b,DK94} in a
different context:
Originally, the DMFT+CPA has been introduced and charactized as the exact theory in the 
limit of infinite spatial dimensions which remains non-trivial for a proper scaling
of the hopping parameters. \cite{MV89}
The presented rederivation places the DMFT+CPA into the broader framework of the SFT.

The typical medium theory (TMT), as suggested recently by Dobrosavljevi\'c{}, Pastor and 
Nikoli\'c, \cite{DPN03} can be seen as a variant of the DMFT+CPA (see also Refs.\ 
\onlinecite{BHV05,Byc05}).
Here a modification of the above-described self-consistency cycle is considered by 
replacing ({\em ad hoc}) the usual arithmetical average of the interacting impurity 
Green's function over the (local) configurations by a {\em geometrical} one.
Actually, this geometrical average is applied to the corresponding (positive definite)
spectral density, and the Green's function $\Gamma'_{\rm loc}(i\omega_n)$ is obtained 
afterwards from this average as the usual Hilbert transform. 
A rederivation of this variant within the SFT does not seem to be possible:
Any approximation generated within the SFT preserves the elementary sum rule
$\int d\omega \, A_{\rm loc}(\omega) = 1$ for the local spectral density while this 
must be violated when averaging geometrically.
Despite this conceptual shortcoming, the TMT can be
motivated physically (see Refs.\ \onlinecite{DPN03,BHV05}) and clearly improves upon the 
mean-field concept as it is able to describe aspects of Anderson localization.
It should be seen as a pragmatic simplification of the statistical DMFT that
allows for practical computations in many cases which are not accessible to the 
statistical DMFT.
As can be shown for a simplified model, \cite{GP06} the results of TMT qualitatively 
agree very well with those obtained from a full evaluation of the statistical DMFT for 
interaction- and disorder-driven metal-insulator transitions.

\section{Cluster extensions}
\label{sec:cluster}

%*********************************************************************************
\begin{figure}[t]
  \includegraphics[width=0.9\columnwidth]{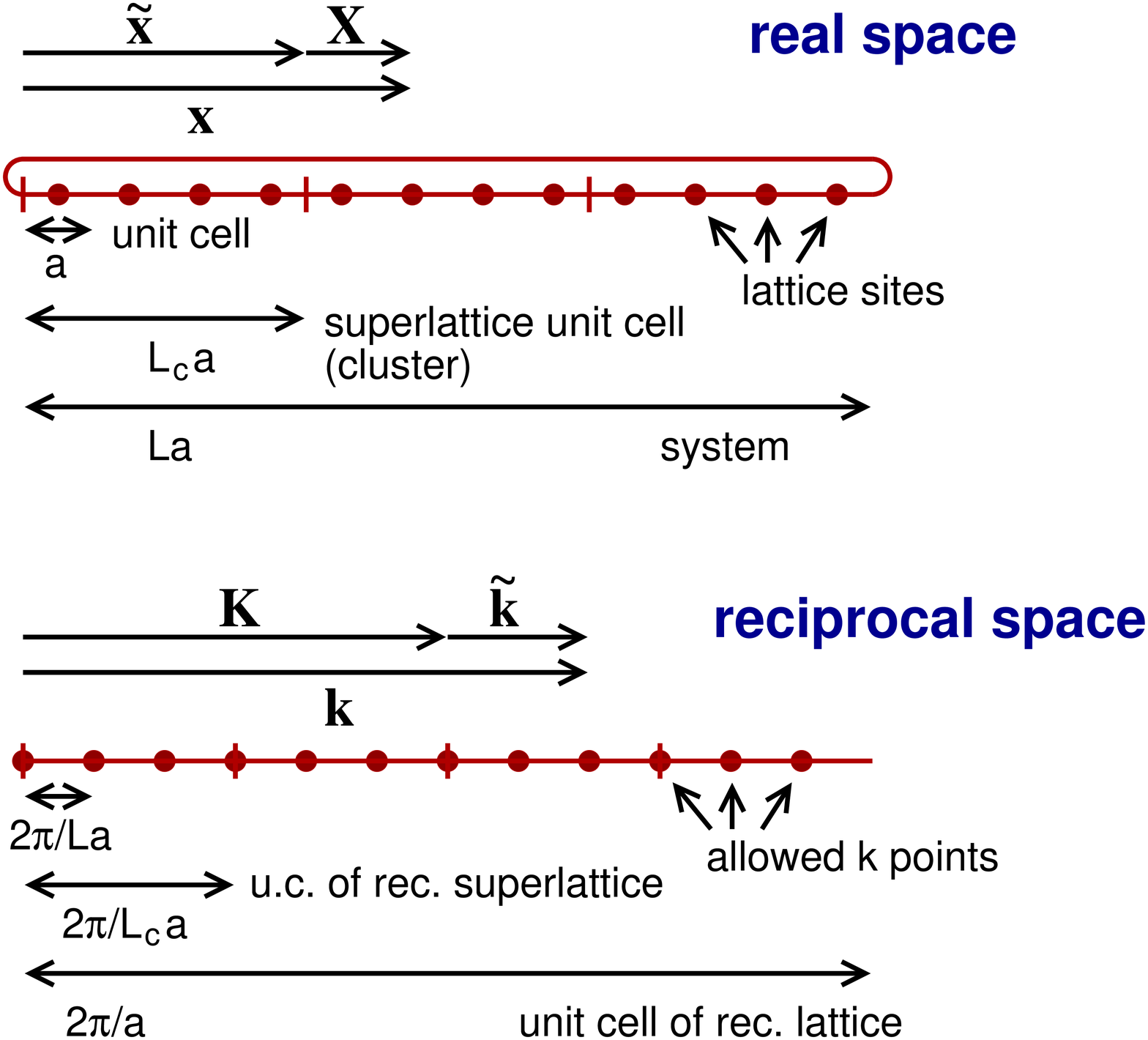}
\caption{(Color online)
Decomposition of real-space lattice vectors, $\ff x = \widetilde{\ff x} + \ff X$,
and reciprocal-space wave vectors, $\ff k = \widetilde{\ff k} + \ff K$, for a 
$D=1$ dimensional lattice (lattice constant $a$) with $L=12$ sites tiled with 
$L/L_c=3$ clusters consisting of $L_c=4$ sites each.
$\ff x$: original lattice.
$\widetilde{\ff x}$: superlattice.
$\ff X$: sites in a cluster.
Reciprocal space:
There are $L$ allowed wave vectors $\ff k$ in the unit cell of the lattice reciprocal to $\ff x$,
and there are $L/L_c$ allowed wave vectors $\widetilde{\ff k}$ in the unit cell of the lattice
reciprocal to the superlattice $\widetilde{\ff x}$. 
$\ff K$ are the reciprocal superlattice vectors, 
$\exp(i \ff K \widetilde{\ff x})=1$.
}
\label{fig:not}
\end{figure}
%*********************************************************************************

A mean-field approximation neglects spatial correlations by considering an effective 
{\em single-site} problem to generate the self-energy.
A straightforward idea to improve upon the mean-field concept is therefore to
replace the effective single-site model by an effective cluster model consisting 
of a finite (small) number of sites.
This should restore the important short-range correlations and provide a systematic
approach with the inverse cluster size as a small parameter.
The goal is to construct a cluster approximation which on the one hand reduces to the 
CPA or to a simpler mean-field approach, which on the other hand becomes exact in 
the infinite-cluster limit and which preserves the translational and point-group
symmetries of the lattice in addition.
This, however, is by no means trivial, and naive theories often suffer from causality 
violations. \cite{Gon92}
In contrast, it is easy to see that all approximations that are constructed within 
the SFT, including the different cluster approximations discussed in the following, 
are manifestly causal.
This is briefly discussed in Appendix \ref{sec:cau}.

Before discussing the different cluster approaches in detail, let us introduce some 
notations (see Fig.\ \ref{fig:not}):
We consider a system on a $D$-dimensional lattice of $L$ sites with periodic boundary 
conditions and $L \to \infty$ in the end.
The position vector to a site in the lattice is denoted by $\ff x$. 
There are $L$ allowed wave vectors in a unit cell of the reciprocal lattice which are 
denoted by $\ff k$.
The lattice is tiled with $L/L_c$ clusters consisting of $L_c$ sites each.
Let $\widetilde{\ff x}$ be the position vector of the cluster origin, and $\ff X$ the 
position vector of a site in a cluster, referring to the cluster origin. 
We then have the unique decomposition $\ff x = \widetilde{\ff x} + \ff X$.
The vectors $\widetilde{\ff x}$ form a superlattice with a unit-cell volume enlarged 
by the factor $L_c$.
In a unit cell of the reciprocal superlattice there are $L/L_c$ allowed wave vectors 
$\widetilde{\ff k}$.
Its volume is reduced by the factor $L_c$ as compared to the volume of the reciprocal
unit cell of the original lattice.
For a given $\ff k$ we have the unique decomposition $\ff k = \widetilde{\ff k} + \ff K$
where $\ff K$ are the vectors of the reciprocal superlattice, i.e.\ 
$\exp(i \ff K \widetilde{\ff x})=1$.
In the reciprocal unit cell of the original lattice, there are $L_c$ vectors $\ff K$.
These can also be interpreted as the allowed cluster wave vectors when imposing periodic 
boundary conditions on the individual cluster.

Consider the $L \times L$ matrix $\ff U$ with elements
\begin{equation}
  U_{\ff x, \ff k} = \frac{1}{\sqrt{L}} e^{i \ff k \ff x} \: ,
\labeq{ftu}
\end{equation}
and the $L/L_c \times L/L_c$ matrix $\ff V$ with elements
\begin{equation}
  V_{\widetilde{\ff x}, \widetilde{\ff k}} = \frac{1}{\sqrt{L/L_c}} e^{i\widetilde{\ff k}\widetilde{\ff x}} \: ,
\labeq{ftv}
\end{equation}
and the $L_c \times L_c$ matrix $\ff W$ with elements
\begin{equation}
  W_{\ff X, \ff K} = \frac{1}{\sqrt{L_c}} e^{i \ff K \ff X} \: .
\labeq{ftw}
\end{equation}
$\ff U$, $\ff V$ and $\ff W$ are unitary and define Fourier transformations between
the respective real and reciprocal spaces. Note that $\ff U \ne \ff V \ff W = \ff W \ff V$.
A quantity $A_{\ff x,\ff x'}$ which is invariant under lattice translations $\ff x_0$, 
i.e.\ $A_{\ff x+\ff x_0,\ff x'+\ff x_0}=A_{\ff x,\ff x'}$, is diagonalized by $\ff U$:
$(\ff U^\dagger \ff A \ff U)_{\ff k\ff k'}=A(\ff k) \delta_{\ff k,\ff k'}$. 
A quantity $A_{\ff x,\ff x'}$ which is invariant under superlattice translations 
$\widetilde{\ff x}_0$ as well as under cluster translations $\ff X_0$ (i.e.\ which is cyclic
on the cluster), 
$A_{\ff x+\widetilde{\ff x}_0,\ff x'+\widetilde{\ff x}_0} =
A_{\ff x+\ff X_0,\ff x'+\ff X_0}=A_{\ff x,\ff x'}$, is diagonalized by $\ff V \ff W$:
$(\ff W^\dagger \ff V^\dagger \ff A \ff V \ff W)_{\widetilde{\ff k} \ff K, \widetilde{\ff k'}\ff K'}
= A(\widetilde{\ff k}, \ff K) \delta_{\widetilde{\ff k},\widetilde{\ff k}'} \delta_{\ff K, \ff K'}$. 

\subsection{Variational cluster approach}

\begin{figure}[t]
  \includegraphics[width=0.85\columnwidth]{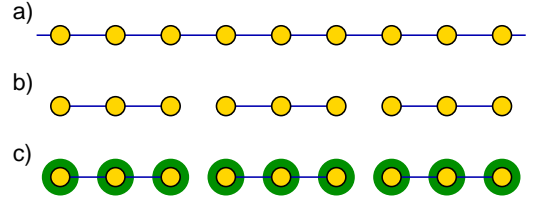}
\caption{(Color online)
a) Representation of the Anderson model \refeq{and}. 
b) Reference system generating the (disorder) variational-cluster 
approximation (VCA) (without variational optimization: cluster-perturbation
theory).
c) Reference system generating the molecular CPA (M-CPA).
In the case of an additional local (Hubbard) interaction, a) represents the 
Anderson-Hubbard model, b) yields a generalized VCA, and c) generates
the C-DMFT+M-CPA approximation.
}
\label{fig:cdmft}
\end{figure}

A straightforward extension of the single-site atomic approximation (see Sec.\ \ref{sec:atomic})
is the (disorder) variational cluster approach (VCA) which is the analog of the VCA
known \cite{PAD03,DAH+04,AEvdLP+04} for the interacting but pure system. 
To be definite, consider the Anderson model \refeq{and} again. 
The model is represented in Fig.\ \ref{fig:cdmft}a. 
A suitable reference system to include short-range correlations in the disorder self-energy 
consists of a set of isolated clusters of $L_c$ sites tiling the original lattice as shown 
by Fig.\ \ref{fig:cdmft}b. 
Since within the SFT, the (only) approximation is to replace the exact self-energy by the
self-energy of the reference system, the VCA must become exact in the limit $L_c \to \infty$,
i.e.\ the VCA is a systematic approach which is controlled by the inverse cluster size as a 
small parameter.

To calculate the cluster self-energy $\ff S_{\ff t',P}$, the Anderson model must be solved
on an individual cluster (by switching off the inter-cluster hopping) for all configurations. 
For large cluster size $L_c$, the computational cost therefore scales exponentially with $L_c$. 
This is, of course, characteristic for all cluster approximations.

The intra-cluster hopping parameters $\ff t'$ including the on-site energies are considered 
as variational parameters.
A non-variational variant of the VCA is obtained if one sets these parameters to the original
(physical) hopping parameters within the cluster: ${\ff t'}={\ff t}_{\rm intra}$.
This constitues the analog of the cluster-perturbation theory (CPT) \cite{GV93,SPPL00} 
which is known for the interacting, pure system. 

Obviously, the CPT is also systematic and controlled by $1/L_c$.
The additional optimization of the variational parameters $\ff t'$ within the VCA is expected 
to speed up the convergence with increasing $L_c$ as it does for interacting but pure 
systems. \cite{PAD03}
One should note, however, that the variational optimization should become less and less
important with increasing $L_c$ as the exact solution is approached for $L_c \to \infty$
with ${\ff t'}={\ff t}_{\rm intra}$ anyway.
It is therefore advisable to restrict the space of variational parameters and to optimize a 
few parameters only as, for example, those parameters that are close to the cluster boundary.

Parameter optimization is beneficial if (local) interactions are considered in addition
(as e.g.\ in the Anderson-Hubbard model).
Fig.\ \ref{fig:cdmft}b defines a generalized VCA for this case.
The additional interactions can drive a spontaneous breaking of a continuous symmetry.
This is signaled within the VCA by a stationary point of the SFT grand potential at a
non-vanishing value of a symmetry-breaking field.
Note that the respective field term can be added to the reference-system Hamiltonian 
if this is given by a one-particle term. 
This ``Weiss'' field is a ficticious one which clearly has to be distinguished from
a physical field and which describes spontaneous opposed to induced symmetry breaking.
For pure systems and spontaneous SU(2) and U(1) symmetry breaking, this concept has
already been applied successfully. \cite{DAH+04,AEvdLP+04,SLMT05,AAPH06a}

%*********************************************************************************
%*********************************************************************************

\subsection{Molecular CPA}

The (disorder) VCA is obtained as the cluster generalization of the atomic approximation.
Likewise, within the framework of the SFT, the cluster generalization of the CPA leads
to the so-called molecular CPA (M-CPA). \cite{Tsu72}
To be definite, we again consider the Anderson model \refeq{and}.
The reference system generating the M-CPA is shown in Fig.\ \ref{fig:cdmft}c.
It consists of a set of isolated clusters of $L_c$ sites each and a continuum of bath
sites attached to each of the original sites.

As in the case of the CPA, this allows to derive a simplified Euler equation.
Analogous to \refeq{localeulerdisonly} and \refeq{eeu} we can derive the following 
stationarity condition:
\ba
0 &=& T \sum_{n} \sum_{\ff x,\ff x'}
    \Big( 
    \frac{1}{i\omega_n + \mu - \ff t - \ff S(i\omega_n)} 
\nonumber \\
    &-& 
    {\ff \Gamma}_{\ff t',P}(i\omega_n)
    \Big)_{\ff x' \ff x, 1 1}
    \frac{\partial S_{\ff x \ff x'}(i\omega_n)}
         {\partial {{\bm t}'}} \: .
\labeq{mcpas}	 
\ea 
$\ff x$ and $\ff x'$ must belong to the same cluster since 
$S_{\ff x \ff x'}(i\omega_n)=0$ and also the ``projector'' 
$\partial S_{\ff x \ff x'}(i\omega_n) / \partial {{\bm t}'} = 0$
if $\ff x, \ff x'$ belong to different clusters as these are decoupled in the 
reference system.
Note that $\ff t'$ is a matrix labeled as $t'_{\ff x\ff x', i i'}$ where $i=1,...,n_s$
and $i \ne 1$ refers to the additional bath sites attached to each original site $\ff x$.
The same holds for ${\ff \Gamma}_{\ff t',P}$ and $\ff S \equiv \ff S_{\ff t',P}$.
As the reference system exhibits local disorder on the original sites $\ff x$, 
there are non-zero elements of the self-energy for $i=i'=1$ only.
We write $S_{\ff x\ff x', 1 1} = S_{\ff x\ff x'}$ for short.
Obviously, 
$S_{\ff x\ff x'} 
= S^{(\widetilde{\ff x})}_{\ff X\ff X'} \delta_{\widetilde{\ff x}\widetilde{\ff x}'}
= S_{\ff X\ff X'} \delta_{\widetilde{\ff x}\widetilde{\ff x}'}$.

Both, the original and the reference system, are invariant under translations
of the superlattice. 
Hence, Fourier transformation given by $V_{\widetilde{\ff x}, \widetilde{\ff k}}$
(\refeq{ftv}) is appropriate.
This yields:
\ba
0 &=& T \sum_{n} \frac{L_c}{L}
    \sum_{\widetilde{\ff k}} \sum_{\ff X\ff X'}
    \Big( 
    \frac{1}{i\omega_n + \mu - \ff t(\widetilde{\ff k}) - \ff S(i\omega_n)} 
\nonumber \\
    &-& 
    {\ff \Gamma}_{\ff t',P}(i\omega_n)
    \Big)_{\ff X' \ff X, 1 1}
    \frac{\partial S_{\ff X \ff X'}(i\omega_n)}
         {\partial {{\bm t}'}} \: .
\labeq{mcpastat}	 
\ea 
Here, $\widetilde{\ff k}$ runs over the $L/L_c$ wave vectors in the reduced Brillouin 
zone.
The trial self-energy $\ff S$ as well as ${\ff \Gamma}_{\ff t',P}$ are 
$\widetilde{\ff k}$ independent matrices in the intra-cluster position vectors
$\ff X,\ff X'$ while $\ff t(\widetilde{\ff k})$ is $\widetilde{\ff k}$ dependent.

The M-CPA self-consistency equation reads: \cite{Gon92}
\be
   \frac{L_c}{L}
   \sum_{\widetilde{\ff k}}
   \Big(
   \frac{1}{i\omega_n + \mu - \ff t(\widetilde{\ff k}) - \ff S} 
   \Big)_{\ff X \ff X'}
   =
   \Big(
   {\ff \Gamma}_{\ff t',P}
   \Big)_{\ff X \ff X'} \: ,
\labeq{mcpaeq}
\ee 
where the frequency dependence has been suppressed for convenience.
This generalizes the CPA self-consistency equation (\ref{eq:cpaeq}) which is obtained
from \refeq{mcpaeq} for $\ff X = \ff X'$ and cluster size $L_c=1$.

Comparing with the stationarity condition \refeq{mcpastat}, we note that the SFT grand 
potential is stationary at the M-CPA self-energy, i.e.\ at the self-energy 
$\ff S(i\omega_n)=\ff S_{\ff t',P}$ if variational parameters $\ff t'$ can be found such 
that \refeq{mcpaeq} is satisfied for any $i\omega_n$, $\ff X$ and $\ff X'$.
Hence, the reference system Fig.\ \ref{fig:cdmft}c generates the M-CPA.

%*********************************************************************************
\begin{figure}[t]
  \includegraphics[width=0.95\columnwidth]{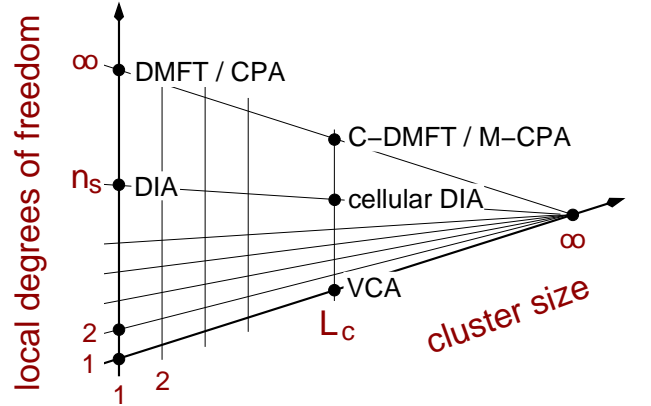}
\caption{(Color online)
Systematics of different approximations that can be constructed within
the SFT framework for pure interacting systems, disordered free systems
or interacting systems with disorder.
The interaction and/or disorder is assumed to be local.
The axis intercepts $L_c$ and $n_s$ characterize the reference system
and thereby the type of approximation.
A local approximation (``dynamical impurity approximation'', DIA) 
is obtained for $L_c=1$.
The optimum local approximation with $n_s \to \infty$ bath sites is
the DMFT, CPA, or DMFT+CPA, respectively.
A cluster (``cellular'' or ``molecular'') extension of DMFT/CPA is obtained for $L_c>1$.
The variational cluster approximation (VCA) is a cluster approximation without the use 
of bath sites ($L_c > 1$, $n_s=1$).
The exact solution would be obtained in the limit $L_c \to \infty$ irrespective of $n_s$.
}
\label{fig:sys}
\end{figure}
%*********************************************************************************

The following self-consistency cycle can be set up:
We start with a guess for the one-particle parameters of the reference system, 
i.e.\ the parameters for a single cluster $t'_{0,\ff X,\ff X'} \equiv 
t'_{\widetilde{\ff x}+\ff X,\widetilde{\ff x}+\ff X',11}$, $\varepsilon_i^{(\ff X)}$ 
and $V_i^{(\ff X)}$ ($i=2,...,n_s$).
For a fixed (intra-cluster) configuration of the disorder potential $\eta_{\ff X}$,
the intra-cluster part of the Green's function $\ff G_{\ff t'_0,\ff \eta}$ is an 
$L_c \times L_c$ matrix $\ff G'$ with elements 
$G'_{\widetilde{\ff x}+\ff X,\widetilde{\ff x}+\ff X'}$.
By solving its equation of motion, one easily verifies
$\ff G'(i\omega_n) = 1 / (i\omega_n + \mu - \ff t'_0 - \ff \eta - \ff \Delta(i\omega_n))$
where the matrix hybridization function $\ff \Delta(i\omega_n)$ is diagonal with
elements $\Delta_{\ff X}(i\omega_n) = \sum_{i=2}^\infty {V_i^{(\ff X)}}^2 / 
(i\omega_n + \mu -\varepsilon_i^{(\ff X)})$.
Averaging over the different configurations (e.g.\ $2^{L_c}$ for a binary
alloy) yields ${\ff \Gamma}_{\ff t',P}(i\omega_n)$ and, by comparison with the
free Green's function, the self-energy $\ff S(i\omega_n)$. 
This fixes the l.h.s.\ of \refeq{mcpaeq}, i.e.\ the averaged Green's function of the
original model. 
\refeq{mcpaeq} is then used to get a new ${\ff \Gamma}_{\ff t',P}(i\omega_n)$ and
thus a new hybridization function via
$\ff \Delta(i\omega_n) = i\omega_n + \mu - \ff t'_0 - \ff \eta - \ff G'(i\omega_n)^{-1}$.
New parameters $\varepsilon_i^{(\ff X)}$ and $V_i^{(\ff X)}$ are obtained as the poles
and weights of $\ff \Delta(i\omega_n)$. 
From the above it is obvious, however, that the self-consistency cycle can be set up 
for $\ff \Delta$ directly. 
The variational determination of $\ff t'_0$ is not a problem: Analogous to the discussion
given in Ref.\ \onlinecite{AAPH06a}, a high-frequency expansion easily shows that, at 
stationarity, $\ff t'_0$ must equal the intra-cluster part of the original hopping $\ff t$.

Note that for the non-interacting but disordered system, the M-CPA is related to the CPA
as is, for the interacting but pure system, the cellular DMFT (C-DMFT) \cite{KSPB01,LK00}
to the DMFT. 
The M-CPA is a conceptually simple and straightforward generalization of the
single-site CPA which includes short-range correlations in the disorder 
self-energy but is mean-field-like on a scale beyond the cluster size.
Like the variational cluster approximation (VCA), the M-CPA is systematic 
and controlled by $1/L_c$ as a small parameter, in principle.
Practical calculations, however, are restricted to comparatively small 
clusters due to the exponential growth of the number of local (intra-cluster)
configurations.
For systems with local interactions and disorder, the structure of the
reference system Fig.\ \ref{fig:cdmft}c generates a combined C-DMFT + M-CPA
approach which is the straightforward cluster extension of the DMFT+CPA
(Sec.\ \ref{sec:dmftcpa}).
The approximations within the SFT for different $n_s$ and $L_c$ are schematically grouped
in Fig.\ \ref{fig:sys}.

\subsection{Translation symmetry}

There is an apparent problem remaining: 
Due to the construction of the reference system as a set of decoupled clusters, 
the trial self-energies do not preserve the translational symmetries of the 
original lattice.
This is a self-evident problem for any cluster approximation which is formulated
in real space and has initiated the development of several further cluster 
approximations in the past. 
It has turned out, \cite{Gon92} however, that the construction of a self-consistent 
and systematic approach which on the one hand (for $L_c=1$) recovers the CPA and on 
the other ($L_c \to \infty$) approaches the exact solution, which respects the 
requirements of causality and which perserves the lattice translational symmetries 
at the same time, is not easy to find.

\begin{figure}[t]
  \includegraphics[width=0.85\columnwidth]{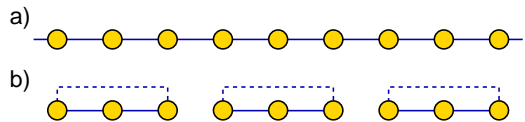}
\caption{(Color online)
a) Anderson model.
b) Possible reference system consisting of isolated clusters with 
periodic boundary conditions generating a variant of the CPT.
Within the VCA, the hopping marked as a dashed line may be treated
as a variational parameter.
}
\label{fig:period}
\end{figure}

A straightforward idea is to use a reference system as displayed by Fig.\ 
\ref{fig:period}b.
Here, isolated clusters are considered again but with a hopping $\ff t'$ satisfying 
periodic boundary conditions. 
This restores translational invariance within the cluster at least.
The idea has been put forward in the context of a pure but interacting system as a
modified cluster-perturbation theory (CPT with periodic boundary conditions) by
Zacher et al.\ \cite{ZEAH00} but was recognized \cite{SPP02} to give less convincing 
results when compared to the usual CPT.
Later on it could be shown \cite{PAD03} within the VCA that periodic boundary conditions 
are in fact unfavorable:
One simply has to treat the hopping integral connecting the edges of a cluster as a 
variational parameter. 
The SFT grand potential turns out to be stationary if this hopping integral vanishes.
This corresponds to open boundary conditions.
In any case, however, it is obvious that the full translational symmetry cannot be
restored with the reference system Fig.\ \ref{fig:period}b.

Another straightforward idea is to formulate the original problem as well as the 
reference system in reciprocal space by using annihilators $c_{\ff k}$ instead 
of $c_{\ff x}$ etc.
In graphical representations of $H$ and $H'$ like the one in Fig.\ \ref{fig:period},
dots would have to be reinterpreted as referring to one-particle states labeled by
$\ff k$.
The advantage is that, by construction, $H'$ always exhibits the full translational
symmetry. 
While there are no principle objections, it appears to be impossible in practice,
however, to generate meaningful approximations in this way.
The reason is that a local interaction in real space transforms into a delocalized one
in reciprocal space: The interaction parameters $U_{\ff k \ff k' \ff k'' \ff k''}$ 
basically couple any $\ff k$ point to any other.
Likewise, a local uncorrelated disorder transforms into a delocalized correlated one.
As $H$ and $H'$ must share the same interaction and disorder, it is unlikely to find
a $\ff t'$ that permits a (simple) solution of the resulting problem.

A pragmatic way out would be to distinguish formally between the self-energy
$\ff S_{\ff t',P}$ that is determined as the stationary point of the SFT grand potential 
on the one hand, and the translationally invariant (``physical'') self-energy on the 
other. 
The latter is obtained from $\ff S = \ff S_{\ff t',P}$ by some periodization procedure
which employs a universal functional $\widehat{\ff T}$ ``periodizing'' the self-energy:
$\ff S \to \widehat{\ff T}[\ff S]$.
This idea has been suggested in the context of the C-DMFT. \cite{KSPB01,BPK04}
One possibility is to Fourier transform the optimized self-energy $\ff S_{\ff t',P}$ 
from real space, $S_{\ff x\ff x'}$, to reciprocal space via $U_{\ff x,\ff k}$, 
\refeq{ftu}.
This yields the self-energy in the representation $S_{\ff k \ff k'}$. 
Substituting $S_{\ff k \ff k'} \to S_{\ff k\ff k} \delta_{\ff k, \ff k'} \equiv 
\widehat{T}[\ff S]_{\ff k\ff k'}$ gives a translationally invariant (``physical'') self-energy
$\widehat{\ff T}[\ff S]$. In real space, this periodization reads:
\begin{equation}
  \widehat{T}[\ff S]_{\ff x\ff x'} \equiv \frac{1}{L} \sum_{\ff y\ff y'}
  \delta_{\ff x - \ff x' , \ff y - \ff y'} S_{\ff y\ff y'} \: .
\end{equation}
$\widehat{\ff T}$ could also be applied to the Green's function calculated from $\ff S$ via
Dyson's equation. 
This is the usual procedure within the CPT to generate translationally invariant Green's 
functions.
While the problem of translational symmetries can be fixed in this or in a similar 
way, the procedure appears to be {\em ad hoc} as it is placed on top of a variational 
(or self-consistent) calculation which itself involves $\ff S$ instead of $\widehat{\ff T}[\ff S]$.

%*********************************************************************************
%*********************************************************************************

\subsection{Periodized M-CPA, periodized VCA}

Koller and Dupuis \cite{KD05} have discussed a modified form of the self-energy
functional for (pure) systems of interacting bosons to get translationally invariant 
trial self-energies.
This method can be adapted to disordered systems as described below.

Consider the Anderson model \refeq{and} and the following self-energy functional:
\ba
  \widehat{\Omega}^{(1)}_{\ff t, P}[\ff S] 
&=&
  \mbox{Tr} \ln \frac{1}{\ff G_{\ff t,0,0}^{-1} - \widehat{\ff T} [\ff S]}
- 
  \mbox{Tr} \ln \widehat{\ff \Gamma}_{P}[\ff S]
\nonumber \\
&+&
 \left\langle \mbox{Tr} \ln 
  \frac{1}{ \widehat{\ff \Gamma}_{P}[\ff S]^{-1}
  + \ff S - \ff \eta 
  }
  \right\rangle_P \: .
\labeq{disonly1}
\ea
As compared to the original functional $\widehat{\Omega}_{\ff t, P}[\ff S]$ given by
\refeq{disonly}, the modified functional $\widehat{\Omega}^{(1)}_{\ff t, P}[\ff S]$ 
differs in the first term as this includes the periodizing functional $\widehat{\ff T}$.
As usual, the hopping term of the Anderson model is supposed to be translationally 
invariant, $\widehat{\ff T}[\ff t] = \ff t$.
The exact self-energy is therefore translationally invariant, too, and satisfies
$\widehat{\ff T}[\ff S_{\ff t,P}] = \ff S_{\ff t,P}$.
Hence, we have:
\begin{equation}
\widehat{\Omega}^{(1)}_{\ff t, P}[\ff S_{\ff t,P}]
=
\widehat{\Omega}_{\ff t, P}[\ff S_{\ff t,P}] \: .
\end{equation}
Using a one-particle basis of Bloch states labeled by wave vectors $\ff k$ to evaluate
the traces, we furthermore have:
\begin{equation}
  \frac{\delta \widehat{\Omega}^{(1)}_{\ff t, P}[\ff S]}{\delta S_{\ff k\ff k'}}
  =
  \left( \frac{1}{\ff G_{\ff t,0,0}^{-1} - \widehat{\ff T} [\ff S]} \right)_{\ff k\ff k}
  \delta_{\ff k',\ff k}
  - 
  \Gamma[\ff S]_{\ff k'\ff k} \: .
\end{equation}
Since $\Gamma[\ff S_{\ff t, P}]_{\ff k\ff k'} = \delta_{\ff k,\ff k'}
\Gamma[\ff S_{\ff t, P}]_{\ff k\ff k}$, 
this shows that the exact (translationally invariant) self-energy is a stationary
point of the modified self-energy functional.
Concludingly, the functionals $\widehat{\Omega}^{(1)}_{\ff t, P}[\ff S]$ and 
$\widehat{\Omega}_{\ff t, P}[\ff S]$ are different in general but coincide and are 
stationary at the exact self-energy $\ff S_{\ff t,P}$.
They are similarly suited for constructing approximations.

Consider now a trial self-energy $\ff S_{\ff t',P}$ from a reference system with the
same disorder (i.e.\ the same distribution function $P$) but with hopping parameters 
$\ff t'$ breaking translational symmetries, as it is the case e.g.\ for the reference 
system given by Fig.\ \ref{fig:cdmft}c.
The condition for stationarity of the modified self-energy functional within the 
restricted set of trial self-energies then reads:
\ba
0 &=& T \sum_{n} \sum_{\ff k,\ff k'}
    \Big[\Big( 
    \frac{1}{i\omega_n + \mu - \ff t - \widehat{\ff T}[\ff S](i\omega_n)} 
    \Big)_{\ff k\ff k} \delta_{\ff k,\ff k'}
\nonumber \\
    &-& 
    \Big(
    {\ff \Gamma}_{\ff t',P}(i\omega_n)
    \Big)_{\ff k' \ff k}
    \Big]
    \frac{\partial S_{\ff k \ff k'}(i\omega_n)}
         {\partial {{\bm t}'}} \: .
\ea 
Here, $\ff S \equiv \ff S_{\ff t', P}$ for short.
This replaces the condition \refeq{mcpas} characteristic for the M-CPA.

Fourier transformation $\ff U$ (\refeq{ftu}) yields
$S_{\ff k \ff k'}=L^{-1} \sum_{\ff x\ff x'} e^{-i\ff k \ff x} e^{i \ff k' \ff x'} 
S_{\ff x \ff x'}$ and thus:
\ba
0 &=& T \sum_{n} \sum_{\ff x,\ff x'}
    \Big[
    \frac{1}{L}
    \sum_{\ff k} e^{-i \ff k(\ff x- \ff x')}
    \Big( 
    \frac{1}{i\omega_n + \mu - \ff t - \widehat{\ff T}[\ff S]} 
    \Big)_{\ff k\ff k} 
\nonumber \\
    &-& 
    \Big(
    {\ff \Gamma}_{\ff t',P}(i\omega_n)
    \Big)_{\ff x' \ff x}
    \Big]
    \frac{\partial S_{\ff x \ff x'}(i\omega_n)}
         {\partial {{\bm t}'}} \: .
\labeq{pc1}
\ea 
For the reference system Fig.\ \ref{fig:cdmft}c with isolated clusters we have
$S_{\ff x \ff x'} = S_{\widetilde{\ff x} + \ff X, \widetilde{\ff x}' + \ff X'} 
= S_{\widetilde{\ff x} + \ff X, \widetilde{\ff x} + \ff X'} 
\delta_{\widetilde{\ff x}\widetilde{\ff x}'}$ (and the same for its $\ff t'$ derivative)
and thus the Euler equation (\ref{eq:pc1}) is satisfied if 
\begin{equation}
    \frac{1}{L}
    \sum_{\ff k} 
    \frac{e^{i \ff k(\ff X- \ff X')}}
    {i\omega_n + \mu - \varepsilon(\ff k) - \widehat{T}[\ff S](\ff k)} 
=
    \Big(
    {\ff \Gamma}_{\ff t',P} 
    \Big)_{\ff X \ff X'} \: .
\labeq{pcdmft}
\end{equation}
As compared to the self-consistency condition of the M-CPA, \refeq{mcpaeq}, the main
difference consists in the fact that its solution is based on a self-consistency cycle 
which involves at each step the periodized self-energy $\widehat{\ff T}[\ff S]$ instead of $\ff S$.
The Green's function of this periodized M-CPA, 
$\ff G \equiv (i\omega_n + \mu - \ff t - \widehat{\ff T}[\ff S])^{-1}$,
is likewise translationally invariant: $\widehat{\ff T}[\ff G]= \ff G$.

The analog of the self-consistency equation (\ref{eq:pcdmft}) for the pure but interacting
system is the self-consistency equation of the periodized C-DMFT (PC-DMFT), see Ref.\
\onlinecite{BPK04}.
Hence, the above derivation also shows how the PC-DMFT can be rederived within the 
SFT framework. 
Furthermore, the construction of a combined PC-DMFT + periodized M-CPA is straightforward.
It is also straightforward to constuct a periodized VCA for disordered or for interacting
systems along the lines above. 

As in the usual VCA (and also in the usual M-CPA) one in principle has the choice between
open and periodic boundary conditions in the reference system.
In the case of the periodized VCA (periodized M-CPA) a translationally invariant self-energy
and Green's function are generated by either choice.
The value of the SFT grand potential at the stationary point decides which kind of 
boundary conditions should be preferred.

With a suitably defined functional $\widehat{\ff T}$ there are no problems to extend the approach
and to restore point-group symmetries of the underlying lattice which could be violated
by the choice of the reference system. 
Essentially the same arguments given for the case of translational symmetries can be
repeated.

There is yet another way to modify the original functional and thereby to construct a 
theory which respects the translational symmetries of the underlying lattice.
This shall be mentioned here for the sake of completeness.
Consider the functional
\ba
  \widehat{\Omega}^{(2)}_{\ff t, P}[\ff S] 
&=&
  \mbox{Tr} \ln \widehat{\ff T} \left[ \frac{1}{\ff G_{\ff t,0,0}^{-1} - \ff S} \right]
- 
  \mbox{Tr} \ln \widehat{\ff \Gamma}_{P}[\ff S]
\nonumber \\
&+&
 \left\langle \mbox{Tr} \ln 
  \frac{1}{ \widehat{\ff \Gamma}_{P}[\ff S]^{-1}
  + \ff S - \ff \eta 
  }
  \right\rangle_P \: .
\labeq{disonly2}
\ea
Again, only the first term on the r.h.s.\ is modified.
Note that this is essential for the construction of approximations:
It ensures that the not explicitly known but universal functional given by the 
remaining terms cancels out when comparing with the functional of a suitably defined 
reference system, as usual. 
$\widehat{\Omega}^{(2)}_{\ff t, P}[\ff S]$ coincides with 
$\widehat{\Omega}_{\ff t, P}[\ff S]$ and
$\widehat{\Omega}^{(1)}_{\ff t, P}[\ff S]$ at the exact self-energy $\ff S_{\ff t,P}$ 
and is stationary there.
With the reference system Fig.\ \ref{fig:cdmft}c the following self-consistency
equation can be derived:
\begin{equation}
    \frac{1}{L}
    \sum_{\ff k} 
    e^{i \ff k(\ff X- \ff X')}
    \left( \frac{1}{i\omega_n + \mu - \ff t - \ff S} \right)_{\ff k \ff k}
=
    \Big(
    {\ff \Gamma}_{\ff t',P} 
    \Big)_{\ff X \ff X'} \: .
\labeq{pcdmft2}
\end{equation}
As compared to the periodized M-CPA, the conceptual disadvantage consists in the fact
that only the propagator 
$\widehat{T} \left[ ({\ff G_{\ff t,0,0}^{-1} - \ff S})^{-1} \right]$ but not the self-energy
is translationally invariant.

\subsection{Dynamical cluster approximation}

Originally, the dynamical cluster approximation (DCA) has been proposed as a cluster 
extension of the DMFT for interacting systems. \cite{HTZ+98,MJPK00,HMJK00}
Essentially the same ideas, however, can also be used to construct a generalization 
of the single-site CPA for the disorder problem. \cite{JK01,MJ02b}
Here, it is shown that the (disorder) DCA can be rederived within the SFT framework
by utilizing the real-space perspective on the DCA first discussed by Biroli et al.
\cite{BPK04} 

While the main idea of the periodized M-CPA to restore translational symmetry is to 
consider a modified but equivalent self-energy functional, one could also keep the
exact functional form $\Omega_{\ff t,P}[\ff S]$ but modify the hopping of the original 
system, i.e.\ $\ff t \to \overline{\ff t}$.
Approximations are then constructed by starting from $\Omega_{\overline{\ff t},P}[\ff S]$
and using a reference system consisting of isolated clusters again.
To ensure that the resulting approximations systematically approach the exact solution
for cluster size $L_c \to \infty$, the replacement $\ff t \to \overline{\ff t}$
must be controlled by $L_c$, i.e.\ it must become exact (up to irrelavant boundary terms)
in the infinite-cluster limit.

\begin{figure}[t]
  \includegraphics[width=0.85\columnwidth]{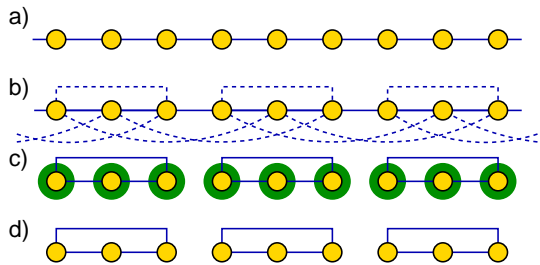}
\caption{(Color online)
a) Anderson model.
b) The original Anderson model but with a modified one-particle part 
$\ff t \to \overline{\ff t}$ which is the starting point for the dynamical cluster
approximation (DCA). 
$\overline{\ff t}$ is invariant under superlattice and cluster translations. 
c) Reference system generating the DCA. 
Note that $\ff t'$ has the same translational symmetries as $\overline{\ff t}$.
d) Reference system generating a simplified DCA (see text). 
}
\label{fig:dca}
\end{figure}

Consider, for example, 
\begin{equation}
  \overline{\ff t} = (\ff V\ff W) \ff U^\dagger \: \ff t \: \ff U (\ff V \ff W)^\dagger \: .
\labeq{tt}
\end{equation}
For clusters of finite size $L_c$, the combined Fourier transformation $\ff V \ff W$ is
different from $\ff U$. 
For $L_c \to \infty$, however, this becomes irrelevant.
With $\varepsilon(\ff k)=(\ff U^\dagger \ff t \ff U)(\ff k)$ we have:
\begin{equation}
  \overline{t}_{\ff x\ff x'} = \frac{1}{L_c} \sum_{\ff K} e^{i\ff K(\ff X - \ff X')}
  \frac{L_c}{L} \sum_{\widetilde{\ff k}} e^{i\widetilde{\ff k}
  (\widetilde{\ff x} - \widetilde{\ff x}')} \varepsilon(\widetilde{\ff k} + \ff K) \: .
\labeq{tt1}
\end{equation}
Obviously, $\overline{\ff t}$ is invariant under superlattice translations
as well as under cluster translations (with periodic cluster boundary conditions).
The original and the modified system with $\overline{H} = H(\overline{\ff t}, \ff \eta, 0)$
are represented by Fig.\ \ref{fig:dca}a, b.
The construction of $\overline{\ff t}$ is such that it exhibits the same translational 
symmetries as the one-particle parameters $\ff t'$ of a reference system consisting of
isolated clusters tiling the original lattice with periodic boundary conditions, 
see Fig.\ \ \ref{fig:dca}c, d.
Since both, $\ff t$ and $\overline{\ff t}$, are invariant under superlattice translations,
we can compare $t_{\ff X\ff X'}(\widetilde{\ff k}) = 
(\ff V^\dagger \ff t \ff V)_{\ff X\ff X'}(\widetilde{\ff k})$ with 
$\overline{t}_{\ff X\ff X'}(\widetilde{\ff k}) = 
(\ff V^\dagger \overline{\ff t} \ff V)_{\ff X\ff X'}(\widetilde{\ff k})$. 
It turns out they are equal up to a phase factor:
\begin{eqnarray}
  \overline{t}_{\ff X\ff X'}(\widetilde{\ff k}) 
  & = &
  \frac{1}{L_c} \sum_{\ff K}
  e^{i \ff K (\ff X - \ff X')}
  \varepsilon(\widetilde{\ff k} + \ff K)
\nonumber \\
  & = &
  \frac{L_c}{L} \sum_{\widetilde{\ff x}\widetilde{\ff x}'} 
  e^{- i \widetilde{\ff k}(\widetilde{\ff x}+\ff X - \widetilde{\ff x}'- \ff X')}
  t_{\widetilde{\ff x}+\ff X , \widetilde{\ff x}' + \ff X'}
\nonumber \\
  & = & 
  e^{-i\widetilde{\ff k}(\ff X - \ff X')}
  t_{\ff X\ff X'}(\widetilde{\ff k}) \: .
\end{eqnarray}

To rederive the disorder DCA within the framework of the SFT, we use the functional 
$\Omega_{\overline{\ff t},P}[\ff S]$ and the reference system Fig.\ \ \ref{fig:dca}c.
Therewith, one formally arrives at the M-CPA self-consistency condition, 
\refeq{mcpaeq}, but with $\ff t$ replaced by $\overline{\ff t}$:
\be
   \frac{L_c}{L}
   \sum_{\widetilde{\ff k}}
   \Big(
   \frac{1}{i\omega_n + \mu - \overline{\ff t}(\widetilde{\ff k}) - \ff S} 
   \Big)_{\ff X \ff X'}
   =
   \Big(
   {\ff \Gamma}_{\ff t',P}
   \Big)_{\ff X \ff X'} \: .
\labeq{mcpaeqpp}
\ee 
The bold symbols are matrices in the cluster variables $\ff X, \ff X'$.
The self-energy and the averaged Green's function of the reference system are
independent of $\widetilde{\ff k}$.
The decisive difference as compared to the M-CPA is that the (modified) hopping 
of the {\em original} system $\overline{\ff t}$ is invariant under cluster 
translations, as it is the case for $\ff t'$.
This is important as it allows to simultaneously diagonalize all matrices in 
\refeq{mcpaeqpp} by the cluster Fourier transformation $\ff W$.
Furthermore, from the definition of $\overline{\ff t}$, the $\ff W$ transformation yields 
$\overline{\ff t}(\widetilde{\ff k}) \to 
\varepsilon(\widetilde{\ff k} + \ff K)=\varepsilon(\ff k)$, i.e.\ we find:
\ba
   &&
   \frac{L_c}{L}
   \sum_{\widetilde{\ff k}}
   \frac{1}{i\omega_n + \mu -  \varepsilon(\widetilde{\ff k} + \ff K) - (\ff W^\dagger \ff S \ff W)_{\ff K}} 
   \nonumber \\
   &&
   =
   \left( \ff W^\dagger {\ff \Gamma}_{\ff t',P} \ff W \right)_{\ff K}
   \: .
\labeq{dcap}
\ea 
This, however, is just the self-consistency equation of the disorder DCA. \cite{JK01}

%*********************************************************************************
%*********************************************************************************

We can thus state that, analogous to the work of Biroli et al.\ \cite{BPK04} for the usual
DCA, it is found that the disorder DCA is equivalent with the M-CPA applied to the system 
with modified hopping \refeq{tt}.
Hence, it has been shown that the (disorder) DCA (as well as the usual DCA and also the 
combined theory for interacting systems with disorder) can be recovered within the 
framework of the SFT when starting from a suitably modified problem.
As already been noticed, \cite{PAD03} however, a strict rederivation starting from the
original system (with $\ff t$) appears to be impossible.

In this context an interesting new approximation suggests itself:
Starting with the modified hopping $\overline{\ff t}$ and using the reference system 
shown in Fig.\ \ref{fig:dca}d, generates a simplified (disorder) DCA without bath
degrees of freedom. 
This simplfied DCA is a systematic (controlled by $1/L_c$) cluster approximation and 
gives a translationally invariant self-energy and propagator.
The simplified DCA is related to the periodized VCA in the same way as the DCA is
related to the periodized M-CPA.
The analogous formulation of a simplified DCA for the pure but interacting system
represents a variational extension of a non-self-consistent approximation 
(``periodic CPT'') recently introduced by Minh-Tien. \cite{MT06}

\section{Summary and discussion}
\label{sec:dis}

An advantageous feature of the self-energy-functional approach for systems of interacting
electrons is that approximations are easily specified by choosing a reference system. 
The reference system helps to span a space of trial self-energies which are optimized
using an appropriate dynamical variational principle.
Since a reference system must share with the original system under consideration the same
interaction part, the number of possible reference systems is severly limited. 
This leads to a straightforward classification of approximations which may be called
``dynamic'' as these are essentially based on an approximation for the self-energy or,
equivalently, the Green's function, i.e.\ quantities characterizing the spectrum of 
one-particle excitations.

For systems with local interactions, the set of dynamical approximations includes the 
dynamical mean-field theory which corresponds to a reference system consisting of 
decoupled single-impurity models.
Modifying this reference system generates DMFT-related dynamical approximations but does
not spoil the main attractive properties of the DMFT, in particular its non-perturbative 
character and its thermodynamical consistency.
This is interesting as it opens up a way to construct approximations which (i) are based
on simpler reference systems that can be solved with less numerical effort or/and (ii)
include spatial correlations beyond the local mean-field concept.
In fact, the classification of dynamical approximations has led to new approximations
which have successfully been employed in the past (see Fig.\ \ref{fig:sys} for pure systems):  
The dynamical-impurity approximation (DIA) as well as the variational cluster approximation 
(VCA).

The main idea of this paper has been to translate this story to the case of disordered
(and interacting) systems.
The disorder is assumed to be local and uncorrelated between different sites.
One possible way is to apply the conventional SFT to treat the interaction part of the 
problem for any disorder configuration and to average subsequently. 
This procedure treats the disorder part of the problem exactly.
Choosing a decoupled set of effective impurity models as a reference system, yields an 
extension of the DMFT for systems with reduced translational symmetries or, after
reinterpretation of the mean-field (Euler) equations as stochastic recursion equations,
to the so-called statistical DMFT.
This is feasible in the case of a Bethe lattice only.
Even then and even with a simpler reference system including a minimum number of bath 
sites, however, the statistical SFT remains a numerically extremely expensive method.

The alternative consists in an {\em approximate} treatment of the disorder on the same
footing as the interaction.
Technically, this requires a reformulation of the SFT for interacting systems within a
functional-integral language which, as a by-product, provides an entirely non-perturbative
construction of the SFT, i.e.\ avoids formal summations of skeleton diagrams.
The functional-integral framework then allows to formulate a disorder SFT in essentially 
the same way as for pure interacting systems; the main corners of the theory are left
unchanged:
The (averaged) grand potential can be expressed as a functional of the (disorder) 
self-energy with the physical self-energy being a stationary point.
The functional still consists of a simple and explicitly known part depending on the 
one-particle parameters as well as of a complicated and basically unknown part which, 
however, is ``universal'', i.e.\ depends on the probability distribution only.
Choosing a reference system with the same distribution function, the universal part can
be eliminated, and an exact evaluation of the self-energy functional becomes possible on 
the space of trial self-energies generated by the reference system.
Restricting the search for the stationary point to this limited subspace generates 
approximations.

It is conceptually very satisfying that within this generalized SFT, the coherent-potential
approximation for the disorder problem takes the place of the DMFT for an interacting system.
As the DMFT, the CPA is a dynamical approximation for the self-energy and is distinguished
by the fact that it becomes formally exact in the limit of infinite spatial dimensions.
``Replacing interaction by disorder'', any reference system for the case of an interacting 
system can also be considered as a reference system for the disorder case.
This establishes a one-to-one mapping of the respective approximations with the 
DMFT corresponding to the CPA and the cellular DMFT corresponding to the molecular CPA.
Via this mapping a disorder DIA and a disorder VCA appear as new approximations as well as
the periodized M-CPA, namely
the disorder pendant of the periodized cellular DMFT, when starting from a different but
equivalent functional involving the periodizing functional $\widehat{\ff T}$.
Starting from the self-energy functional with a suitably modified hopping $\ff t \to 
\overline{\ff t}$, one also recovers the disorder analog of the dynamical cluster
approximation.
Finally, a simplified DCA without bath degrees of freedom can be set up.

The case of non-interacting disordered systems could actually be treated by specialization 
of a more general functional $\widehat{\Omega}_{\ff t, P, \ff U}
[\ff S,\{ \ff \Sigma_{\ff \eta} \}]$ depending on the full {\em and} on the 
configuration-dependent interaction self-energies.
This is applicable to interacting {\em and} disorderd systems.
Though formally more complicated, the not explicitly known part of the functional is 
universal, i.e.\ depends on $\ff U$ and $P$ only.
The classification of dynamical approximations extends accordingly (see Fig.\ \ref{fig:sys}).

Summing up, the type of dynamical approximations that can be constructed for systems with 
local interactions or/and local disorder are mean-field and cluster approximations which 
differ with respect to the number of local variational degrees of freedom included in the 
reference system.
All approximations fulfill the requirements of causality.
They are non-perturbative as, contrary to truncating diagrammtic approximations, the
exact functional form is retained. 
Thermodynamical consistency is ensured by the existence of an approximate but explicit
expression for the grand potential, i.e.\ for a thermodynamical potential, from which 
the physical quantities are derived.
Additional but systematic modifications of the self-energy functional have been shown to 
generate (cluster) approximations that respect the symmetries of the underlying lattice, 
particularly the translational symmetries.

Convergence properties of the different quantum-cluster schemes with a continuum of bath 
sites have been discussed in Refs.\ \onlinecite{MJ02a,BK02,AMJ05,BK05} and apply to the 
schemes with $n_s=1$ in an analogous way. 
For $L_c \to \infty$, {\em local} quantities generally converge exponentially fast within 
the C-DMFT / M-CPA and within the (disorder) VCA.  
This has to be compared with the $1 / l_c^2$ 
behavior obtained within the DCA and within the simplified DCA (with $L_c=l_c^D$). 
For practical purposes, however, the quality of a given approximation for {\em small} 
$L_c$ is more important and can apparently be estimated {\em a posteriori} only.
It is, for example, an open question whether cluster schemes with or without bath degrees 
of freedom should be preferred in this respect (see, however, Ref.\ \onlinecite{PAD03}).

The application of the SFT has been restricted to systems with local interactions and 
local disorder. 
This is consistent with the spirit of the cluster-mean-field approximations discussed 
above.
One should note, however, that non-local or even long-ranged interactions or disordered 
nearest-neighbor or longer-ranged hopping or spatially correlated disorder in the on-site 
energies, for example, pose difficulties. 
While the general self-energy functional can be set up as usual, it appears to be impossible
to find a suitable reference system and thereby usable approximations in most cases:
As the reference system should have the same interaction and disorder, it is hard to find a
decoupling of its degrees of freedom by modifying the hopping part only.
This problem could be handled either pragmatically by additional mean-field decouplings of non-local
terms connecting different clusters \cite{AEvdLP+04} or, more thoroughly by considering more 
complex functionals involving two-particle correlation functions. \cite{Ton05}

One should be aware that the various local approximations and their different cluster 
extensions all neglect long-range correlations beyond the linear scale of the cluster.
This is typical for any cluster mean-field approach. 
To our knowledge there is no possibility to systematically restore long-range correlations 
by a suitable embedding of an isolated cluster in the environment within the presented
formalism (apart from simply enlarging the cluster).
The same holds for cluster schemes formulated within reciprocal space.

This implies that a proper description of transport properties is hardly possible.
Effects like Anderson localization, for example, cannot be captured within a  
mean-field approach like CPA and are obviously difficult to restore by cluster schemes
extending CPA (see Ref.\ \onlinecite{JK01}, for an example). 
Anyway, it is basically impossible to access two-particle correlation functions, and the 
conductivity in particular, within an approach that places one-partible excitations in the 
center of interest. 
The statistical DMFT (statistical SFT) represents an exception as here the full distribution 
of the (one-particle) density of states can be used to discriminate between extended and 
localized states, for example.

Besides the spectrum of one-particle excitations, however, the SFT derives an approximate but 
explicit expression for a thermodynamical potential and thus provides a consistent picture of 
the entire thermodynamics and of static expectation values.
This includes spontaneous symmetry breaking, i.e.\ the determination of order parameters. 
The formal framework presented here should therefore be ideally suited to study the effects
of disorder on different types of long-range order, such as magnetism or superconductivity. 
For different material classes, such as diluted magnetic semiconductors, cuprate-based 
high-temperature superconductors, manganites, rare-earth compounds etc.\ these are central
questions.

\acknowledgments
This work is supported by the Deutsche Forschungsgemeinschaft within the 
Sonderforschungsbereich SFB 410 and the Forschergruppe FOR 538.

\appendix

\section{Disorder in the interaction}
\label{sec:udis}

It is formally straightforward to extend the theory to the case of disorder in the interaction
part of the Hamiltonian. 
The theory can be constructed without major modifications:

Consider a system with Hamiltonian 
$H = H_0(\ff t) + H_{\rm dis}(\ff \eta) + H_{\rm int}(\ff U)$
where $H_0(\ff t)$ and $H_{\rm int}(\ff U)$ describe the one-particle
and the interaction part as before while 
\be
  H_{\rm dis}(\ff \eta) = \frac{1}{2} 
  \sum_{\alpha\beta\gamma\delta} \eta_{\alpha\beta\delta\gamma} \:
  c^\dagger_\alpha c^\dagger_\beta c_\gamma c_\delta
\ee
is an interaction term with random parameters $\ff \eta$ distributed
according to some $P(\ff \eta)$.
The definition of the free Green's function \refeq{gt00},
the interacting Green's function \refeq{intg}, and the averaged
Green's function \refeq{defg} is formally unchanged.
The self-energies are defined as
\be
  \ff \Sigma_{\ff t, \ff \eta, \ff U} = 
  \ff G_{\ff t,0,0}^{-1} - \ff G_{\ff t, \ff \eta, \ff U}^{-1}
\; , \quad
  \ff S_{\ff t, P, \ff U} = 
  \ff G_{\ff t,0,0}^{-1} - \ff \Gamma_{\ff t, P, \ff U}^{-1} \: .
\labeq{newsdefs}
\ee
The reasoning in section \ref{sec:funcint} is unchanged.
The functionals 
$\widehat{A}_{\ff U,\ff \eta,\xi\xi^\ast} [\ff G_0^{-1}]$,
$\widehat{\Omega}_{\ff U,\ff \eta} [\ff G_0^{-1}]$,
$\widehat{\cal \ff G}_{\ff U,\ff \eta}\left[\ff G_0^{-1}\right]$,
$\widehat{\ff G}_{\ff U,\ff \eta}[\ff \Sigma]$, and
$\widehat{F}_{\ff U,\ff \eta}[\ff \Sigma]$,
however, additionally depend on the parameters $\ff \eta$.
The same holds for section \ref{sec:funcdis} except for
\refeq{disfunc} which has to be replaced by
\be
  \left\langle
  \frac{1}
  {\ff \Gamma^{-1} + \ff S - \ff \Sigma_{\ff \eta}}
  \right\rangle_P = \ff \Gamma \: .
\ee
The final self-energy functional thus reads
\ba
\widehat{\Omega}_{\ff t, P, \ff U}[\ff S , \ff \Sigma_{\ff \eta}]
  &=& \mbox{Tr} \ln \frac{1}{\ff G_{\ff t,0,0}^{-1} - \ff S}
  - \mbox{Tr} \ln \widehat{\ff \Gamma}_{P}[
  \ff S,
  \ff \Sigma_{\ff \eta}]
\nonumber \\  
  &+&
 \left\langle \mbox{Tr} \ln 
  \frac{1}{ \widehat{\ff \Gamma}_{P}[\ff S , \ff \Sigma_{\ff \eta}]^{-1}
  + \ff S - \ff \Sigma_{\ff \eta}
  }
  \right\rangle_P
\nonumber \\ 
&+& \left\langle \widehat{F}_{\ff U,\ff \eta}[\ff \Sigma_{\ff \eta}] \right\rangle_P
\: .
\ea
Approximations are constructed in the same way as described in Sec.\ \ref{sec:app}
by making contact with a reference system with the same interaction and disorder.

\section{Thermodynamical consistency}
\label{sec:tdconsis}

Here, we briefly discuss the thermodynamical consistency of approximations within the 
SFT generalized to disordered systems.
To be definite, we concentrate on the particle number as an example.
The reasoning closely follows Ref.\ \onlinecite{AAPH06a}.

There are two ways by which the configurational average of the quantum-statistical 
expectation value of the particle number 
\begin{equation}
  N_{\ff t, P, \ff U} = \int d\ff \eta P(\ff \eta) N_{\ff t, \ff \eta, \ff U}
  = \int d\ff \eta P(\ff \eta) \tr ( \rho_{\ff t, \ff \eta, \ff U} N )
\end{equation}
can be obtained: 
(i) $N_{\ff t, P, \ff U}$ is calculated, on the ``zero-particle level'' as the $\mu$ 
derivative of the averaged grand potential:
\begin{equation}
  N_{\ff t, P, \ff U} = - \frac{\partial \Omega_{\ff t,P,\ff U}}{\partial \mu} \: ,
\labeq{n1}
\end{equation}
or (ii) $N_{\ff t, P, \ff U}$ is calculated, on the ``one-particle level'' by frequency
integration of the one-particle averaged retarded Green's function:
\begin{equation}
  N_{\ff t, P, \ff U} = - \frac{1}{\pi} \int_{-\infty}^\infty d\omega f(\omega) 
  \mbox{Im} \, \tr \, \Gamma_{\ff t,P,\ff U}(\omega+i0^+) \: ,
\labeq{n2}
\end{equation}
where $f(\omega) = 1/(e^{\omega/T} + 1)$ is the Fermi function.
Thermodynamical consistency means that both ways must yield the same result.
This is not clear {\em a priori} as within the SFT $\Omega_{\ff t,P,\ff U}$ and $
\Gamma_{\ff t,P,\ff U}$ are approximate quantities. 

As in Ref.\ \onlinecite{AAPH06a}, however, it can be argued that there is a twofold 
$\mu$ dependence of $\Omega_{\ff t,P,\ff U}$: an {\em explicit} $\mu$ dependence 
which in \refeq{res} is due to the free Green's function 
$\ff G_{\ff t, 0 ,0}^{-1}=\omega + \mu - \ff t$, and an {\em implicit} $\mu$ dependence 
due to the $\mu$ dependence of the disorder self-energy at the stationary point.
Now, if and only if an overall shift $\varepsilon'$ of the on-site energies in the
reference system is treated as a variational parameter, the derivative w.r.t.\ the
implicit $\mu$ dependence vanishes, and one is left with the explicit one:
\begin{eqnarray}
  N_{\ff t, P, \ff U} 
  & = &
  - \frac{\partial \Omega_{\ff t,P,\ff U}}{\partial \mu} 
  = 
  - \frac{\partial \Omega_{\ff t,P,\ff U}}{\partial \mu_{\rm expl.}} 
\nonumber \\  
  & = &
  - \frac{\partial}{\partial \mu_{\rm expl.}} 
  \mbox{Tr} \ln \frac{1}{\omega + \mu_{\rm expl.} - \ff t - \ff S_{\ff t', P, \ff U}} 
\nonumber \\  
\end{eqnarray}
according to \refeq{res}. 
Carrying out the differentiation and using
$\Gamma_{\ff t,P,\ff U} = (\ff G_{\ff t,0,0}^{-1} - \ff S_{\ff t', P, \ff U})^{-1}$,
one immediately arrives at \refeq{n2} which proves the equivalence with \refeq{n1}.

Typically, only a distinguished set of parameters are treated as variational parameters
in a practical calculation. 
The argument shows the necessity to include $\varepsilon'$ in the set of variational
parameters if thermodynamical consistency is required for the particle number. 
The argument straightforwardly generalizes to all one-particle operators coupling 
linearly with a parameter $\lambda'$ to the reference-system Hamiltonian. 

\section{Causality}
\label{sec:cau}

The causality of all (approximate) dynamical quantities (configuration-dependent 
as well as averaged Green's functions and configuration-dependent/independent 
self-energies) is inevitable for having well-defined self-energy functionals and 
expressions involving $\Tr \ln ( \cdots )$ in particular. 
A frequency-dependent quantity is termed to be causal if it is analytical in the 
entire complex $\omega$ plane except for first-order poles on the real axis with 
positive residues.
Equivalently, we can demand that, after retardation $\omega \to \omega + i0^+$ 
($\omega$ real), the imaginary part be negative definite.

Causality is easily verified, for example, for all quantities which appear in the 
self-energy functional \refeq{func} when evaluated on its domain (see Sec.\
\ref{sec:lim}).
The last three terms on the r.h.s.\ only involve {\em exact} and therewith causal 
quantities as arguments of $\Tr \ln ( \cdots )$ (as these are taken from the 
reference system).
Note that the configuration average of a configuration-dependent exact quantity 
is causal:

Let $\ff G(\omega)$ be the exact Green's function of a model $H$.
We have the Lehmann representation \cite{FW71}
$\ff G(\omega) = \ff Q \ff g(\omega) \ff Q^\dagger$ with $\omega$-independent
matrices $\ff Q$ ($\ff Q$ not quadratic, see Ref.\ \onlinecite{AAPH06b}) and a 
diagonal matrix $\ff g(\omega)$ with elements $g_{nn}(\omega) = 1/(\omega - \omega_n)$, 
where $\omega_n$ are the poles of $\ff G$. 
Consider $\ff \Gamma(\omega) = \langle \ff G(\omega) \rangle$ averaged over only
two configurations, for simplicity, i.e.\ consider
$\ff \Gamma(\omega) = p_1 \ff Q_1 \ff g_1(\omega) \ff Q_1^\dagger +
p_2 \ff Q_2 \ff g_2(\omega) \ff Q_2^\dagger$ with $p_1,p_2\ge 0, p_1+p_2=1$.
With $\ff Q \equiv (\sqrt{p_1} \ff Q_1 , \sqrt{p_2} \ff Q_2)$ and $\ff \gamma(\omega)
\equiv \left( 
\begin{array}{cc}
\ff g_1(\omega) & 0 \\
0 & \ff g_2(\omega) 
\end{array}
\right)$ we then immediately have the representation 
$\ff \Gamma(\omega) = \ff Q \ff \gamma(\omega) \ff Q^\dagger$ 
from which the causality of $\ff \Gamma(\omega)$ is easily verified.

The only possible source of causality violation is left for the first term on 
the r.h.s.\ of \refeq{func}.
However, it is easily shown that (i) a causal $\ff \Gamma'(\omega)$ (the exact averaged 
Green's function of the reference system) implies a causal $\ff S(\omega)$.
The main point is that the retarded quantity 
$\ff S(\omega + i0^+) = \omega + \mu - \ff t' - \ff \Gamma'(\omega+i0^+)^{-1}$
has a negative definite imaginary part since the imaginary part of 
$\ff \Gamma'(\omega+i0^+)^{-1}$ is positive definite according to a lemma given in 
Ref.\ \onlinecite{Pot03b}. 
Furthermore (ii), it can be shown that the causality of $\ff S(\omega)$ implies
the causality of $\ff \Gamma(\omega) \equiv ( \ff G_0(\omega)^{-1} - \ff S(\omega))^{-1}$
for {\em arbitrary} $\ff G_0 = (\omega + \mu - \ff t)^{-1}$.
Namely, with $\ff S(\omega) = \ff G'_0(\omega)^{-1} - \ff \Gamma'(\omega)^{-1}$ and 
$\ff \Gamma'(\omega) = \ff Q' \ff \gamma'(\omega) {\ff Q'}^\dagger$ we have the representation 
$\ff \Gamma(\omega) = 1/((\ff Q' \ff \gamma'(\omega) {\ff Q'}^\dagger)^{-1}  - ({\ff t - \ff t'}))$
and thus (see Ref.\ \onlinecite{AAPH06b}) 
$\ff \Gamma(\omega) = \ff Q' (\ff \gamma'(\omega)^{-1} - {\ff Q'}^\dagger (\ff t - \ff t') \ff Q')^{-1} 
{\ff Q'}^\dagger$ from which is it obvious that $\ff \Gamma(\omega)$ is analytical with the
exception of first-order real poles with positive residues.


\begin{thebibliography}{93}
\expandafter\ifx\csname natexlab\endcsname\relax\def\natexlab#1{#1}\fi
\expandafter\ifx\csname bibnamefont\endcsname\relax
  \def\bibnamefont#1{#1}\fi
\expandafter\ifx\csname bibfnamefont\endcsname\relax
  \def\bibfnamefont#1{#1}\fi
\expandafter\ifx\csname citenamefont\endcsname\relax
  \def\citenamefont#1{#1}\fi
\expandafter\ifx\csname url\endcsname\relax
  \def\url#1{\texttt{#1}}\fi
\expandafter\ifx\csname urlprefix\endcsname\relax\def\urlprefix{URL }\fi
\providecommand{\bibinfo}[2]{#2}
\providecommand{\eprint}[2][]{\url{#2}}

\bibitem[{\citenamefont{Dietl}((2002))}]{Die02}
\bibinfo{author}{\bibfnamefont{T.}~\bibnamefont{Dietl}},
  \bibinfo{journal}{Semicond. Sci. Technol.} \textbf{\bibinfo{volume}{17}},
  \bibinfo{pages}{377} (\bibinfo{year}{2002}).

\bibitem[{\citenamefont{Imada et~al.}(1998)\citenamefont{Imada, Fujimori, and
  Tokura}}]{IFT98}
\bibinfo{author}{\bibfnamefont{M.}~\bibnamefont{Imada}},
  \bibinfo{author}{\bibfnamefont{A.}~\bibnamefont{Fujimori}}, \bibnamefont{and}
  \bibinfo{author}{\bibfnamefont{Y.}~\bibnamefont{Tokura}},
  \bibinfo{journal}{Rev. Mod. Phys.} \textbf{\bibinfo{volume}{68}},
  \bibinfo{pages}{13} (\bibinfo{year}{1998}).

\bibitem[{\citenamefont{Lee and Ramakrishnan}((1985))}]{LR85}
\bibinfo{author}{\bibfnamefont{P.~A.} \bibnamefont{Lee}} \bibnamefont{and}
  \bibinfo{author}{\bibfnamefont{T.}~\bibnamefont{Ramakrishnan}},
  \bibinfo{journal}{Rev. Mod. Phys.} \textbf{\bibinfo{volume}{57}},
  \bibinfo{pages}{287} (\bibinfo{year}{1985}).

\bibitem[{\citenamefont{Vollhardt and W\"olfle}((1992))}]{VW92}
\bibinfo{author}{\bibfnamefont{D.}~\bibnamefont{Vollhardt}} \bibnamefont{and}
  \bibinfo{author}{\bibfnamefont{P.}~\bibnamefont{W\"olfle}},
  \emph{\bibinfo{title}{{\rm In:} Electronic Phase Transitions}}, Ed. by W.
  Hanke and Yu.V. Kopaev (\bibinfo{publisher}{Elsevier},
  \bibinfo{year}{1992}).

\bibitem[{\citenamefont{Belitz and Kirkpatrick}((1994))}]{BK94}
\bibinfo{author}{\bibfnamefont{D.}~\bibnamefont{Belitz}} \bibnamefont{and}
  \bibinfo{author}{\bibfnamefont{T.~R.} \bibnamefont{Kirkpatrick}},
  \bibinfo{journal}{Rev. Mod. Phys.} \textbf{\bibinfo{volume}{66}},
  \bibinfo{pages}{261} (\bibinfo{year}{1994}).

\bibitem[{\citenamefont{Giamarchi and Orignac}((2003))}]{GO03}
\bibinfo{author}{\bibfnamefont{T.}~\bibnamefont{Giamarchi}} \bibnamefont{and}
  \bibinfo{author}{\bibfnamefont{E.}~\bibnamefont{Orignac}},
  \emph{\bibinfo{title}{{\rm in:} Theoretical Methods for Strongly Correlated
  Electrons}}, Ed. by D. Senechal et al. (\bibinfo{publisher}{Springer},
  \bibinfo{address}{New York}, \bibinfo{year}{2003}).

\bibitem[{\citenamefont{Miranda and Dobrosavljevic}((2005))}]{MD05}
\bibinfo{author}{\bibfnamefont{E.}~\bibnamefont{Miranda}} \bibnamefont{and}
  \bibinfo{author}{\bibfnamefont{V.}~\bibnamefont{Dobrosavljevic}},
  \bibinfo{journal}{Rep. Prog. Phys.} \textbf{\bibinfo{volume}{68}},
  \bibinfo{pages}{2337} (\bibinfo{year}{2005}).

\bibitem[{\citenamefont{Mermin and Wagner}((1966))}]{MW66}
\bibinfo{author}{\bibfnamefont{N.~D.} \bibnamefont{Mermin}} \bibnamefont{and}
  \bibinfo{author}{\bibfnamefont{H.}~\bibnamefont{Wagner}},
  \bibinfo{journal}{Phys. Rev. Lett.} \textbf{\bibinfo{volume}{17}},
  \bibinfo{pages}{1133} (\bibinfo{year}{1966}).

\bibitem[{\citenamefont{Anderson}((1958))}]{And58}
\bibinfo{author}{\bibfnamefont{P.~W.} \bibnamefont{Anderson}},
  \bibinfo{journal}{Phys. Rev.} \textbf{\bibinfo{volume}{109}},
  \bibinfo{pages}{1492} (\bibinfo{year}{1958}).

\bibitem[{\citenamefont{Fetter and Walecka}(1971)}]{FW71}
\bibinfo{author}{\bibfnamefont{A.~L.} \bibnamefont{Fetter}} \bibnamefont{and}
  \bibinfo{author}{\bibfnamefont{J.~D.} \bibnamefont{Walecka}},
  \emph{\bibinfo{title}{Quantum Theory of Many-Particle Systems}}
  (\bibinfo{publisher}{McGraw-Hill}, \bibinfo{address}{New York},
  \bibinfo{year}{1971}).

\bibitem[{\citenamefont{Gonis}((1992))}]{Gon92}
\bibinfo{author}{\bibfnamefont{A.}~\bibnamefont{Gonis}},
  \emph{\bibinfo{title}{Green's functions for ordered and disordered systems}}
  (\bibinfo{publisher}{North-Holland}, \bibinfo{address}{Amsterdam},
  \bibinfo{year}{1992}).

\bibitem[{\citenamefont{Metzner and Vollhardt}(1989)}]{MV89}
\bibinfo{author}{\bibfnamefont{W.}~\bibnamefont{Metzner}} \bibnamefont{and}
  \bibinfo{author}{\bibfnamefont{D.}~\bibnamefont{Vollhardt}},
  \bibinfo{journal}{Phys. Rev. Lett.} \textbf{\bibinfo{volume}{62}},
  \bibinfo{pages}{324} (\bibinfo{year}{1989}).

\bibitem[{\citenamefont{Vlaming and Vollhardt}((1992))}]{VV92}
\bibinfo{author}{\bibfnamefont{R.}~\bibnamefont{Vlaming}} \bibnamefont{and}
  \bibinfo{author}{\bibfnamefont{D.}~\bibnamefont{Vollhardt}},
  \bibinfo{journal}{Phys. Rev. B} \textbf{\bibinfo{volume}{45}},
  \bibinfo{pages}{4637} (\bibinfo{year}{1992}).

\bibitem[{\citenamefont{Georges and Kotliar}(1992)}]{GK92a}
\bibinfo{author}{\bibfnamefont{A.}~\bibnamefont{Georges}} \bibnamefont{and}
  \bibinfo{author}{\bibfnamefont{G.}~\bibnamefont{Kotliar}},
  \bibinfo{journal}{Phys. Rev. B} \textbf{\bibinfo{volume}{45}},
  \bibinfo{pages}{6479} (\bibinfo{year}{1992}).

\bibitem[{\citenamefont{Jarrell}(1992)}]{Jar92}
\bibinfo{author}{\bibfnamefont{M.}~\bibnamefont{Jarrell}},
  \bibinfo{journal}{Phys. Rev. Lett.} \textbf{\bibinfo{volume}{69}},
  \bibinfo{pages}{168} (\bibinfo{year}{1992}).

\bibitem[{\citenamefont{Georges et~al.}(1996)\citenamefont{Georges, Kotliar,
  Krauth, and Rozenberg}}]{GKKR96}
\bibinfo{author}{\bibfnamefont{A.}~\bibnamefont{Georges}},
  \bibinfo{author}{\bibfnamefont{G.}~\bibnamefont{Kotliar}},
  \bibinfo{author}{\bibfnamefont{W.}~\bibnamefont{Krauth}}, \bibnamefont{and}
  \bibinfo{author}{\bibfnamefont{M.~J.} \bibnamefont{Rozenberg}},
  \bibinfo{journal}{Rev. Mod. Phys.} \textbf{\bibinfo{volume}{68}},
  \bibinfo{pages}{13} (\bibinfo{year}{1996}).

\bibitem[{\citenamefont{Kotliar and Vollhardt}((2004))}]{KV04}
\bibinfo{author}{\bibfnamefont{G.}~\bibnamefont{Kotliar}} \bibnamefont{and}
  \bibinfo{author}{\bibfnamefont{D.}~\bibnamefont{Vollhardt}},
  \bibinfo{journal}{Physics Today} \textbf{\bibinfo{volume}{57}},
  \bibinfo{pages}{53} (\bibinfo{year}{2004}).

\bibitem[{\citenamefont{Hubbard}(1963)}]{Hub63}
\bibinfo{author}{\bibfnamefont{J.}~\bibnamefont{Hubbard}},
  \bibinfo{journal}{Proc. R. Soc. London A} \textbf{\bibinfo{volume}{276}},
  \bibinfo{pages}{238} (\bibinfo{year}{1963}).

\bibitem[{\citenamefont{Gutzwiller}((1963))}]{Gut63}
\bibinfo{author}{\bibfnamefont{M.~C.} \bibnamefont{Gutzwiller}},
  \bibinfo{journal}{Phys. Rev. Lett.} \textbf{\bibinfo{volume}{10}},
  \bibinfo{pages}{159} (\bibinfo{year}{1963}).

\bibitem[{\citenamefont{Kanamori}((1963))}]{Kan63}
\bibinfo{author}{\bibfnamefont{J.}~\bibnamefont{Kanamori}},
  \bibinfo{journal}{Prog. Theor. Phys. (Kyoto)} \textbf{\bibinfo{volume}{30}},
  \bibinfo{pages}{275} (\bibinfo{year}{1963}).

\bibitem[{\citenamefont{Soven}((1967))}]{Sov67}
\bibinfo{author}{\bibfnamefont{P.}~\bibnamefont{Soven}},
  \bibinfo{journal}{Phys. Rev.} \textbf{\bibinfo{volume}{156}},
  \bibinfo{pages}{809} (\bibinfo{year}{1967}).

\bibitem[{\citenamefont{Taylor}((1967))}]{Tay67}
\bibinfo{author}{\bibfnamefont{D.~W.} \bibnamefont{Taylor}},
  \bibinfo{journal}{Phys. Rev.} \textbf{\bibinfo{volume}{156}},
  \bibinfo{pages}{1017} (\bibinfo{year}{1967}).

\bibitem[{\citenamefont{Velicky et~al.}((1968))\citenamefont{Velicky,
  Kirkpatrick, and Ehrenreich}}]{VKE68}
\bibinfo{author}{\bibfnamefont{B.}~\bibnamefont{Velicky}},
  \bibinfo{author}{\bibfnamefont{S.}~\bibnamefont{Kirkpatrick}},
  \bibnamefont{and}
  \bibinfo{author}{\bibfnamefont{H.}~\bibnamefont{Ehrenreich}},
  \bibinfo{journal}{Phys. Rev.} \textbf{\bibinfo{volume}{175}},
  \bibinfo{pages}{747} (\bibinfo{year}{1968}).

\bibitem[{\citenamefont{Elliott et~al.}((1974))\citenamefont{Elliott,
  Krumhansl, and Leath}}]{EKL74}
\bibinfo{author}{\bibfnamefont{R.~J.} \bibnamefont{Elliott}},
  \bibinfo{author}{\bibfnamefont{J.~A.} \bibnamefont{Krumhansl}},
  \bibnamefont{and} \bibinfo{author}{\bibfnamefont{P.~L.} \bibnamefont{Leath}},
  \bibinfo{journal}{Rev. Mod. Phys.} \textbf{\bibinfo{volume}{46}},
  \bibinfo{pages}{465} (\bibinfo{year}{1974}).

\bibitem[{\citenamefont{Maier et~al.}((2005))\citenamefont{Maier, Jarrell,
  Pruschke, and Hettler}}]{MJPH05}
\bibinfo{author}{\bibfnamefont{T.}~\bibnamefont{Maier}},
  \bibinfo{author}{\bibfnamefont{M.}~\bibnamefont{Jarrell}},
  \bibinfo{author}{\bibfnamefont{T.}~\bibnamefont{Pruschke}}, \bibnamefont{and}
  \bibinfo{author}{\bibfnamefont{M.~H.} \bibnamefont{Hettler}},
  \bibinfo{journal}{Rev. Mod. Phys.} \textbf{\bibinfo{volume}{77}},
  \bibinfo{pages}{1027} (\bibinfo{year}{2005}).

\bibitem[{\citenamefont{Potthoff}(2003{\natexlab{a}})}]{Pot03a}
\bibinfo{author}{\bibfnamefont{M.}~\bibnamefont{Potthoff}},
  \bibinfo{journal}{Euro. Phys. J. B} \textbf{\bibinfo{volume}{32}},
  \bibinfo{pages}{429} (\bibinfo{year}{2003}{\natexlab{a}}).

\bibitem[{\citenamefont{Potthoff}(2003{\natexlab{b}})}]{Pot03b}
\bibinfo{author}{\bibfnamefont{M.}~\bibnamefont{Potthoff}},
  \bibinfo{journal}{Euro. Phys. J. B} \textbf{\bibinfo{volume}{36}},
  \bibinfo{pages}{335} (\bibinfo{year}{2003}{\natexlab{b}}).

\bibitem[{\citenamefont{Potthoff et~al.}(2003)\citenamefont{Potthoff, Aichhorn,
  and Dahnken}}]{PAD03}
\bibinfo{author}{\bibfnamefont{M.}~\bibnamefont{Potthoff}},
  \bibinfo{author}{\bibfnamefont{M.}~\bibnamefont{Aichhorn}}, \bibnamefont{and}
  \bibinfo{author}{\bibfnamefont{C.}~\bibnamefont{Dahnken}},
  \bibinfo{journal}{Phys. Rev. Lett.} \textbf{\bibinfo{volume}{91}},
  \bibinfo{pages}{206402} (\bibinfo{year}{2003}).

\bibitem[{\citenamefont{Potthoff}((2005))}]{Pot05}
\bibinfo{author}{\bibfnamefont{M.}~\bibnamefont{Potthoff}},
  \bibinfo{journal}{Adv. Solid State Phys.} \textbf{\bibinfo{volume}{45}},
  \bibinfo{pages}{135} (\bibinfo{year}{2005}).

\bibitem[{\citenamefont{Jani\u{s} and Vollhardt}((1992))}]{JV92b}
\bibinfo{author}{\bibfnamefont{V.}~\bibnamefont{Jani\u{s}}} \bibnamefont{and}
  \bibinfo{author}{\bibfnamefont{D.}~\bibnamefont{Vollhardt}},
  \bibinfo{journal}{Phys. Rev. B} \textbf{\bibinfo{volume}{46}},
  \bibinfo{pages}{15712} (\bibinfo{year}{1992}).

\bibitem[{\citenamefont{Dobrosavljevi\'c and Kotliar}((1993))}]{DK93b}
\bibinfo{author}{\bibfnamefont{V.}~\bibnamefont{Dobrosavljevi\'c}}
  \bibnamefont{and} \bibinfo{author}{\bibfnamefont{G.}~\bibnamefont{Kotliar}},
  \bibinfo{journal}{Phys. Rev. Lett.} \textbf{\bibinfo{volume}{71}},
  \bibinfo{pages}{3218} (\bibinfo{year}{1993}).

\bibitem[{\citenamefont{Dobrosavljevi\'c and Kotliar}((1994))}]{DK94}
\bibinfo{author}{\bibfnamefont{V.}~\bibnamefont{Dobrosavljevi\'c}}
  \bibnamefont{and} \bibinfo{author}{\bibfnamefont{G.}~\bibnamefont{Kotliar}},
  \bibinfo{journal}{Phys. Rev. B} \textbf{\bibinfo{volume}{50}},
  \bibinfo{pages}{1430} (\bibinfo{year}{1994}).

\bibitem[{\citenamefont{Byczuk et~al.}((2004))\citenamefont{Byczuk, Hofstetter,
  and Vollhardt}}]{BHV04}
\bibinfo{author}{\bibfnamefont{K.}~\bibnamefont{Byczuk}},
  \bibinfo{author}{\bibfnamefont{W.}~\bibnamefont{Hofstetter}},
  \bibnamefont{and}
  \bibinfo{author}{\bibfnamefont{D.}~\bibnamefont{Vollhardt}},
  \bibinfo{journal}{Phys. Rev. B} \textbf{\bibinfo{volume}{69}},
  \bibinfo{pages}{045112} (\bibinfo{year}{2004}).

\bibitem[{\citenamefont{Ulmke et~al.}((1995))\citenamefont{Ulmke, Jani\u{s},
  and Vollhardt}}]{UJV95}
\bibinfo{author}{\bibfnamefont{M.}~\bibnamefont{Ulmke}},
  \bibinfo{author}{\bibfnamefont{V.}~\bibnamefont{Jani\u{s}}},
  \bibnamefont{and}
  \bibinfo{author}{\bibfnamefont{D.}~\bibnamefont{Vollhardt}},
  \bibinfo{journal}{Phys. Rev. B} \textbf{\bibinfo{volume}{51}},
  \bibinfo{pages}{10411} (\bibinfo{year}{1995}).

\bibitem[{\citenamefont{Byczuk et~al.}((2003))\citenamefont{Byczuk, Ulmke, and
  Vollhardt}}]{BUV03}
\bibinfo{author}{\bibfnamefont{K.}~\bibnamefont{Byczuk}},
  \bibinfo{author}{\bibfnamefont{M.}~\bibnamefont{Ulmke}}, \bibnamefont{and}
  \bibinfo{author}{\bibfnamefont{D.}~\bibnamefont{Vollhardt}},
  \bibinfo{journal}{Phys. Rev. Lett.} \textbf{\bibinfo{volume}{90}},
  \bibinfo{pages}{196403} (\bibinfo{year}{2003}).

\bibitem[{\citenamefont{Byczuk and Ulmke}((2005))}]{BU05}
\bibinfo{author}{\bibfnamefont{K.}~\bibnamefont{Byczuk}} \bibnamefont{and}
  \bibinfo{author}{\bibfnamefont{M.}~\bibnamefont{Ulmke}},
  \bibinfo{journal}{Euro. Phys. J. B} \textbf{\bibinfo{volume}{45}},
  \bibinfo{pages}{449} (\bibinfo{year}{2005}).

\bibitem[{\citenamefont{Abou-Chacra et~al.}((1973))\citenamefont{Abou-Chacra,
  Anderson, and Thouless}}]{AAT73}
\bibinfo{author}{\bibfnamefont{R.}~\bibnamefont{Abou-Chacra}},
  \bibinfo{author}{\bibfnamefont{P.~W.} \bibnamefont{Anderson}},
  \bibnamefont{and} \bibinfo{author}{\bibfnamefont{D.~J.}
  \bibnamefont{Thouless}}, \bibinfo{journal}{J. Phys. C}
  \textbf{\bibinfo{volume}{6}}, \bibinfo{pages}{1734} (\bibinfo{year}{1973}).

\bibitem[{\citenamefont{Alvermann and Fehske}((2004))}]{AF04}
\bibinfo{author}{\bibfnamefont{A.}~\bibnamefont{Alvermann}} \bibnamefont{and}
  \bibinfo{author}{\bibfnamefont{H.}~\bibnamefont{Fehske}},
  \bibinfo{journal}{preprint cond-mat} \textbf{\bibinfo{volume}{0411516}}
  (\bibinfo{year}{2004}).

\bibitem[{\citenamefont{Dobrosavljevi\'c and Kotliar}((1997))}]{DK97}
\bibinfo{author}{\bibfnamefont{V.}~\bibnamefont{Dobrosavljevi\'c}}
  \bibnamefont{and} \bibinfo{author}{\bibfnamefont{G.}~\bibnamefont{Kotliar}},
  \bibinfo{journal}{Phys. Rev. Lett.} \textbf{\bibinfo{volume}{78}},
  \bibinfo{pages}{3943} (\bibinfo{year}{1997}).

\bibitem[{\citenamefont{Dobrosavljevi\'c and Kotliar}((1998))}]{DK98}
\bibinfo{author}{\bibfnamefont{V.}~\bibnamefont{Dobrosavljevi\'c}}
  \bibnamefont{and} \bibinfo{author}{\bibfnamefont{G.}~\bibnamefont{Kotliar}},
  \bibinfo{journal}{Phil. Trans. R. Soc. London A}
  \textbf{\bibinfo{volume}{356}}, \bibinfo{pages}{57} (\bibinfo{year}{1998}).

\bibitem[{\citenamefont{Bronold et~al.}((2004))\citenamefont{Bronold,
  Alvermann, and Fehske}}]{BAF04}
\bibinfo{author}{\bibfnamefont{F.~X.} \bibnamefont{Bronold}},
  \bibinfo{author}{\bibfnamefont{A.}~\bibnamefont{Alvermann}},
  \bibnamefont{and} \bibinfo{author}{\bibfnamefont{H.}~\bibnamefont{Fehske}},
  \bibinfo{journal}{Philos. Mag.} \textbf{\bibinfo{volume}{84}},
  \bibinfo{pages}{673} (\bibinfo{year}{2004}).

\bibitem[{\citenamefont{Kotliar et~al.}(2001)\citenamefont{Kotliar, Savrasov,
  P\'alsson, and Biroli}}]{KSPB01}
\bibinfo{author}{\bibfnamefont{G.}~\bibnamefont{Kotliar}},
  \bibinfo{author}{\bibfnamefont{S.~Y.} \bibnamefont{Savrasov}},
  \bibinfo{author}{\bibfnamefont{G.}~\bibnamefont{P\'alsson}},
  \bibnamefont{and} \bibinfo{author}{\bibfnamefont{G.}~\bibnamefont{Biroli}},
  \bibinfo{journal}{Phys. Rev. Lett.} \textbf{\bibinfo{volume}{87}},
  \bibinfo{pages}{186401} (\bibinfo{year}{2001}).

\bibitem[{\citenamefont{Lichtenstein and Katsnelson}((2000))}]{LK00}
\bibinfo{author}{\bibfnamefont{A.~I.} \bibnamefont{Lichtenstein}}
  \bibnamefont{and} \bibinfo{author}{\bibfnamefont{M.~I.}
  \bibnamefont{Katsnelson}}, \bibinfo{journal}{Phys. Rev. B}
  \textbf{\bibinfo{volume}{62}}, \bibinfo{pages}{R9283}
  (\bibinfo{year}{2000}).

\bibitem[{\citenamefont{Jarrell and Krishnamurthy}((2001))}]{JK01}
\bibinfo{author}{\bibfnamefont{M.}~\bibnamefont{Jarrell}} \bibnamefont{and}
  \bibinfo{author}{\bibfnamefont{H.~R.} \bibnamefont{Krishnamurthy}},
  \bibinfo{journal}{Phys. Rev. B} \textbf{\bibinfo{volume}{63}},
  \bibinfo{pages}{125102} (\bibinfo{year}{2001}).

\bibitem[{\citenamefont{Maier and Jarrell}((2002){\natexlab{a}})}]{MJ02b}
\bibinfo{author}{\bibfnamefont{T.~A.} \bibnamefont{Maier}} \bibnamefont{and}
  \bibinfo{author}{\bibfnamefont{M.}~\bibnamefont{Jarrell}},
  \bibinfo{journal}{Phys. Rev. Lett.} \textbf{\bibinfo{volume}{89}},
  \bibinfo{pages}{077001} (\bibinfo{year}{2002}{\natexlab{a}}).

\bibitem[{\citenamefont{Minh-Tien}((2006))}]{MT06}
\bibinfo{author}{\bibfnamefont{T.}~\bibnamefont{Minh-Tien}},
  \bibinfo{journal}{Phys. Rev. B} \textbf{\bibinfo{volume}{74}},
  \bibinfo{pages}{155121} (\bibinfo{year}{2006}).

\bibitem[{\citenamefont{Dobrosavljevi\'c
  et~al.}((2003))\citenamefont{Dobrosavljevi\'c, Pastor, and
  Nikoli\'c}}]{DPN03}
\bibinfo{author}{\bibfnamefont{V.}~\bibnamefont{Dobrosavljevi\'c}},
  \bibinfo{author}{\bibfnamefont{A.~A.} \bibnamefont{Pastor}},
  \bibnamefont{and} \bibinfo{author}{\bibfnamefont{B.~K.}
  \bibnamefont{Nikoli\'c}}, \bibinfo{journal}{Europhys. Lett.}
  \textbf{\bibinfo{volume}{62}}, \bibinfo{pages}{76} (\bibinfo{year}{2003}).

\bibitem[{\citenamefont{Byczuk et~al.}((2005))\citenamefont{Byczuk, Hofstetter,
  and Vollhardt}}]{BHV05}
\bibinfo{author}{\bibfnamefont{K.}~\bibnamefont{Byczuk}},
  \bibinfo{author}{\bibfnamefont{W.}~\bibnamefont{Hofstetter}},
  \bibnamefont{and}
  \bibinfo{author}{\bibfnamefont{D.}~\bibnamefont{Vollhardt}},
  \bibinfo{journal}{Phys. Rev. Lett.} \textbf{\bibinfo{volume}{94}},
  \bibinfo{pages}{056404} (\bibinfo{year}{2005}).

\bibitem[{\citenamefont{Byczuk}((2005))}]{Byc05}
\bibinfo{author}{\bibfnamefont{K.}~\bibnamefont{Byczuk}},
  \bibinfo{journal}{Phys. Rev. B} \textbf{\bibinfo{volume}{71}},
  \bibinfo{pages}{205105} (\bibinfo{year}{2005}).

\bibitem[{\citenamefont{Biroli et~al.}((2004))\citenamefont{Biroli, Parcollet,
  and Kotliar}}]{BPK04}
\bibinfo{author}{\bibfnamefont{G.}~\bibnamefont{Biroli}},
  \bibinfo{author}{\bibfnamefont{O.}~\bibnamefont{Parcollet}},
  \bibnamefont{and} \bibinfo{author}{\bibfnamefont{G.}~\bibnamefont{Kotliar}},
  \bibinfo{journal}{Phys. Rev. B} \textbf{\bibinfo{volume}{69}},
  \bibinfo{pages}{205108} (\bibinfo{year}{2004}).

\bibitem[{\citenamefont{Gros and Valenti}((1993))}]{GV93}
\bibinfo{author}{\bibfnamefont{C.}~\bibnamefont{Gros}} \bibnamefont{and}
  \bibinfo{author}{\bibfnamefont{R.}~\bibnamefont{Valenti}},
  \bibinfo{journal}{Phys. Rev. B} \textbf{\bibinfo{volume}{48}},
  \bibinfo{pages}{418} (\bibinfo{year}{1993}).

\bibitem[{\citenamefont{S\'en\'echal et~al.}(2000)\citenamefont{S\'en\'echal,
  P\'erez, and Pioro-Ladri\`ere}}]{SPPL00}
\bibinfo{author}{\bibfnamefont{D.}~\bibnamefont{S\'en\'echal}},
  \bibinfo{author}{\bibfnamefont{D.}~\bibnamefont{P\'erez}}, \bibnamefont{and}
  \bibinfo{author}{\bibfnamefont{M.}~\bibnamefont{Pioro-Ladri\`ere}},
  \bibinfo{journal}{Phys. Rev. Lett.} \textbf{\bibinfo{volume}{84}},
  \bibinfo{pages}{522} (\bibinfo{year}{2000}).

\bibitem[{\citenamefont{Dahnken et~al.}(2004)\citenamefont{Dahnken, Aichhorn,
  Hanke, Arrigoni, and Potthoff}}]{DAH+04}
\bibinfo{author}{\bibfnamefont{C.}~\bibnamefont{Dahnken}},
  \bibinfo{author}{\bibfnamefont{M.}~\bibnamefont{Aichhorn}},
  \bibinfo{author}{\bibfnamefont{W.}~\bibnamefont{Hanke}},
  \bibinfo{author}{\bibfnamefont{E.}~\bibnamefont{Arrigoni}}, \bibnamefont{and}
  \bibinfo{author}{\bibfnamefont{M.}~\bibnamefont{Potthoff}},
  \bibinfo{journal}{Phys. Rev. B} \textbf{\bibinfo{volume}{70}},
  \bibinfo{pages}{245110} (\bibinfo{year}{2004}).

\bibitem[{\citenamefont{Aichhorn et~al.}(2004)\citenamefont{Aichhorn, Evertz,
  von~der Linden, and Potthoff}}]{AEvdLP+04}
\bibinfo{author}{\bibfnamefont{M.}~\bibnamefont{Aichhorn}},
  \bibinfo{author}{\bibfnamefont{H.~G.} \bibnamefont{Evertz}},
  \bibinfo{author}{\bibfnamefont{W.}~\bibnamefont{von~der Linden}},
  \bibnamefont{and} \bibinfo{author}{\bibfnamefont{M.}~\bibnamefont{Potthoff}},
  \bibinfo{journal}{Phys. Rev. B} \textbf{\bibinfo{volume}{70}},
  \bibinfo{pages}{235107} (\bibinfo{year}{2004}).

\bibitem[{\citenamefont{Negele and Orland}(1988)}]{NO88}
\bibinfo{author}{\bibfnamefont{J.~W.} \bibnamefont{Negele}} \bibnamefont{and}
  \bibinfo{author}{\bibfnamefont{H.}~\bibnamefont{Orland}},
  \emph{\bibinfo{title}{Quantum Many-Particle Systems}}
  (\bibinfo{publisher}{Addison-Wesley}, \bibinfo{address}{Redwood City},
  \bibinfo{year}{1988}).

\bibitem[{\citenamefont{Luttinger and Ward}(1960)}]{LW60}
\bibinfo{author}{\bibfnamefont{J.~M.} \bibnamefont{Luttinger}}
  \bibnamefont{and} \bibinfo{author}{\bibfnamefont{J.~C.} \bibnamefont{Ward}},
  \bibinfo{journal}{Phys. Rev.} \textbf{\bibinfo{volume}{118}},
  \bibinfo{pages}{1417} (\bibinfo{year}{1960}).

\bibitem[{\citenamefont{Potthoff}((2006))}]{Pot06}
\bibinfo{author}{\bibfnamefont{M.}~\bibnamefont{Potthoff}},
  \emph{\bibinfo{title}{{\rm In:} Effective models for low-dimensional strongly
  correlated systems}}, Ed. by G. Batrouni and D. Poilblanc
  (\bibinfo{publisher}{AIP proceedings}, \bibinfo{address}{Melville},
  \bibinfo{year}{2006}).

\bibitem[{\citenamefont{Potthoff and Nolting}((1999){\natexlab{a}})}]{PN99c}
\bibinfo{author}{\bibfnamefont{M.}~\bibnamefont{Potthoff}} \bibnamefont{and}
  \bibinfo{author}{\bibfnamefont{W.}~\bibnamefont{Nolting}},
  \bibinfo{journal}{Phys. Rev. B} \textbf{\bibinfo{volume}{59}},
  \bibinfo{pages}{2549} (\bibinfo{year}{1999}{\natexlab{a}}).

\bibitem[{\citenamefont{Potthoff and Nolting}((1999){\natexlab{b}})}]{PN99a}
\bibinfo{author}{\bibfnamefont{M.}~\bibnamefont{Potthoff}} \bibnamefont{and}
  \bibinfo{author}{\bibfnamefont{W.}~\bibnamefont{Nolting}},
  \bibinfo{journal}{Euro. Phys. J. B} \textbf{\bibinfo{volume}{8}},
  \bibinfo{pages}{555} (\bibinfo{year}{1999}{\natexlab{b}}).

\bibitem[{\citenamefont{Potthoff and Nolting}((1999){\natexlab{c}})}]{PN99d}
\bibinfo{author}{\bibfnamefont{M.}~\bibnamefont{Potthoff}} \bibnamefont{and}
  \bibinfo{author}{\bibfnamefont{W.}~\bibnamefont{Nolting}},
  \bibinfo{journal}{Phys. Rev. B} \textbf{\bibinfo{volume}{60}},
  \bibinfo{pages}{7834} (\bibinfo{year}{1999}{\natexlab{c}}).

\bibitem[{\citenamefont{Aichhorn
  et~al.}((2006){\natexlab{a}})\citenamefont{Aichhorn, Arrigoni, Potthoff, and
  Hanke}}]{AAPH06a}
\bibinfo{author}{\bibfnamefont{M.}~\bibnamefont{Aichhorn}},
  \bibinfo{author}{\bibfnamefont{E.}~\bibnamefont{Arrigoni}},
  \bibinfo{author}{\bibfnamefont{M.}~\bibnamefont{Potthoff}}, \bibnamefont{and}
  \bibinfo{author}{\bibfnamefont{W.}~\bibnamefont{Hanke}},
  \bibinfo{journal}{Phys. Rev. B} \textbf{\bibinfo{volume}{74}},
  \bibinfo{pages}{024508} (\bibinfo{year}{2006}{\natexlab{a}}).

\bibitem[{\citenamefont{Haydock and Mookerjee}((1974))}]{HM74}
\bibinfo{author}{\bibfnamefont{R.}~\bibnamefont{Haydock}} \bibnamefont{and}
  \bibinfo{author}{\bibfnamefont{A.}~\bibnamefont{Mookerjee}},
  \bibinfo{journal}{J. Phys. C} \textbf{\bibinfo{volume}{7}},
  \bibinfo{pages}{3001} (\bibinfo{year}{1974}).

\bibitem[{\citenamefont{Bulla and Potthoff}((2000))}]{BP00}
\bibinfo{author}{\bibfnamefont{R.}~\bibnamefont{Bulla}} \bibnamefont{and}
  \bibinfo{author}{\bibfnamefont{M.}~\bibnamefont{Potthoff}},
  \bibinfo{journal}{Euro. Phys. J. B} \textbf{\bibinfo{volume}{13}},
  \bibinfo{pages}{257} (\bibinfo{year}{2000}).

\bibitem[{\citenamefont{Potthoff}((2001))}]{Pot01}
\bibinfo{author}{\bibfnamefont{M.}~\bibnamefont{Potthoff}},
  \bibinfo{journal}{Phys. Rev. B} \textbf{\bibinfo{volume}{64}},
  \bibinfo{pages}{165114} (\bibinfo{year}{2001}).

\bibitem[{\citenamefont{Inaba et~al.}((2005){\natexlab{a}})\citenamefont{Inaba,
  Koga, Suga, and Kawakami}}]{IKSK05a}
\bibinfo{author}{\bibfnamefont{K.}~\bibnamefont{Inaba}},
  \bibinfo{author}{\bibfnamefont{A.}~\bibnamefont{Koga}},
  \bibinfo{author}{\bibfnamefont{S.-I.} \bibnamefont{Suga}}, \bibnamefont{and}
  \bibinfo{author}{\bibfnamefont{N.}~\bibnamefont{Kawakami}},
  \bibinfo{journal}{Phys. Rev. B} \textbf{\bibinfo{volume}{72}},
  \bibinfo{pages}{085112} (\bibinfo{year}{2005}{\natexlab{a}}).

\bibitem[{\citenamefont{Inaba et~al.}((2005){\natexlab{b}})\citenamefont{Inaba,
  Koga, Suga, and Kawakami}}]{IKSK05b}
\bibinfo{author}{\bibfnamefont{K.}~\bibnamefont{Inaba}},
  \bibinfo{author}{\bibfnamefont{A.}~\bibnamefont{Koga}},
  \bibinfo{author}{\bibfnamefont{S.-I.} \bibnamefont{Suga}}, \bibnamefont{and}
  \bibinfo{author}{\bibfnamefont{N.}~\bibnamefont{Kawakami}},
  \bibinfo{journal}{J. Phys. Soc. Jpn.} \textbf{\bibinfo{volume}{74}},
  \bibinfo{pages}{2393} (\bibinfo{year}{2005}{\natexlab{b}}).

\bibitem[{\citenamefont{Balzer and Potthoff}((2005))}]{BP05}
\bibinfo{author}{\bibfnamefont{M.}~\bibnamefont{Balzer}} \bibnamefont{and}
  \bibinfo{author}{\bibfnamefont{M.}~\bibnamefont{Potthoff}},
  \bibinfo{journal}{Physica B} \textbf{\bibinfo{volume}{359-361}},
  \bibinfo{pages}{768} (\bibinfo{year}{2005}).

\bibitem[{\citenamefont{Pozgajcic}((2004))}]{Poz04}
\bibinfo{author}{\bibfnamefont{K.}~\bibnamefont{Pozgajcic}},
  \bibinfo{journal}{preprint cond-mat} \textbf{\bibinfo{volume}{0407172}}
  (\bibinfo{year}{2004}).

\bibitem[{\citenamefont{Schwartz and Siggia}((1972))}]{SS72}
\bibinfo{author}{\bibfnamefont{L.}~\bibnamefont{Schwartz}} \bibnamefont{and}
  \bibinfo{author}{\bibfnamefont{E.}~\bibnamefont{Siggia}},
  \bibinfo{journal}{Phys. Rev. B} \textbf{\bibinfo{volume}{5}},
  \bibinfo{pages}{383} (\bibinfo{year}{1972}).

\bibitem[{\citenamefont{Hubbard}((1964))}]{Hub64b}
\bibinfo{author}{\bibfnamefont{J.}~\bibnamefont{Hubbard}},
  \bibinfo{journal}{Proc. R. Soc. London A} \textbf{\bibinfo{volume}{281}},
  \bibinfo{pages}{401} (\bibinfo{year}{1964}).

\bibitem[{\citenamefont{Herrmann and Nolting}((1996))}]{HN96}
\bibinfo{author}{\bibfnamefont{T.}~\bibnamefont{Herrmann}} \bibnamefont{and}
  \bibinfo{author}{\bibfnamefont{W.}~\bibnamefont{Nolting}},
  \bibinfo{journal}{Phys. Rev. B} \textbf{\bibinfo{volume}{53}},
  \bibinfo{pages}{10579} (\bibinfo{year}{1996}).

\bibitem[{\citenamefont{Hirooka and Shimizu}((1977))}]{HS77}
\bibinfo{author}{\bibfnamefont{S.}~\bibnamefont{Hirooka}} \bibnamefont{and}
  \bibinfo{author}{\bibfnamefont{M.}~\bibnamefont{Shimizu}},
  \bibinfo{journal}{J. Phys. Soc. Jpn.} \textbf{\bibinfo{volume}{43}},
  \bibinfo{pages}{70} (\bibinfo{year}{1977}).

\bibitem[{\citenamefont{Jani\u{s}}((1989))}]{Jan89}
\bibinfo{author}{\bibfnamefont{V.}~\bibnamefont{Jani\u{s}}},
  \bibinfo{journal}{Phys. Rev. B} \textbf{\bibinfo{volume}{40}},
  \bibinfo{pages}{11331} (\bibinfo{year}{1989}).

\bibitem[{\citenamefont{Kakehashi}((1992))}]{Kak92}
\bibinfo{author}{\bibfnamefont{Y.}~\bibnamefont{Kakehashi}},
  \bibinfo{journal}{Phys. Rev. B} \textbf{\bibinfo{volume}{45}},
  \bibinfo{pages}{7196} (\bibinfo{year}{1992}).

\bibitem[{\citenamefont{Kakehashi}((2002))}]{Kak02}
\bibinfo{author}{\bibfnamefont{Y.}~\bibnamefont{Kakehashi}},
  \bibinfo{journal}{Phys. Rev. B} \textbf{\bibinfo{volume}{66}},
  \bibinfo{pages}{104428} (\bibinfo{year}{2002}).

\bibitem[{\citenamefont{Brandt and Mielsch}((1989))}]{BM89}
\bibinfo{author}{\bibfnamefont{U.}~\bibnamefont{Brandt}} \bibnamefont{and}
  \bibinfo{author}{\bibfnamefont{C.}~\bibnamefont{Mielsch}},
  \bibinfo{journal}{Z. Phys. B} \textbf{\bibinfo{volume}{75}},
  \bibinfo{pages}{365} (\bibinfo{year}{1989}).

\bibitem[{\citenamefont{Brandt and Mielsch}((1990))}]{BM90}
\bibinfo{author}{\bibfnamefont{U.}~\bibnamefont{Brandt}} \bibnamefont{and}
  \bibinfo{author}{\bibfnamefont{C.}~\bibnamefont{Mielsch}},
  \bibinfo{journal}{Z. Phys. B} \textbf{\bibinfo{volume}{79}},
  \bibinfo{pages}{295} (\bibinfo{year}{1990}).

\bibitem[{\citenamefont{Brandt and Mielsch}((1991))}]{BM91}
\bibinfo{author}{\bibfnamefont{U.}~\bibnamefont{Brandt}} \bibnamefont{and}
  \bibinfo{author}{\bibfnamefont{C.}~\bibnamefont{Mielsch}},
  \bibinfo{journal}{Z. Phys. B} \textbf{\bibinfo{volume}{82}},
  \bibinfo{pages}{37} (\bibinfo{year}{1991}).

\bibitem[{\citenamefont{Gutberlet and Potthoff}((2006))}]{GP06}
\bibinfo{author}{\bibfnamefont{M.}~\bibnamefont{Gutberlet}} \bibnamefont{and}
  \bibinfo{author}{\bibfnamefont{M.}~\bibnamefont{Potthoff}},
  \bibinfo{journal}{unpublished}.

\bibitem[{\citenamefont{S\'en\'echal et~al.}((2005))\citenamefont{S\'en\'echal,
  Lavertu, Marois, and Tremblay}}]{SLMT05}
\bibinfo{author}{\bibfnamefont{D.}~\bibnamefont{S\'en\'echal}},
  \bibinfo{author}{\bibfnamefont{P.-L.} \bibnamefont{Lavertu}},
  \bibinfo{author}{\bibfnamefont{M.-A.} \bibnamefont{Marois}},
  \bibnamefont{and} \bibinfo{author}{\bibfnamefont{A.-M.~S.}
  \bibnamefont{Tremblay}}, \bibinfo{journal}{Phys. Rev. Lett.}
  \textbf{\bibinfo{volume}{94}}, \bibinfo{pages}{156404}
  (\bibinfo{year}{2005}).

\bibitem[{\citenamefont{Tsukada}((1972))}]{Tsu72}
\bibinfo{author}{\bibfnamefont{M.}~\bibnamefont{Tsukada}}, \bibinfo{journal}{J.
  Phys. Soc. Jpn.} \textbf{\bibinfo{volume}{32}}, \bibinfo{pages}{1475}
  (\bibinfo{year}{1972}).

\bibitem[{\citenamefont{Zacher et~al.}((2000))\citenamefont{Zacher, Eder,
  Arrigoni, and Hanke}}]{ZEAH00}
\bibinfo{author}{\bibfnamefont{M.~G.} \bibnamefont{Zacher}},
  \bibinfo{author}{\bibfnamefont{R.}~\bibnamefont{Eder}},
  \bibinfo{author}{\bibfnamefont{E.}~\bibnamefont{Arrigoni}}, \bibnamefont{and}
  \bibinfo{author}{\bibfnamefont{W.}~\bibnamefont{Hanke}},
  \bibinfo{journal}{Phys. Rev. Lett.} \textbf{\bibinfo{volume}{85}},
  \bibinfo{pages}{2585} (\bibinfo{year}{2000}).

\bibitem[{\citenamefont{S\'en\'echal et~al.}(2002)\citenamefont{S\'en\'echal,
  P\'erez, and Plouffe}}]{SPP02}
\bibinfo{author}{\bibfnamefont{D.}~\bibnamefont{S\'en\'echal}},
  \bibinfo{author}{\bibfnamefont{D.}~\bibnamefont{P\'erez}}, \bibnamefont{and}
  \bibinfo{author}{\bibfnamefont{D.}~\bibnamefont{Plouffe}},
  \bibinfo{journal}{Phys. Rev. B} \textbf{\bibinfo{volume}{66}},
  \bibinfo{pages}{075129} (\bibinfo{year}{2002}).

\bibitem[{\citenamefont{Koller and Dupuis}((2005))}]{KD05}
\bibinfo{author}{\bibfnamefont{W.}~\bibnamefont{Koller}} \bibnamefont{and}
  \bibinfo{author}{\bibfnamefont{N.}~\bibnamefont{Dupuis}},
  \bibinfo{journal}{preprint cond-mat} \textbf{\bibinfo{volume}{0511294}}
  (\bibinfo{year}{2005}).

\bibitem[{\citenamefont{Hettler et~al.}((1998))\citenamefont{Hettler,
  Tahvildar-Zadeh, Jarrell, Pruschke, and Krishnamurthy}}]{HTZ+98}
\bibinfo{author}{\bibfnamefont{M.~H.} \bibnamefont{Hettler}},
  \bibinfo{author}{\bibfnamefont{A.~N.} \bibnamefont{Tahvildar-Zadeh}},
  \bibinfo{author}{\bibfnamefont{M.}~\bibnamefont{Jarrell}},
  \bibinfo{author}{\bibfnamefont{T.}~\bibnamefont{Pruschke}}, \bibnamefont{and}
  \bibinfo{author}{\bibfnamefont{H.~R.} \bibnamefont{Krishnamurthy}},
  \bibinfo{journal}{Phys. Rev. B} \textbf{\bibinfo{volume}{58}},
  \bibinfo{pages}{R7475} (\bibinfo{year}{1998}).

\bibitem[{\citenamefont{Maier et~al.}((2000))\citenamefont{Maier, Jarrell,
  Pruschke, and Keller}}]{MJPK00}
\bibinfo{author}{\bibfnamefont{T.}~\bibnamefont{Maier}},
  \bibinfo{author}{\bibfnamefont{M.}~\bibnamefont{Jarrell}},
  \bibinfo{author}{\bibfnamefont{T.}~\bibnamefont{Pruschke}}, \bibnamefont{and}
  \bibinfo{author}{\bibfnamefont{J.}~\bibnamefont{Keller}},
  \bibinfo{journal}{Euro. Phys. J. B} \textbf{\bibinfo{volume}{13}},
  \bibinfo{pages}{613} (\bibinfo{year}{2000}).

\bibitem[{\citenamefont{Hettler et~al.}((2000))\citenamefont{Hettler,
  Mukherjee, Jarrell, and Krishnamurthy}}]{HMJK00}
\bibinfo{author}{\bibfnamefont{M.~H.} \bibnamefont{Hettler}},
  \bibinfo{author}{\bibfnamefont{M.}~\bibnamefont{Mukherjee}},
  \bibinfo{author}{\bibfnamefont{M.}~\bibnamefont{Jarrell}}, \bibnamefont{and}
  \bibinfo{author}{\bibfnamefont{H.~R.} \bibnamefont{Krishnamurthy}},
  \bibinfo{journal}{Phys. Rev. B} \textbf{\bibinfo{volume}{61}},
  \bibinfo{pages}{12739} (\bibinfo{year}{2000}).

\bibitem[{\citenamefont{Maier and Jarrell}((2002){\natexlab{b}})}]{MJ02a}
\bibinfo{author}{\bibfnamefont{T.~A.} \bibnamefont{Maier}} \bibnamefont{and}
  \bibinfo{author}{\bibfnamefont{M.}~\bibnamefont{Jarrell}},
  \bibinfo{journal}{Phys. Rev. B} \textbf{\bibinfo{volume}{65}},
  \bibinfo{pages}{041104(R)} (\bibinfo{year}{2002}{\natexlab{b}}).

\bibitem[{\citenamefont{Biroli and Kotliar}((2002))}]{BK02}
\bibinfo{author}{\bibfnamefont{G.}~\bibnamefont{Biroli}} \bibnamefont{and}
  \bibinfo{author}{\bibfnamefont{G.}~\bibnamefont{Kotliar}},
  \bibinfo{journal}{Phys. Rev. B} \textbf{\bibinfo{volume}{65}},
  \bibinfo{pages}{155112} (\bibinfo{year}{2002}).

\bibitem[{\citenamefont{Aryanpour et~al.}((2005))\citenamefont{Aryanpour,
  Maier, and Jarrell}}]{AMJ05}
\bibinfo{author}{\bibfnamefont{K.}~\bibnamefont{Aryanpour}},
  \bibinfo{author}{\bibfnamefont{T.~A.} \bibnamefont{Maier}}, \bibnamefont{and}
  \bibinfo{author}{\bibfnamefont{M.}~\bibnamefont{Jarrell}},
  \bibinfo{journal}{Phys. Rev. B} \textbf{\bibinfo{volume}{71}},
  \bibinfo{pages}{037101} (\bibinfo{year}{2005}).

\bibitem[{\citenamefont{Biroli and Kotliar}((2005))}]{BK05}
\bibinfo{author}{\bibfnamefont{G.}~\bibnamefont{Biroli}} \bibnamefont{and}
  \bibinfo{author}{\bibfnamefont{G.}~\bibnamefont{Kotliar}},
  \bibinfo{journal}{Phys. Rev. B} \textbf{\bibinfo{volume}{71}},
  \bibinfo{pages}{037102} (\bibinfo{year}{2005}).

\bibitem[{\citenamefont{Tong}((2005))}]{Ton05}
\bibinfo{author}{\bibfnamefont{N.-H.} \bibnamefont{Tong}},
  \bibinfo{journal}{Phys. Rev. B} \textbf{\bibinfo{volume}{72}},
  \bibinfo{pages}{115104} (\bibinfo{year}{2005}).

\bibitem[{\citenamefont{Aichhorn
  et~al.}((2006){\natexlab{b}})\citenamefont{Aichhorn, Arrigoni, Potthoff, and
  Hanke}}]{AAPH06b}
\bibinfo{author}{\bibfnamefont{M.}~\bibnamefont{Aichhorn}},
  \bibinfo{author}{\bibfnamefont{E.}~\bibnamefont{Arrigoni}},
  \bibinfo{author}{\bibfnamefont{M.}~\bibnamefont{Potthoff}}, \bibnamefont{and}
  \bibinfo{author}{\bibfnamefont{W.}~\bibnamefont{Hanke}},
  \bibinfo{journal}{preprint cond-mat} \textbf{\bibinfo{volume}{0607271}}
  (\bibinfo{year}{2006}{\natexlab{b}}).

\end{thebibliography}
\end{document}